\documentclass[article]{JHEP3}
\usepackage{amssymb,amsfonts,bm}
\usepackage{epsfig}
\usepackage{amsmath}
\usepackage{amssymb,amsfonts}
\renewcommand{\theequation}{\arabic{section}.\arabic{equation}}
\def\II{\relax{\rm I\kern-.18em I}}
\def\be{\begin{equation}}
\def\ee{\end{equation}}
\def\bs{\begin{subequations}}
\def\es{\end{subequations}}
\def\bc{\begin{center}}
\def\ec{\end{center}}

\def\lab{\label}

\newcommand{\<}{\langle}
\renewcommand{\>}{\rangle}
\newcommand{\een}{\end{subequations}}
\newcommand{\ben}{\begin{subequations}}
\def\beq{\begin{equation}}
\def\eeq{\end{equation}}

\def\hre#1#2{\href{http://arxiv.org/abs/#1/#2}{[ArXiv:#1/#2]}}

\def\hri#1#2{\href{http://arxiv.org/abs/#1}{[ArXiv:#1]#2}}

\def\g{\gamma}

\def\l{\lambda}
\def\m{\mu}
\def\n{\nu}

\def\a{\alpha}
\def\b{\beta}

\def\d{\partial}
\def\p{\partial}

\def\sp{\;\;\;,\;\;\;}

\def\e{\epsilon}

\def\pa{\partial}

\newcommand\fverb{\setbox\pippobox=\hbox\bgroup\verb}
\newcommand\fverbdo{\egroup\medskip\noindent%
                        \fbox{\unhbox\pippobox}\ }
\newcommand\fverbit{\egroup\item[\fbox{\unhbox\pippobox}]}
\newbox\pippobox

\def\beq{\begin{equation}}
\def\eeq{\end{equation}}
\newcommand{\bea}{\begin{eqnarray}}
\newcommand{\eea}{\end{eqnarray}}
\def\nn{\nonumber}
\def\d{\delta}

\def\hre#1#2{\href{http://arxiv.org/abs/#1/#2}{[ArXiv:#1/#2]}}

\newcommand{\f}{{h}} 

\title{Holographic Models for QCD in the Veneziano Limit}
\author{Matti J\"arvinen$^a$ and Elias Kiritsis$^{a,b}$\\
~\\
$^a$ \href{http://hep.physics.uoc.gr}{Crete Center for Theoretical Physics},
Department of Physics, University of Crete, 71003 Heraklion, Greece\\
~\\
$^b$ \href{http://www.apc.univ-paris7.fr}{APC, Universit\'e Paris 7}, \\
B\^atiment Condorcet, F-75205, Paris Cedex 13, France (UMR du CNRS 7164).}

\preprint{CCTP-2011-30  }

\abstract{We construct a class of bottom-up holographic models with physics
comparable
to the one expected from QCD in the Veneziano limit of large $N_f$ and $N_c$
with fixed $x={N_f\over N_c}$. The models
capture the holographic dynamics of the dilaton (dual to the YM coupling)
 and a tachyon (dual to the chiral condensate), and are parametrized
by the real parameter $x$, which can take values within the range $0\leq x <
{11\over 2}$.
We analyze the saddle point solutions, and draw the phase diagram at zero
temperature and density. 
The backreaction of flavor on the glue is fully included.
We find the conformal window for $x\geq x_c$, and the
QCD-like phase with chiral symmetry
breaking at $x<x_c$, where the critical value $x_c$ lies close to four.
We also find
Miransky scaling as $x\to x_c$ as well as Efimov-like saddle points.
By calculating the holographic $\beta$-functions, we demonstrate the ``walking''
behavior of the coupling in the region near and below $x_c$.     }

\begin{document}

\section{Introduction}

Holographic techniques have been used recently in order to understand the strong
dynamics of gauge theories. QCD is one of the obvious targets of such a program,
but strongly coupled gauge theories may also emerge in other contexts, namely in
the physics beyond the Standard model (non-perturbative electroweak symmetry
breaking is an example), as well as in condensed matter contexts.

An interesting mild generalization of QCD, involves $SU(N_c)$ YM coupled to
$N_f$ Dirac fermions transforming in the fundamental representation (that we
will still call quarks).
This is a theory, that can be studied in the $(N_c,N_f)$ plane.

As usual simplifications arise in the large-$N_c$ limit. The standard 't Hooft
large-$N_c$ limit, \cite{hooft},  lets $N_c\to \infty$ keeping $N_f$ and
$\lambda=g_\mathrm{YM}^2N_c$ finite. In this limit,  the effect of the quarks
are suppressed by powers of ${N_f\over N_c}\to 0$, and therefore it corresponds
to the ``quenched'' limit.
In particular interesting dynamical effects as the conformal window, and exotic
phases at finite density, driven by the presence of the quarks are not expected
to be visible in the 't Hooft  large-$N_c$ limit.

In \cite{v} Veneziano introduced an alternative large-$N_c$ limit in which
\be
N_c\to \infty\sp N_f\to \infty\sp {N_f\over N_c}=x~~~~{\rm fixed}\sp
\lambda=g_\mathrm{YM}^2N_c~~~~{\rm fixed}
\label{v1}\ee
in order to make the chiral U(1) anomaly visible to leading order in the $1/N_c$
expansion.
This is the large-$N_c$ limit we will study in this paper.

There are several interesting issues that are accessible in the Veneziano limit.
\begin{itemize}

\item The ``conformal window'' with an IR fixed point. The window extends from
$x={11\over 2}$ to smaller values of $x$, and includes the  Banks-Zaks (BZ)
weakly-coupled region as $x\to {11\over 2}$ \cite{bz}.

\item The phase transition at a critical $x=x_c$ from the conformal window to
theories with chiral symmetry breaking in the IR.

\item A transition region near and below $x_c$, where the theory is expected to
exhibit ``walking behavior''. The theory flows towards the IR fixed point but
misses it ending up with chiral symmetry breaking, so that the coupling constant
varies slowly over a long range of energies.

\item New phenomena at finite density, involving color superconductivity
\cite{cs} and flavor-color locking \cite{cfl}.

\end{itemize}

The point $x=x_c$ is the point of a quantum phase transition where the theory
passes from an IR Conformal theory (above $x_c$) to a theory with a non-trivial
chiral condensate (below $x_c$). Such transitions were termed conformal phase
transitions in \cite{conf}, and have been recently argued to be due to the
fusion of a UV and IR fixed points \cite{Kaplan}.
The scaling of the condensate is similar to that of the 2D
Berezinskii-Kosterlitz-Thouless (BKT) transition
\cite{Kosterlitz:1974sm}, and is also known as Miransky scaling,
\cite{Miransky}.
Several such quantum phase transitions have been described recently in
holographic theories in \cite{Jensen,BKT,kutasov}.

The location of the lower edge of the conformal window is determined by
nonperturbative dynamics. Several estimates of the value for $x_c$, and for the
critical value of $N_f$ at finite $N_c$ have been put forward by using different
methods \cite{ds,cw,antipin}, and the boundary of the conformal window is also
being studied actively on the lattice (see, e.g., \cite{lattice}).

The walking region near $x_c$ as well as the value of $x_c$, have been of
interest for a while, due to their potential relevance for the realization of
walking technicolor, \cite{holdom}.
Technicolor has been used as a generic name for non-perturbative electroweak
symmetry breaking mimicking the one induced by QCD,
\cite{review}. Although non-perturbative effects in a new  strongly coupled
gauge theory can induce electroweak symmetry breaking, generating the quark and
lepton masses is an extra problem. A new interaction (extended technicolor) is
usually invoked to generate the requisite couplings. However their magnitude
(and therefore the Standard model masses) are controlled by the dimension of the
scalar operator that breaks the electroweak symmetry. In free field theory its
dimension is 3, as it is a fermion bilinear. However, for the SM masses to have
realistic values the dimension must be reduced to at least around 2, i.e., the
anomalous dimension of the operator should be at least one. This is indeed
expected to be generated by a ``walking'' theory \cite{ds,Holdom2}. Notice that
coupling between the technicolor sector and the standard model may decrease the
dimension substantially \cite{fukano}.

There are issues that so far have made such non-perturbative approaches to be in
apparent conflict with data, like the value of the $S$ parameter, \cite{peskin}.
It was argued that in cases where the strongly coupled theory is near a
conformal transition the $S$-parameter can be quite different potentially
evading the experimental constraints\footnote{Recent studies that extrapolate
from the Banks-Zaks region
suggest that the modification of the S-parameter near the conformal window may
be modest, \cite{sannino2}.}, \cite{sannino}.

There have been several bottom-up models of technicolor \cite{butc},  mostly
inspired from the hard wall models for mesons, \cite{hard}.
Lately there have also been top-down holographic models of walking behavior,
\cite{nunez,pare,angel,kutasov}. They use several contexts like favored MN solutions,
\cite{nunez}, $D_7-\overline{D_7}$ pairs, \cite{angel} or $D_3-D_7$ systems,
\cite{kutasov}.

A theory that can be compared is
 $\mathcal{N}=1$ supersymmetric QCD with $N_f$ flavors.
The ground states of this theory have
been found by Seiberg, \cite{seiberg}, and we understand several issues
associated to low energy dynamics, including the Seiberg duality.
Such a theory gives already several important hints on the structure expected in
non-supersymmetric QCD, \cite{strassler}.
Defining again $x={N_f\over N_c}$, we have the following regimes:

\begin{itemize}

\item At $x=0$ the theory has confinement, a mass gap and $N_c$ distinct
 vacua associated with a spontaneous braking of the leftover R
symmetry $Z_{N_c}$.

\item At $0<x<1$, the theory has a runaway ground state.

\item At $x=1$, the theory has a quantum moduli space with no singularity.
 This reflects confinement with chiral symmetry breaking.

\item At $x=1+{1\over N_c} $, the moduli space is classical (and singular).
 The theory confines, but there is no chiral symmetry breaking.

\item At $1+{2\over N_c}<x<{3\over 2}$ the theory is in the non-abelian
magnetic IR-free phase, with the magnetic gauge group $SU(N_f-N_c)$ IR free.

\item At ${3\over 2}<x<3$, the theory flows to a CFT in the IR. Near $x=3$
 this is the Banks-Zaks region where the original theory has an IR fixed point
at weak coupling. Moving to lower values, the coupling of the IR $SU(N_c)$
 gauge theory grows. However near $x={3\over 2}$ the dual magnetic $SU(N_f-N_c)$
is in its Banks-Zaks region, and provides a weakly coupled description of
 the IR fixed point theory.

\item At $x>3$, the theory is IR free.

\end{itemize}

The region ${3\over 2}<x<3$ is comparable to the conformal window that is
 expected in (non-supersymmetric) QCD.
Indeed, the IR coupling of the original gauge theory
is becoming stronger as $x$ decreases, but for x above $3/2$, a new set
of IR states becomes weakly coupled, namely the magnetic gluons and quarks.
These states have been interpreted as the $\rho$ mesons and their supersymmetric
 avatars, \cite{komar}. They are massless and weakly coupled in this region.
The regime $1+{2\over N_c}<x<{3\over 2}$ does not seem to have an analogue
 in QCD. The IR theory is again an IR non-abelian gauge theory,
  therefore trivially scale
invariant in the far IR, but also free.

It was suggested already in \cite{KM,BCCKP} that the presence of a conformal
 window in ${\cal N}=1$ supersymmetric QCD is associated with 
the violation of the Breitenlohner-Freedman (BF) bound. 
A recent attempt to describe related physics was done in \cite{russo}.

\subsection{Bottom-up models for QCD in the quenched approximation}

To construct a bottom-up holographic model for QCD in the Veneziano
limit\footnote{We will call such a model V-QCD from now on.} we need to first
understand pure YM. The simplest bottom model for pure YM in four dimensions is
the hard-wall model, first introduced in \cite{ps}. Despite its simplicity it
could capture a few qualitative features of the strong interaction.
More sophisticated models accounted for the running of the YM coupling constant,
incorporating therefore the dilaton into the gravitational action,
\cite{ihqcd,ihreview}.
By simply adjusting a dilaton potential they could exhibit many of the
properties of large-$N_c$ YM including confinement, a mas gap, asymptotic linear
trajectories and realistic glueball spectra at zero temperature. Moreover they
fared rather well at finite temperature, \cite{gkmn}, and after the tuning of
two phenomenological parameters in the dilaton potential\footnote{The bottom-up
Einstein-dilaton model for large-$N_c$ YM was termed Improved Holographic QCD
(IHQCD).}, \cite{data}, they could agree with lattice data both at zero and
finite temperature, \cite{panero}.
The properties of IHQCD at finite temperature were further explored
in \cite{kajantieFT}.
Alternative Einstein-dilaton models exhibiting a cross-over rather than a first
order deconfining transition, and matching YM finite temperature dynamics were
also developed in \cite{gubser}.
Such bottom up models,  were used to compute transport properties of YM, like
the bulk viscosity and the diffusion properties of heavy quarks,
\cite{transport}.  Backgrounds having an IR fixed point or a ``walking'' region,
where the system flows close to an fixed point, were studied within in the IHQCD
model in \cite{antipin,Jarvinen:2009fe,kajantieIRFP,kajantie}.
In these studies the fixed
point was introduced via the input beta function, without proper modeling of the
dynamics of the quarks, even though $N_f/N_c$ was large.

To go beyond YM a new ingredient is needed, namely the flavor branes. An
important field in this context is the order parameter for chiral symmetry
breaking, dual to a complex bifundamental field $T$. In the hard wall
\cite{hard} and soft wall \cite{soft} models for mesons, such a field was added
using a quadratic action, and chiral symmetry breaking proceeded by giving to
such a field a vev by hand.

In \cite{ckp} it was remarked that as the flavor sector of gauge theories in
string theory arises from D-brane-antibrane pairs, the bifundamental field $T$
could be naturally be identified with the brane-antibrane tachyon field that
had been studied profusely (around flat space) in string theory by Sen and
others, \cite{senreview}. The non-linear action proposed by Sen, \cite{sen}
could be therefore used as a well-motivated starting point in order to study the
holographic dynamics of chiral symmetry breaking. Several general features of
this approach were explored in \cite{ckp},
\begin{itemize}

\item Chiral symmetry breaking is dynamical and is induced/controlled by the
tachyon Dirac-Born-Infeld (DBI) action.

\item Confining asymptotics of the geometry were shown to trigger chiral
symmetry breaking.

\item A Gell-Mann-Oakes-Renner relation is generically satisfied.

\item The Sen DBI tachyon action induces linear Regge trajectories or mesons.

\item The Wess-Zumino (WZ) terms of the tachyon action, computed in string theory
\cite{tach1}-\cite{TTU}, produce the appropriate flavor anomalies, include the
axial $U(1)$ anomaly and $\eta'$-mixing, and implement a holographic version of
the Coleman-Witten theorem.

\end{itemize}

In the context above, the analysis was done in the quenched approximation: the
flavor sector does not backreact on the metric and dilaton. 
Similar results were also obtained by considering tachyon condensation in 
the Sakai-Sugimoto model \cite{sstachyon}. 

In \cite{ikp} an implementation of these ideas was performed by choosing a
concrete confining background, that is simple and asymptotically AdS. This was
the Kuperstein-Sonnenschein background, \cite{ks2},  with a constant dilaton and
an AdS$_6$ soliton. In this background the tachyon DBI action was analyzed with
the following results
\begin{itemize}

\item The model incorporates  confinement in the sense that the quark-antiquark
potential
computed with the usual AdS/CFT prescription confines.
Moreover, magnetic quarks are screened.

\item The string theory nature of the bulk fields dual to the quark bilinear
currents is readily identified:
they are low-lying modes living in a brane-antibrane pair.

\item Chiral symmetry breaking is realized dynamically and consistently, because
of the tachyon dynamics. The dynamics determines the chiral condensate uniquely
a s function of the bare quark mass.

\item The mass of the $\rho$-meson grows with increasing quark mass, or,
more physically, with increasing pion mass.

\item By adjusting the same parameters as in QCD ($\Lambda_\mathrm{QCD}$,
$m_{ud}$) a good fit can be obtained of the light meson masses.

\end{itemize}

\subsection{Holographic models in the Veneziano limit}

To construct V-QCD we will put together the experience from IHQCD and the
tachyon
implementation in the quenched approximation.
Putting
 the two together we will see that, under reasonable assumptions,
 we obtain a phase diagram which is
  qualitatively in agreement with what to expect from QCD in the Veneziano
  limit. Moreover we will verify that changes of the bulk tachyon and dilaton
   potentials that are mild give the same qualitative physics.
In this sense we can state confidence in our results.

The bulk action we will consider is
 \be
S=S_g+S_f\sp  S_{g}=M^3N_c^2\int d^5x\sqrt{g}\left[R-{4\over 3}
{(\p \l)^2\over \l^2}+V_g(\l)\right]
\label{i1} \ee
 with $\l$ the 't Hooft coupling (exponential of the dilaton $\phi$)
  and the flavor action is
  \be
  S_{f}=-xM^3N_c^2\int d^5x~ V_f(\l,T)\sqrt{\det(g_{\m\n}
  +h(\l)\partial_{\m}T\partial_{\nu}T^{\dagger})}
\label{i2}  \ee
To find the vacuum (saddle point) solution we must  set the gauge fields
 $A_{\m}^{L,R}$ to zero, as they are not expected 
to have vacuum expectation values at zero density.
We also take the tachyon field $T$ to be diagonal  and suppressed the WZ 
terms as they also do not contribute to the vacuum solution.

The pure glue potential $V_g$ has been determined from previous studies,
 \cite{data} and we will use the same here.
The tachyon potential $V_f(\l,T)$ must satisfy some basic properties,
that are determined by the dual theory or general properties of tachyons
 in string theory: (a) To provide the proper dimension for the dual
 operator near the boundary (b) To exponentially vanish like
$\log V_f\sim -T^2+\cdots$ for $T\to \infty$.
The function $h(\l)$ captures the transformation from the string frame to the
Einstein frame in five dimensions and will be chosen appropriately.

As with IHQCD, we will arrange that the theory is logarithmically asymptotically
 AdS, and will implement the two-loop
 $\beta$-function plus one-loop anomalous dimension for the chiral condensate.
 Although the geometrical picture is not expected to be reliable near the
boundary,  the renormalization group (RG) flows that emerge are reliable at
least in the IR.
 The UV boundary conditions we choose can be thought of as a convenient way of
  anchoring the theory in UV. We can always define a finite cutoff and evolve
  the theory from there in the
 IR.

 We first analyze the fixed points of the bulk theory. Choosing  a potential
that
  implements the Banks-Zaks fixed point, its presence exists for a range of the
parameter $x$.
 We will make choices where this is the whole range: $0<x<{11\over 2}$.
 In such a fixed point the dilaton is constant and the tachyon vanishes
identically.
  We have also checked that choices of potential for which the fixed point
exists for
   $x_*<x<{11\over 2}$ , have qualitatively similar physics.

We define appropriate $\beta$-functions for the YM coupling and the quark mass.
We
then rewrite the equations following \cite{ihqcd} as first order equations
that specify the flow of the couplings, as well as non-linear first order
equations
that determine the $\beta$-functions in terms of the potentials that appear in
the bulk action
(\ref{i1}), (\ref{i2}).

 We can calculate the dimension of the chiral condensate in
the IR fixed point theory from the bulk equations. We find that it decreases
monotonically with $x$
for reasonably chosen potentials.
It crosses the value 2 at $x=x_c$
where $x_c$ corresponds to the end of the conformal window as argued in
\cite{Kaplan}.
We make the following observations which are relevant for
technicolor studies:
\begin{itemize}

 \item The lower edge of the conformal window $x_c$ lies in the vicinity of 4.
Requiring the holographic $\beta$-functions to match with QCD in the UV, we find
that quite in general
\beq
 3.7 \lesssim ~~x_c~~ \lesssim 4.2
\eeq
which is in good agreement with other estimates \cite{ds,cw,antipin}.

 \item The fact that the dimension of the chiral condensate at the IR fixed
point approaches
two (and the anomalous dimension approaches unity) as $x \to x_c$ is in
line with the standard expectation from field theory
approaches \cite{ds,Holdom2}. It is also to a large extent
independent of the details of the model.

\end{itemize}

It is important to stress that in the full analysis of this paper, 
the backreaction of the flavor sector on the glue sector 
is fully  included. This is very important for the ``walking'' 
region, in the vicinity of $x=4$, where we expect the backreaction to be 
important. Indeed we do not expect to see a Conformal Phase Transition 
in the quenched limit of QCD.

Apart from $x$, there is a single parameter in the theory, namely ${m\over
\Lambda_\mathrm{QCD}}$ where $m$ is the UV value of the (common) quark mass.
For
each value of $x$, we solve the bulk equations with fixed sources corresponding
to fixed
$m, \Lambda_\mathrm{QCD}$, and determine the vevs so that the solution is
``regular''
in the IR. The notion of regularity is tricky even in the case of  IHQCD (pure
glue),
 as there is a naked singularity in the far IR. For the dilaton this has been
 resolved in \cite{ihqcd,gkmn}. For the tachyon the notion of regularity is
different and has been studied in detail
 in \cite{ikp}.

Implementing the regularity condition in the IR and solving the equations
 from the IR to the UV (this has been done mostly numerically), there is a
single parameter that determines the
 solutions as well as the UV
coupling constants and vevs, and this is the a real number $T_0$
controlling the value of the Tachyon in the IR. This reflects the single
dimensionless parameter ${m\over
\Lambda_\mathrm{QCD}}$ of the theory.

For different values of $x$ and $m$ we find the following qualitatively
different regions:
\begin{itemize}

 \item When $x_c \leq x <11/2$ and $m=0$, the theory flows to an IR fixed
point.
The IR CFT is weakly coupled near $x={11\over 2}$ and strongly coupled in the
vicinity of $x_c$.
Chiral symmetry is unbroken in this regime (this is known as the conformal
window).

 \item When $x_c \leq x <11/2$ and $m\not=0$, the tachyon has a non-trivial
profile,
 and there is a single solution with the given source, which is ``regular'' in
the IR.

 \item When $0<x<x_c$ and $m=0$, there is an infinite number of regular
solutions with non-trivial tachyon profile, and a special solution with an
identically vanishing tachyon and an IR fixed point.

 \item When $0<x<x_c$ and $m\not=0$, the theory has vacua with nontrivial
profile for the tachyon. For every non-zero $m$, there is a finite number of
regular solutions that grows as $m$ approaches zero.

\end{itemize}

In the region $x<x_c$ where several solutions exist, there is a interesting
relation between the IR value $T_0$
controlling
 the regular solutions, and the UV parameters, namely $m$.
This is determined numerically, and a relevant plot describing the
 relation between $m$ and $T_0$ at fixed $x$ is in Fig.~\ref{mT0} (left).
As $m$ and $-m$ are related by a chiral rotation by $\pi$, we can take $m\geq
0$.

The solutions are characterized by the number of times $n$ the tachyon field
changes sign
 as it evolves from the UV to the IR. For all values of $m$ there is a single
solution with no tachyon zeroes. In addition, for each positive $n$ there are
two solutions which exist within a finite  range $0<m<m_n$, where the limiting
value $m_n$ decreases with increasing $n$, and one solution for $m=0$.
In particular, for large enough fixed $m$, we find that only the
solution without tachyon zeroes exists.

For $m\not =0$, out of all regular solutions, the ``first'' one
without tachyon zeroes
has the smallest free energy.
The same is true for $m=0$, namely the solution with non-trivial tachyon without
zeroes is energetically favored over the solutions with positive $n$ as well as
over the special solution with identically vanishing tachyon, which appears only
for $m=0$ and would leave chiral symmetry unbroken.
Therefore, chiral symmetry is broken for $x<x_c$.

The multiplicity of regular solutions is closely related to
 the regime where the IR dimension of the chiral condensate is smaller than 2,
and the associated Efimov vacua. They seem to be associated with
the fixed point theory that here exists for all values of $x$
but is not reachable by flowing from the UV of QCD for $x<x_c$.
On the other hand,  the presence of a fixed-point theory in the
 landscape of possible theories does not seem necessary for the appearance
of multiple saddle points.
Indeed, in \cite{ikp} which employed the quenched approximation
and where no such fixed points exist, a second saddle point was
 found that provided a regular
tachyon solution. It was verified however that this second saddle
 point was perturbatively unstable as meson fluctuations were tachyonic.

In the region just below $x_c$ we find Miransky or BKT scaling for the chiral
condensate.
As $x \to x_c$, we obtain
\be
 \sigma  \sim \Lambda_\mathrm{QCD}^3
\exp\left(-\frac{2 \hat K}{\sqrt{x_c-x}}\right) \ .
 \label{i4}\ee
For $x \geq x_c$, let $m_\mathrm{IR}(x)$ be the mass of the tachyon at the IR
fixed point and  $\ell_\mathrm{IR}(x)$ the IR AdS radius.
The coefficient $\hat K$ is then fixed as
\be
\hat K = \frac{\pi}{\sqrt{\frac{d }{d x}\big[m^2_\mathrm{IR}\ell^2_\mathrm{IR}
\big]_{x=x_c}}} \ .
 \label{i5}\ee

The construction of the holographic V-QCD model opens the road for addressing
several interesting questions.
\begin{enumerate}

\item The calculation of the spectrum of mesons and glueballs. This is
 in principle a straightforward albeit tedious exercise, \cite{mesons}. In the
Veneziano limit,
mixing is expected between glueballs and mesons to leading order in
 $1/N_c$. This will affect the $0^{++}$ glueball that will mix with the
 $0^{++}$ flavor-singlet $\sigma$-mesons.
On the other hand the $2^{++}$ glueballs, the $1^{--}$ and $1^{++}$
vector mesons and the $0^{+-}$ mesons do not mix, with the exception
 of the flavor singlet  $0^{+-}$ meson (analogous to $\eta'$) that will
  mix with the $0^{+-}$ glueball due to the axial anomaly.
A particularly interesting question here is the behavior of the mass of
the lightest $0^{++}$ state (the technidilaton, \cite{yamawaki}) as $x\to x_c$.

\item The structure and phase diagram of the theory at finite temperature,
\cite{finiteT}. In the quenched approximation
\cite{ikp} the restoration of chiral symmetry was seen above the (first order)
deconfinement transition.
The expected structure is not clear here and several options exist.

\item The calculation of the energy loss of heavy quarks in a quark-gluon plasma
with non-negligible percentage of quarks.

\item The construction of the baryon states in this theory and the calculation
of their properties.

\item The structure of the phase diagram at finite density and the search for
exotic phases namely color superconductivity and color-flavor locking.

\end{enumerate}

It is plausible that the setup may provide a model for high-$T_c$
superconductors by interpreting the $x$ parameter as a ``doping'' parameter.
The reason is that $x$ controls the IR dimension of the ``Cooper pair"
associated with a quark-antiquark
boundstate, charged under the axial charge. As $x$ decreases, the IR
dimension of this operator decreases,
and the bound state becomes more and more deeply bound. At $x=x_c$,
there is an onset of ``axial" superconductivity (at zero temperature),
that persists down
to $x=0$.

At finite temperature, this picture suggests that the system might
resemble the overdoped regime of strange metals, with $x=x_c$ the start
of the superconduction dome and $x=0$ the optimal doping.
The connection between the value of $x$ and doping in real systems may
not be so far fetched as the changes in the system associated with the
change of carriers, is accompanied by a change
in the effective number of flavors of strongly interacting effective
degrees of freedom.

The structure of this paper is as follows. In Sec.~\ref{SecQCD} we give a brief
review on QCD in the Veneziano limit and its phase structure. In
Sec.~\ref{SecBottomup} we review the IHQCD model, and discuss the earlier
results on mass spectra and the phase structure at finite temperature. Adding
the flavor branes is discussed in detail in Sec.~\ref{flavor}. The V-QCD model
is finally introduced in Sec.~\ref{ssmodel}. Analysis of the model is started by
studying the fixed points in Sec.~\ref{SecFP}. We go on transforming the
equations of motion (EoMs) to
equations for the holographic beta functions, and discuss their UV/IR
asymptotics and solutions in Sec.~\ref{SecBetas}. In Sec.~\ref{SecBG} we analyze
the background, in particular how the UV expansions map to perturbation theory
of QCD, and where the edge of the conformal window appears. We also construct
and present the numerical solutions for the background in the physically
interesting regions. In Sec.~\ref{SecFE} we find the vacua with lowest free
energy and check that they support the expected phase diagram. In
Sec.~\ref{secBKT} we study the system near but below the conformal window, and
show that the chiral condensate, as well as many other observables, obey the BKT
scaling law. In particular, we check the scaling by comparing numerical results
to formulas, that are derived analytically. Finally, we conclude and summarize
the main results in Sec.~\ref{SecConcl}. Technical details are presented in
Appendices~\ref{AppBeta}-\ref{AppBKT}.

\section{QCD in the Veneziano limit} \label{SecQCD}

The conventional large-$N_c$ limit of QCD involves a large number of colors
$N_c\to\infty$, but a fixed number of flavors, $N_f\to$ finite, \cite{hooft}.
In this limit, fermion loops are suppressed, and the dominant diagrams are
classified by the  genus of the associated Riemann surface. Fundamentals
(quarks) are associated with open strings and boundaries, and the number of
flavors is measuring the Chan-Paton factors of the open strings.

There is an alternative large-$N_c$ in QCD, in which
\be
N_c\to \infty\sp N_f\to \infty\sp {N_f\over N_c}=x~~~~{\rm fixed}
\label{vl}\ee
This was first introduced by Veneziano in \cite{v} in order to have the QCD
axial anomaly, of order ${\cal O}(N_cN_f)$ appear in the leading order in the
large-$N_c$ expansion.

This alternative large-$N_c$ limit is very interesting in order to preserve
important effects due to quarks.
In the conventional 't Hooft limit such effects are subleading, and this is
known as the quenched limit for flavor.
Many efforts have been made in the last few years to consider unquenched flavor,
in order to estimate the contribution of quarks to the physics of the
quark-gluon plasma. Such efforts are summarized in the recent review,
\cite{npr}.

The following effects are not easily visible in the conventional 't Hooft limit:
\begin{itemize}

\item The ``conformal window'' with a non-trivial fixed point, that extends from
$x={11\over 2}$ to smaller values of $x$.
      The region $x\to {11\over 2}$ has an IR fixed point while the theory is
still weakly coupled,  as was  analyzed by Banks and Zaks, \cite{bz}.

\item It is expected that at critical $x_c$, the conformal window will end, and
for $x<x_c$, the theory will exhibit chiral symmetry breaking in the IR. This
behavior is expected to persist down to $x=0$.
   Above $x>x_c$ the IR theory is CFT, at strong coupling that progressively
becomes weak as $x\to {11\over 2}$.

\item Near and below $x_c$, there is the transition region to conventional QCD
IR behavior. In this region the theory is expected to be ``walking'', so that
the theory flows towards the IR fixed point but misses it ending up with chiral
symmetry breaking. But the approach to the fixed point
involves a slow variation of the YM coupling constant for a long range of
energies. This has been correlated with a nontrivial dimension for the quark
mass operator near two, rather than three (the free field value).

\item The existence of this ``walking'' region makes the theory  extremely
interesting for applications to strong-couplings solutions to the hierarchy
problem (technicolor).

\item New phenomena are expected to appear at finite density driven by strong
coupling and the presence of quarks. These involve color superconductivity
\cite{cs} and flavor-color locking \cite{cfl}.

\end{itemize}

To discuss the structure expected as a function of the finite ratio $x$, defined
in (\ref{vl}) we write the two-loop QCD $\beta$-function.
With  $N_f$ (non-chiral) flavors in the fundamental, the $\beta$-function reads
\begin{equation}
\beta(g)=-{g^3\over (4\pi)^2}\left\{{11\over 3}N_c-{2\over
3}N_f\right\} -{g^5\over (4\pi)^4}\left\{{34\over 3}N_c^2-{N_f\over
N_c}\left[{13\over 3}N_c^2-1\right]\right\}+\cdots
\end{equation}

Using  the 't Hooft
coupling, and setting  ${N_f\over N_c}\to x$ we obtain
\be \label{QCDbetaVen}
\l\equiv {g^2 N_c}\sp \dot \l=-b_0\l^2+b_1\l^3+{\cal O}(\l^4)
\ee
with
\be
b_0={2\over 3}{(11-2x)\over (4\pi)^2}\sp {b_1\over b_0^2}=-{3\over
2}{(34-13x)\over (11-2x)^2}
\ee

The Banks-Zaks region is $x=11/2-\e$ with $\e\ll 1$ and positive, \cite{bz}.
We obtain a fixed point of the $\beta$-function at
\be
\l_{*}={(8\pi)^2\over 75}\e
\ee
which is trustable in perturbation theory, as $\l_*$ can be made arbitrarily
small.

The infrared fixed point has properties that are computable in perturbation
theory.
In particular the low-lying operators consist of the conserved stress tensor,
$Tr[F^2]$ that is now slightly irrelevant,
and the $L,R$ currents that are still conserved with the exception of the $U(1)$
axial current that its conservation is broken by the anomaly.

The mass operator, $\bar \psi_L\psi_R$
has now dimension slightly smaller
than  three, as attested by its perturbative anomalous dimension
\be
-{d \log m\over d\log \mu}\equiv \gamma={a_0\over 4\pi}g^2+{a_1\over
(4\pi)^2}g^4+\cdots
\ee
\be
a_0={3\over 4\pi}{N_c^2-1\over N_c}\sp a_1={1\over
2\, (4\pi)^2}\left[3{(N_c^2-1)^2\over 2 N_c^2}
-{10\over 3}{N_c^2-1\over N_c}N_f+{97\over 3}(N_c^2-1)\right]
\ee
At large $N_c$ this becomes
\be \label{QCDgamma}
\gamma\simeq {3\over (4\pi)^2}\l+{(203-10x)\over 12\, (4\pi)^4}\l^2+{\cal O}(\l^3,
N_c^{-2})
\ee

One can still perturb this theory by the $U(N_c)$-invariant mass operator
(assuming all quarks have the same mass), and the theory is expected now to flow
to the trivial  (QCD-like) theory in the IR.

It is believed that there is also a value $x_c$ with  $0<x_c<{11\over 2}$ so
that for $x<x_c$ the theory flows to a trivial theory (with a mass gap) in the
IR, with chiral
 symmetry breaking and physics isomorphic to that of standard YM.
 For ${11\over 2}>x>x_c$, the theory is expected to flow to a non-trivial IR
fixed point, and chiral symmetry to remain unbroken, as happens in the BZ
region. Generically,
 the IR theory is strongly coupled except in the region $x\to {11\over 2}$ where
the fixed point theory is weakly coupled (Banks-Zaks fixed point).
 For $x>{11\over 2}$ the theory is IR free.

\section{A step back: Bottom-up models for large-N$_c$ YM} \label{SecBottomup}

The holographic dual of large $N_c$ Yang-Mills theory,  proposed in
\cite{ihqcd},  is
based on a five-dimensional Einstein-dilaton model, \index{Einstein-dilaton
gravity} with the action\footnote{Similar models of Einstein-dilaton gravity
were proposed
 independently in \cite{gubser} to describe the finite temperature physics of
  large $N_c$ YM. They differ in the UV as the dilaton corresponds to a relevant
operator
 instead of the marginal case we study here. The gauge coupling $e^{\Phi}$
  also asymptotes to a constant instead of zero in such models.}:
\begin{equation}
S_5=-M^3_pN_c^2\int d^5x\sqrt{g}
\left[R-{4\over 3}(\partial\Phi)^2+V(\Phi) \right]+2M^3_pN_c^2\int_{\partial
M}d^4x \sqrt{h}~K.
 \label{kira1}\end{equation}
Here, $M_p$ is the  five-dimensional Planck scale and $N_c$ is the number of
colors.
The last term is the Gibbons-Hawking term, \index{Gibbons-Hawking term} with $K$
being the extrinsic curvature
of the boundary. The effective five-dimensional Newton constant
is $G_5 = 1/(16\pi M_p^3 N_c^2)$, and it is small in the large-$N_c$ limit.

Of the 5D coordinates $\{x_i, r\}_{i=0\ldots 3}$, $x_i$ are identified with the
4D space-time coordinates, whereas  the  radial coordinate $r$ roughly
corresponds to the 4D RG scale.
We identify $\l\equiv e^\Phi$ with the  running 't Hooft  coupling $\l_t\equiv
N_cg_\mathrm{YM}^2$,
up to an {\it a priori} unknown multiplicative factor\footnote{ This relation is
well motivated
in the UV, although it may be modified at strong coupling (see
\cite{ihqcd}). The
quantities we will calculate do not depend on the explicit relation between $\l$
and $\l_t$.
}, $\l = \kappa \l_t$.

The dynamics is encoded in the dilaton potential\footnote{With a slight abuse of
notation we will denote $V(\l)$  the
function $V(\Phi)$ expressed as a function of  $\l\equiv e^\Phi$.},  $V(\l)$.
The small-$\l$ and large-$\l$ asymptotics of $V(\l)$ determine the solution in
the UV
and  the IR of the geometry
respectively. For a detailed but concise description of the UV and IR
properties of the solutions the reader
is referred to Section 2 of \cite{gkmn}. Here we will only mention the most
relevant information:
\begin{enumerate}
\item For small $\l$,   $V(\l)$  is required to have a power-law expansion of
the form:
\be \label{kirUVexp}
V(\l) \sim {12\over \ell^2}(1+ v_1 \l + v_2 \l^2 +\ldots), \qquad \l\to 0 \;.
\ee
The value at $\l=0$ is constrained to be finite and positive, and sets the UV
AdS scale $\ell$.
 The coefficients
of the other terms in the expansion fix the  $\beta$-function coefficients for
the
running coupling $\l(E)$. If we identify the energy scale with the metric scale
factor in the Einstein frame, denoted by $e^A$ below, we obtain \cite{ihqcd}:
\be\label{kirbetafunc}
\beta(\l) \equiv {d \lambda \over d\log E} = -b_0\l^2 +b_1 \l^3 +\ldots\sp b_0 =
{9\over 8} v_1, \quad \; \; b_1 = -\frac94 v_2 + \frac{207}{256}v_1^2 \;.
\ee
\item For large $\l$,   confinement and the absence of \index{bad singularity}
bad singularities\footnote{For a description of the notion of  ``bad versus good
 singularities'' and their resolution the reader is referred  to \cite{cgkkm}.}
require:
\be\label{kirIRexp}
 V(\l) \sim \l^{2Q}(\log \l)^P \quad \l\to \infty, \quad \left\{
\begin{array}{l} 2/3 < Q < 2\sqrt{2}/3, \quad P\; {\rm arbitrary}\\ Q = 2/3,
\quad P\geq 0 \end{array} \right. .
\ee
In particular, the values $Q=2/3, P=1/2$ reproduce an asymptotically-linear
glueball spectrum, \index{glueball spectrum!linear}
$m_n^2\sim n$, besides confinement.
We will restrict ourselves to this case in what follows.
\end{enumerate}

In \cite{ihqcd}, the single phenomenological parameters of the potential was
fixed
by looking at the zero-temperature spectrum,
 i.e. by computing various glueball mass ratios and comparing
them  to the corresponding lattice results.
The masses are computed by deriving the effective action for the quadratic
fluctuations around the background,
\cite{nitti} and subsequently reducing the dynamics to four dimensions.

The glueball spectrum is obtained holographically as the spectrum
of normalizable fluctuations around the zero-temperature
background.
In IHQCD the relevant fields are  the 5D  metric, one scalar field
(the dilaton), and one pseudoscalar field (the axion that is subleading in
$N_c$).
 As a consequence, the only normalizable
fluctuations above the vacuum correspond to spin 0 and spin 2 glueballs (more
precisely, states
 with $J^{PC} = 0^{++}, 0^{-+}, 2^{++}$), each species containing an infinite
discrete tower of excited
states.

We only compare the  mass spectrum
obtained in our model to the  lattice results for the lowest $0^{++}, 0^{-+},
2^{++}$ glueballs and their
available excited states. These are limited to one for each spin 0 species, and
none for the spin 2,
 in the study of \cite{chenetal}, which is the one we use for our  comparison.
This provides  two mass ratios in the CP-even sector and two  in the CP-odd
sector.

The glueball masses are computed by first  solving numerically
Einstein's equations, and using the resulting metric
and dilaton to setup an  analogous  Schr\"odinger problem for the
fluctuations, \cite{ihqcd}.
The results for the parity-conserving sector are shown in Table
\ref{masses++}, and are in good agreement with lattice data  for $N_c=3$.

\begin{table}[h!]
\begin{center}
\caption{Glueball Masses}\label{masses++}
\begin{tabular}{|r|c|c|c|}
\hline
& ~~~~IHQCD~~~~~ & ~~~~~~$N_c=3$~~~~~~ & ~~~~~~~$N_c=\infty$ ~~~~~ \\
\hline\hline
$~~~m_{0^{*++}}/m_{0^{++}}~~~$ & 1.61 & 1.56(11) & 1.90(17)  \\
\hline
$m_{2^{++}}/m_{0^{++}}$ & 1.36 & 1.40(4) & 1.46(11)   \\
\hline \hline
\end{tabular}
\end{center}
\end{table}

Unlike the various  mass ratios,
the value of  any given mass in AdS-length units (e.g. $m_{0++} \ell$)  {\em
does  depend}
 on the choice of integration constants in the UV.
Therefore its numerical
value does not have an intrinsic meaning. However it can be used
as a benchmark against which all other dimension-full quantities can
be measured (provided one always uses the same UV boundary conditions).  On the
other hand, given a fixed set of initial conditions, asking that
$m_{0++}$ matches the physical value (in MeV) obtained on the
lattice, fixes the value of $\ell$ hence the energy unit.

The holographic renormalization of such Einstein-dilaton theories is quite
intricate as the AdS boundary conditions on the dilaton is unusual
($\phi\to-\infty$ near the boundary). It has been derived recently in
\cite{papadimitriou}.

\subsection{Finite temperature}

In the large $N_c$ limit,
the  canonical ensemble partition function of the model just described, can be
approximated by a sum over saddle points, each given by a classical solution of
the Einstein-dilaton
field equations:
\be
{\cal Z}(\beta) \simeq e^{-{\cal S}_1(\beta)}  +   e^{-{\cal S}_2(\beta)} +
\ldots
\ee
where ${\cal S}_i$ are the euclidean actions evaluated on each  classical
solution with a fixed
 temperature $T=1/\beta$, i.e. with euclidean time compactified on a circle of
length $\beta$.
There are two possible types of Euclidean solutions which preserve 3-dimensional
rotational invariance.
In conformal coordinates these are:
\begin{enumerate}
\item {\bf Thermal gas solution,}
\be\label{kirthermal}
ds^2 = b^2_o(r)\left(dr^2 + dt^2 +  dx_mdx^m\right), \qquad \Phi = \Phi_o(r),
\ee
with $r\in (0, \infty)$ for the values of $P$ and $Q$ we are using;
\item {\bf Black-hole solutions,}\index{black hole}
\be
ds^2=b(r)^2\left[{dr^2\over f(r)}+f(r)dt^2+dx_mdx^m\right], \qquad \Phi =
\Phi(r),
 \label{kira7}\ee
with  $r\in (0,r_h)$,  such that $f(0)=1$, and $f(r_h)=0$.
\end{enumerate}
In both cases Euclidean time is periodic with period $\b_o$ and $\b$
respectively
 for the thermal gas and black-hole solution,
and 3-space is
taken to be a torus with volume $V_{3o}$ and $V_3$ respectively,
so that the black-hole mass and entropy are finite\footnote{The periods
and 3-space volumes of the thermal gas solution are related to the black-hole
solution values by requiring that the geometry of the two solutions are
the same on the (regulated) boundary. See \cite{gkmn} for details.}.

The black holes are dual to a deconfined phase, since the string
tension vanishes at the horizon, and the Polyakov loop has\index{Polyakov loop}
non-vanishing expectation value. On the
other hand, the thermal  gas  background is confining.

The thermodynamics of the deconfined phase\index{deconfined phase} is dual to
the  5D
black-hole thermodynamics. The  free energy, defined as\index{black hole}
 \be\label{kirfirst law}
{\cal F} = E - T S,
\ee
is identified with the  black-hole on-shell
action; as usual, the energy $E$ and entropy $S$ are identified  with the
black-hole mass, and one
fourth of the horizon area in Planck units,  respectively.

The thermal gas and black-hole solutions with the same temperature differ at
$O(r^4)$:
\be
b(r) = b_o(r)\left[1 + \,{\cal G}\, {r^4\over \ell^3} +\ldots\right],  \qquad
f(r) = 1 -{C\over 4} {r^4\over \ell^3} + \ldots \qquad r \to 0,
\label{kirb-bo}
\ee where ${\cal G}$ and $C$ are constants with units of energy.
As shown in \cite{gkmn} they  are related to the enthalpy $TS$ and
the gluon condensate $\<{\rm tr} F^2\>$ :
\be
\label{kirCG} C = {T S \over
M_p^3 N_c^2 V_3} , \qquad \qquad {\cal G} ={22 \over 3 (4\pi)^{2}}
{\langle {\rm tr} ~F^2 \rangle_T - \langle {\rm tr} ~F^2 \rangle_o \over240
M_p^3 N_c^2}.
\ee
 Although they appear as coefficients in the UV
expansion, $C$ and ${\cal G}$ are determined by regularity at the
black-hole horizon. For $T$ and $S$ the relation is the usual one,
\be\label{kirTS}
 T = - {\dot{f}(r_h) \over 4\pi}, \qquad S  = {\mathrm{Area} \over 4 G_5} =
4\pi\, (M_p^3 N_c^2 V_3) \, b^3(r_h).
\ee
For ${\cal G}$ the relation with the horizon quantities is more complicated and
cannot be put in a simple analytic form. However, as discussed in \cite{gkmn},
for each temperature
there exist only specific values of ${\cal G}$ (each corresponding to a
different black hole)
such that the horizon is regular.

At any given temperature there can be one or more solutions: the thermal gas
is always present, and there can be different black holes with the same
temperature. The solution  that dominates
the partition function at a certain $T$ is the one with smallest free energy.
The free energy difference between the black hole  and
thermal gas was calculated in \cite{gkmn}  to be:
\be\label{kirF}
\frac{\cal F}{M_p^3 N_c^2 V_3}={{\cal F}_{BH} - {\cal F}_{th}\over M_p^3 N_c^2
V_3}  = 15 {\cal G} - {C\over 4}.
\ee
For a dilaton potential corresponding to a confining theory, like the one we
will assume,
the phase structure is the following \cite{gkmn}:
\begin{enumerate}
\item There exists a minimum temperature $T_{min}$ below which the only solution
is the thermal gas.
\item Two branches of black holes (``large'' and ``small'')  appear for $T\geq
T_{min}$, but the ensemble
is still dominated by the confined phase up to a temperature $T_c >
T_{min}$\index{critical temperature}
\item At $T=T_c$ there is a first order phase transition\index{phase transition}
to the {\em large} black-hole phase\index{black hole!large}. The system remains
in the black-hole (deconfined) phase for all $T>T_c$.
\end{enumerate}

The holographic mode  has also been confronted successfully with recent lattice
data \cite{panero} at finite temperature, \cite{data}.

\section{Adding Flavor\label{flavor}}

A number $N_f$ of quark flavors can be included
in our setup by adding space-time filling ``flavor-branes''.
In this case they are pairs of space-filling $D_4-\overline{D_4}$ branes.

To motivate the setup it is important to revisit the low-dimension operators
(dimension=3) in the flavor sector and their realization in string theory.
At the spin-zero level we have the (complex) mass operator
\be
\bar\psi^i_R\psi^j_L\leftrightarrow  T_{ij}
\ee
dual to a complex scalar transforming as $(N_f,\bar N_f)$ under the flavor
symmetry $U(N_f)_R\times U(N_f)_L$.
At the spin-one level we have the two classically conserved currents
\be
\bar\psi^i_L\sigma^{\m}\psi^j_L\leftrightarrow A^{\m}_{L,ij}\sp
\bar\psi^i_R\bar\sigma^{\m}\psi^j_R\leftrightarrow
 A^{\m}_{R,ij}
\ee
They transform in the adjoint of the $U(N_f)_R$ respectively the $U(N_f)_L$
symmetry.
The flavor symmetry is expected to arise in string theory from $N_f$ flavor
branes (R) and $N_f$ flavor antibranes (L).
Due to the quantum numbers, the vectors are the lowest modes of the fluctuations
of the open strings with both ends on the D branes
($A^{\m}_{R}$), or the anti-D branes, $A^{\m}_{L}$.

The bifundamental  scalar $T$, on the other hand,  is the lowest mode of the
$D-\bar D$ strings, compatible with its quantum numbers.
Its holographic dynamics is dual to the dynamics of the chiral condensate.
This is precisely the scalar that in a brane-antibrane system in flat space is
the tachyon whose dynamics has been studied
profusely in string theory, \cite{sen}.  It has been proposed that the
non-linear DBI-like actions proposed by Sen and others
are the proper setup in order to study the holographic dynamics of chiral
symmetry breaking, \cite{ckp}.
This dynamics was analyzed in a toy example, \cite{ikp}, improving several
aspects of the hard \cite{hard}, and soft wall models, \cite{soft}. We will keep
referring to $T$ as the ``tachyon'', as it indeed corresponds to a relevant
operator in the UV.

The tachyon dynamics is
captured holographically  by the open string DBI+WZ action, which schematically
reads, in the
string frame,
\be\lab{actiongauge1}
S[T,A^L,A^R]  =S_{DBI}+S_{WZ}
\ee
where the DBI action for the $D-\bar D$ pair is
\be\label{DBI}
S_{DBI}=\int dr d^4 x  ~{N_c\over \l} ~
{\bf Str}\left[V(T) \bigg(\sqrt{-\det \left(g_{\mu\nu} + D_{\{\mu} T^{\dagger}
D_{\nu\}} T +
F^L_{\mu\nu}\right)}+
\right.
\ee
$$
\left.+\sqrt{-\det \left(g_{\mu\nu} + D_{\{\mu} T^{\dagger} D_{\nu\}} T +
F^R_{\mu\nu}\right)}\bigg)\right]
$$
Here $T$ is the tachyon, a complex $N_f\times N_f$ matrix. $A^{L,R}_{\mu}$ are
the world-volume gauge fields
of the $U(N_f)_L\times U(N_f)_R$ flavor symmetry, under which the tachyon is
transforming as the $(N_f,\bar N_f)$,
a fact reflected in the presence of the covariant derivatives\footnote{We are
using the conventions of \cite{ckp}.}
\be
D_{\m}T\equiv \partial_{\mu}T-iTA^L_{\m}+iA^R_{\m}T\sp D_{\m}T^{\dagger}\equiv
\partial_{\mu}T^{\dagger}-iA^L_{\m}T^{\dagger}+iT^{\dagger}A^R_{\m}
\ee
transforming covariantly under
\be
T\to V_RTV_L^{\dagger}\sp A^L\to V_L(A^L-iV_L^{\dagger}dV_L )V_L^{\dagger}\sp
A^R\to V_R(A^R-iV_R^{\dagger}dV_R )V_R^{\dagger}
\ee
as well as the field strengths $F^{L,R}=dA_{L,R}-iA_{L,R}\wedge A_{L,R}$ of the
$A^{L,R}$ gauge fields.
$\l\equiv e^{\Phi}=N_c e^{\phi}$ is as usual the 't Hooft coupling.
We have also used the symmetric trace ($\equiv Str$) prescription although
higher order terms of the non-abelian DBI action
are not known. It turns out that such a prescription is not relevant for the
vacuum structure in the flavor sector
(as determined by the classical solution of the tachyon) neither for the mass
spectrum.
The reason is that we may treat the light quark masses as equal to the first
approximation and then in the vacuum,
$T=\tau {\bf 1}$ with $\tau$ real, and this is insensitive to non-abelian
ramifications.
Expanding around this solution, the non-abelian ambiguities in the higher order
terms do not enter at quadratic order.
Therefore, for the spectrum we might as well replace $Str\to Tr$.

The WZ action on the other hand is given by\footnote{This expression was
proposed in \cite{tach1}
and proved in \cite{KrausLarsen,TTU} using boundary string field theory}:
\be\label{WZtach}
S_{WZ}=T_4 \int_{M_5} C\wedge \mathrm{str} ~\exp\left[{i
2\pi\a'\mathcal{F}}\right]
\ee
where $M_5$ is the world-volume of the
${\rm D}_4\,$-$\overline{{\rm D}_4}$ branes that coincides with the full
space-time. Here,
$C$ is a formal sum of the RR potentials $C=\sum_n (-i)^{\frac{5-n}{2}}C_n$,
and $\mathcal{F}$ is the curvature of a superconnection
${\cal A}$.
Note also that $\mathrm{str}$ in (\ref{WZtach}) stands for 
supertrace and not symmetric trace. 
It acts on the space of $D$ and $\bar D$ branes and is defined in 
Appendix C of \cite{ckp}.

In terms of the tachyon field matrix $T$
and the gauge fields  $A^L$ and  $A^R$
living respectively on the branes and
antibranes, they are
(We will set $2\pi \alpha'=1$
and use the notation of
\cite{KrausLarsen}):
\be
i\mathcal{A}=\left(\begin{array}{cc} iA_L & T^\dagger\\
T & iA_R\end{array}\right)\,,\qquad
i\mathcal{F}=\left(\begin{array}{cc} iF_L-T^\dagger T & DT^\dagger\\
DT & iF_R-TT^\dagger\end{array}\right)
\label{AFdef}
\ee
The curvature of the superconnection is defined as:
\be
{\cal F} = d{\cal A} - i {\cal A} \wedge {\cal A}\sp
d{\cal F} - i {\cal A} \wedge {\cal F} + i {\cal F} \wedge {\cal A} = 0
\ee
Note that under (flavor) gauge transformation it transforms homogeneously
\be
{\cal F}\to \left(\begin{array}{cc} V_L& 0\\
0 & V_R\end{array}\right)~{\cal F}~\left(\begin{array}{cc} V^{\dagger}_L& 0\\
0 & V^{\dagger}_R\end{array}\right)
\ee
In \cite{ckp} the relevant definitions
and properties of this {\it supermatrix}
formalism can be found.

By expanding we obtain
\be
S_{WZ}=T_4\int C_5\wedge Z_0+C_3\wedge Z_2+C_1\wedge  Z_4+C_{-1}\wedge Z_6
\ee
where $Z_{2n}$ are appropriate forms coming from the expansion of the
exponential of the superconnection.
In particular, $Z_0=0$, signaling the global cancelation of 4-brane charge,
which is equivalent to the cancelation
of the gauge anomaly in QCD.
Further, as was shown in \cite{ckp}
\be
Z_2=d\Omega_1\sp \Omega_1= i str(V(T^{\dagger}T))Tr(A_L-A_R)-\log\det (T)d (Str
V(T^{\dagger}T))
\ee
This term provides the Stuckelberg mixing between $Tr[A^L_{\m}-A^R_{\m}]$ and
the QCD axion that is dual to $C_3$. Unlike the 't Hooft limit, in the Veneziano
limit  this mixing happens at leading order in $1/N_c$, \cite{v}.
Dualizing the full action we obtain
\be
S_{CP-odd}={M^3\over 2N_c^2}\int d^5x\sqrt{g}Z(\l)\left(\pa a+i\Omega_1\right)^2
\ee
$$
={M^3\over 2N_c^2}\int d^5x\sqrt{g}Z(\l)\left(\pa_{\mu} a+x\zeta
\pa_{\m}V(\tau)-xV(\tau)A^A_{\m}\right)^2
$$
with
\be
\zeta ={1\over N_f}\mathrm{Im} \log\det T\sp A_{L}-A_{R}\equiv {1\over
2N_f}A^A\II+(A^a_{L}-A^a_{R})\lambda^a
\ee
and where we have set the tachyon to its vev $T=\tau {\bf 1}$ .
This term is invariant under the $U(1)_A$ transformations
\be
\zeta\to \zeta+\e\sp A^A_{\mu}\to  A^A_{\mu}-\partial_{\mu}\e\sp a\to a-x\e
V(\tau)
\ee
reflecting the QCD $U(1)_A$ anomaly.
It is this Stuckelberg term together with the kinetic term of the
tachyon field that is responsible for the mixing between the QCD axion
and the $\eta'$. In terms of degrees of freedom, we have two scalars
$a,\zeta$ and an (axial)  vector, $A^A_{\m}$.
We can use gauge invariance to remove the longitudinal components of
 $A^A$. Then an appropriate linear combination of the two scalars will become
 the $0^{-+}$ glueball field
while the other will be the $\eta'$. The transverse (5D) vector will
provide the tower of $U(1)_A$ vector mesons.

The next term in the WZ expansion couples the baryon density to a one-form RR
field $C_1$. There is no known operator expected to be dual to this
bulk form.
However its presence and coupling to baryon density can be understood as
follows.
Before decoupling the $N_c$ $D_3$ branes, its dual form $C_2$ couples to the
$U(1)_B$ on the $D_3$ branes via the standard $C_2\wedge F_B$ WZ coupling.
This is dual to a free field, the doubleton, living only at the boundary of the
bulk.
Once we add the probe $D_4+\bar D_4$ branes the free field is now a linear
combination
of $A^B$ and an $N_f/N_c$ admixture of $A^V$ originating
on the flavor branes. The orthogonal combination is the baryon number current on
the
flavor branes and it naturally couples to $C_1$.
Therefore the $C_1$ field is expected to be dual to the topological baryon
current at
 the boundary.

Finally the form of the last term requires some explanation.
By writing $Z_6=d\Omega_5$  we may rewrite this term as
\be
\int F_0\wedge \Omega_5\sp F_0=dC_{-1}
\ee
$F_0\sim N_c$ is nothing else but the dual of the five-form field strength.
This term then provides the correct Chern-Simons form
that reproduces the flavor anomalies of QCD. Its explicit form in terms of the
gauge fields $A_{L,R}$
and the tachyon was given in equation (3.13) in \cite{ckp}.

The action as described is based on the flat space Sen action for the $D-\bar D$
brane-antibrane system. In the presence of curvature and other non-trivial
background fields, like the dilaton we expect corrections to the DBI action.
Such corrections may affect the tachyon potential as well as the kinetic terms
of the vectors and the tachyon.
Some generic properties are expected to remain though, and these include the
tachyonic nature of the scalar near the AdS boundary and the exponential
asymptotics of the potential at large $T$.

\section{The bottom-up models}
\label{ssmodel}

 At $x={N_f\over N_c}=0$ the IHQCD model, \cite{ihqcd},  is described by the
action
 \be
 S_{g}=M^3N_c^2\int d^5x\sqrt{g}\left[R-{4\over 3}{(\p \l)^2\over
\l^2}+V_g(\l)\right]
 \ee
 with $\l$ the 't Hooft coupling (exponential of the dilaton) and a potential
that has the following asymptotics.
 \be
 \lim_{\l\to 0} V_g(\l)={12\over \ell^2}\left[1+v_1\l+v_2\l^2+\cdots\right]\sp
\lim_{\l\to\infty}V_g(\l)\sim \l^{4\over 3}\sqrt{\log \l}
 \ee

 At finite $x$ we must add the flavor sector. For the vacuum structure, and with
all masses of the quarks being equal it is enough
  to add the $U(1)$ part of the tachyon DBI action,
  \be
  S_{f}=-xM^3N_c^2\int d^5x~
V_f(\l,T)\sqrt{\det(g_{\m\n}+h(\l,T)\partial_{\m}T\partial_{\nu}T^{\dagger})}
\label{sf}  \ee
  where we have set the gauge fields to zero.
  The total action is $S=S_{g}+S_{f}$. Note that the overall sign of the DBI
action is negative.
  The function $V_g$ and its asymptotics has been discussed in detail in
\cite{ihqcd}. We will consider it known, and when needed we will use the form
that was in agreement with YM data, \cite{data}.
  The tachyon potential $V_f(\l,T)$ should satisfy some basic principles.
  For flat space D-branes, $V_s\sim {1\over \l}e^{-\mu^2 T^2}$.
  In our case, near the boundary, where $T\to 0, \l\to 0$, we expect, in analogy
with $V_g$, a regular series expansion in $\l,T$
  \be
  V_{f}\simeq V_0(\l)+V_1(\l)T^2+{\cal O}(T^4)
  \ee
  with $V_{0,1}(\l)$ having regular power series expansions in $\l$.
  As we will see later, the functions $V_{0,1}(\l)$ may be mapped into the
perturbative $\beta$-functions for the gauge coupling constant, and the
anomalous dimension of the quark mass operator.

Near the condensation point, $T\to\infty$ we expect the potential $V_f$ to
vanish exponentially. This is based on very general arguments due to Sen that
guarantee that the brane gauge fields disappear beyond that point.

Finally the function $h(\l,T)$ was introduced to accommodate the fact that the
action in (\ref{sf}) is written in the Einstein frame. In flat space, this
factor is unity in the string frame but becomes nontrivial ($h\sim \l^{-{4\over
3}}$) in the Einstein frame.

\subsection{The equations of motion}

Collecting the action of the glue and flavor sectors together,
\beq
{\cal
L}=(M^3\,N_c^2)\left[\sqrt{-g}\left(R-{4\over3}{
(\partial\lambda)^2\over\lambda^2}+V_g(\lambda)\right)
- x
\,V_f(\lambda, T)\sqrt{\det\left(g_{ab}+\f(\lambda, T) \partial_a T\,\partial_b
T\right)}\right]\,.
\label{lagrang}
\eeq
We shall take the following Lorentz-invariant Ansatz for the metric:
\beq
ds^2=e^{2A}\left(dx_{1,3}^2+dr^2\right)\,.
\label{metric}
\eeq
In our Ansatz the warp factor $A$, the scalar $\lambda$ and the tachyon $T$
depend only on the radial coordinate $r$.

The Einstein equations take the form:
\beq
R_{ab}-{1\over2}g_{ab}\,R=T_{ab}^g+T_{ab}^f\,,
\label{einsteq}
\eeq
where $T^g_{ab}$ and $T^f_{ab}$ are the energy momentum tensors of the glue and
flavor sectors, respectively.
These equations translate into
\be
6A''+6(A')^2=-{4\over3}{(\lambda')^2\over\lambda^2}+e^{2A}V_g(\lambda)- x
\,V_f(\lambda, T)\,e^{2A}\sqrt{1+e^{-2A}\f(\lambda, T)\ (T')^2}\,;
\label{e1}
\ee
\be
12(A')^2={4\over3}{(\lambda')^2\over\lambda^2}+e^{2A}V_g(\lambda)- x\,
V_f(\lambda, T)\,{e^{2A}\over\sqrt{1+e^{-2A}\f(\lambda, T)\ (T')^2}}\,,
\label{e2}
\ee
where primes stand for r-derivatives.
For $x=0$ these equations agree with
\cite{ihqcd}.
Finally, the equations of motion for the dilaton and the tachyon are given by:
\bea\label{e4}
\lambda''-{(\lambda')^2\over\lambda}+3A'\,\lambda'&=&{3\over8}\,\lambda^2
\,e^{2A}\bigg[-{d\,V_g\over d\lambda}+ x\,{\partial V_f\over \partial \lambda}
\,\sqrt{1+e^{-2A}\f(\lambda, T)\ (T')^2}\ \\\nn
&&+ { x \over 2}\ {\partial \f\over \partial \lambda} {e^{-2A}V_f \ (T')^2\over
\sqrt{1+e^{-2A}\f(\lambda, T)\ (T')^2}}\ \bigg]\,;
\eea
\be
T''+{e^{-2A}}\left(4 \f \,A'+{\partial \,V_f\over \partial \lambda}\,{\f
\lambda'\over V_f}+{ \lambda'\over 2} {\partial \f \over \partial
\lambda}\right)(T')^3+\left({1 \over 2\f}{\partial \f \over \partial T}-{1 \over
V_f}{\partial \,V_f\over \partial T}\right)(T')^2+\label{e3}
\ee
$$
+\left(3\,A'+{\lambda' \over V_f}{\partial V_f \over \partial \lambda}+{\lambda'
\over \f}{\partial \f \over \partial \lambda}\right)\,T'-{e^{2A} \over \f \
V_f}\,{\partial V_f \over \partial T}=0\,.
$$

\section{Conformal fixed-point solutions} \label{SecFP}

There are solutions to the equations above that are conformal, and are related
to the fixed points of an ``effective potential''.
To find them we must set $\l'=T'=0$, and $\l''=T''=0$ in the equations
(\ref{e4}), (\ref{e3}) which imply that we must be at a critical point of the
effective potential,
\be
V_{\rm eff}=V_g(\l)-xV_f(\l,T)\sp \pa_T V_{\rm eff}=\pa_{\l} V_{\rm eff}=0
\label{s1}\ee
For any solution $\l_*,T_*$ of the above conditions, we obtain an AdS$_5$ space
with
\be
{12\over \ell^2}=V_{\rm eff}(\l_*,T_*)
\label{s2}\ee
as is obvious from equations (\ref{e1}) and (\ref{e2}).

When $V_f$ has the standard dependence on the tachyon \cite{sen},
there are two solutions to the condition  $\pa_T V_{\rm eff}=0$, namely $T=0$
(chiral symmetry unbroken) and $T\to\infty$, (chiral symmetry broken).
\begin{itemize}

 \item $T=0$. In this case the second condition of extremality (\ref{s1}) is
$\pa_{\l}(V_g(\l)-xV_{f}(\l,0))=0$. In the UV, $\l\to 0$, this potential is
constructed to emulate the perturbative $\beta$-function, and therefore has a
free-field  theory fixed point. This is the UV theory.
In the IR, it will also have a fixed point at $\l=\l_*$. In the BZ region
$\l_*\ll 1$. A priori there are two possibilities.

\begin{enumerate}

\item The fixed point disappears for $x\leq x_*$ for a given $x_*$.

\item The fixed point exists for all $0 < x <{11\over 2}$.

\end{enumerate}

We will discuss these two options later on in this paper.

 \item $T\to \infty$. In this case as $\lim_{T\to \infty}V_f(\l,T)=0$ we obtain
that the second extremality condition is $\pa_{\l}V_g=0$.
This is equivalent to the existence  of a fixed point in large-N YM theory. It
has been argued however in \cite{ihqcd} that this is not true.
Therefore, there is no fixed point with $T\to \infty$.

\end{itemize}

\section{Holographic $\beta$-functions} \label{SecBetas}

In holographic theories we may define non-perturbative $\beta$-functions that
capture the dependence of coupling constants on the RG scale. This concept was
developed and used first in \cite{ihqcd}, in order to explore the physics of
Einstein dilaton theories and their holographic relation to YM theory.
In particular in \cite{ihqcd} it was shown that the $\beta$ function is
intimately related to the generalized superpotential.
It was shown later in \cite{gkmn} that such defined $\beta$-functions are indeed
related to the quantum breaking of scale invariance, and appear in the trace of
the stress-tensor. Such relations were shown in full generality in
\cite{papadimitriou} and have been confirmed also in \cite{johana}.

To define the $\beta$-functions we need a notion of energy scale.
At the two-derivative level, such a function is the scale factor $e^A$, and
indeed near the AdS boundary it can be identified as the energy scale.
It remains always a decreasing function, and becomes zero in the ultimate IR.
It can therefore be taken as the energy scale in the whole of the bulk space.

We therefore define the $\beta$-function and ``anomalous'' dimension $\gamma$
as\footnote{Notice that, as we shall see later, it is the ratio $\gamma(\l,T)/T$
(rather than $\gamma(\l,T)$) which corresponds closely to the anomalous
dimension of the quark mass in QCD. Excluding the extra $T$ in the definition
simplifies many of the equations below.}
\be
{d\l\over dA}\equiv \beta(\l,T) \sp {dT\over dA}\equiv \gamma (\l,T)
\ee
The equations of motion provide equations for the $\beta$-functions.
To obtain them we must convert radial derivatives to derivatives with respect to
$A$,
\be
\l'=A'\beta\sp T'=A'\gamma
\sp \l''=\beta A''+(\g\beta_T+\b\b_{\l})A'^2
\sp T''=\g A''+(\g\g_T+\b\g_{\l})A'^2
\ee
where $\b_T\equiv {\partial\beta\over \p T}$ etc.
Substituting in (\ref{e2}) we obtain
\be
12\left(1-{\b^2\over 9\l^2}\right){A'^2\over e^{2A}}=V_g-{x V_f\over
\sqrt{1+\f\g^2{A'^2\over e^{2A}}}}
\label{e6}\ee
which can in principle be solved for $A'^2\over e^{2A}$ as a function of
$\l,V_g,V_f,\f,\beta,\gamma$. In general, the solution is not
unique.\footnote{Standard ``linear'' scalar theories with action given by a
kinetic term plus a potential have a unique solution for  $A'^2\over e^{2A}$.}
Similarly from (\ref{e1}) we obtain
\be
{A''\over e^{2A}}=-\left(1+{2\b^2\over 9\l^2}\right){A'^2\over e^{2A}}+{V_g-{x
V_f\sqrt{1+\f\g^2{A'^2\over e^{2A}}}}\over 6}
\label{e7}\ee

Then (\ref{e4}) and (\ref{e3}) become
\be
\left(\g{\p \b\over \p T}+\b{\p \b\over \p \l}-{\b^2\over \l}\right){A'^2\over
e^{2A}}+{\b\over 6}\left(2V_g-{x V_f\over
\sqrt{1+\f\g^2{A'^2\over e^{2A}}}}-x V_f\sqrt{1+\f\g^2{A'^2\over e^{2A}}}\right)
\label{e8}\ee
$$=
{3\over 8}\l^2\left[-{\p V_g\over \p \l}+x {\p V_f\over \p \l}
\sqrt{1+\f\g^2{A'^2\over e^{2A}}}+{x\over 2}{\p \f\over
\p\l}{V_f\g^2e^{-2A}A'^2\over \sqrt{1+\f\g^2{A'^2\over e^{2A}}}}\right]
$$
\be
-{1\over \f }\,{\partial\log  V_f \over \partial T}+\left(V_g-{x
V_f\sqrt{1\!+\!\f\g^2{A'^2\over e^{2A}}}}\right){\g\over 6}
+\left(4 +\b~{\partial \,\log V_f\over \partial \lambda}\,+{ \b\over 2}
{\partial \log \f \over \partial \lambda}\right)
\f\g^3{A'^4\over e^{4A}}
\label{e9}\ee
$$
+\left[\g{\p\gamma\over \p T}+\b{\p \g\over \p {\l}}+
\left(2\left(1-{\b^2\over 9 \l^2}\right)+\b {\partial \log( \f V_f) \over
\partial \lambda}\right)\g+\left({1 \over 2}{\partial \log \f \over \partial
T}-{\partial \,\log V_f\over \partial T}\right)\g^2\right]{A'^2\over e^{2A}}=0
$$
where we eliminated $A''$ by using (\ref{e7}).
This is a system of two first-order partial differential equations for $\b,\g$,
with inputs $V_f,V_g,\f$. The equations are highly non-linear.
Setting $x=0$, the system reduces to
\be
12\left(1-{\b^2\over 9\l^2}\right){A'^2\over e^{2A}}=V_g
\label{e10}\ee
\be
\left(\g{\p \b\over \p T}+\b{\p \b\over \p \l}-{\b^2\over \l}
\right){A'^2\over e^{2A}}=
-{3\over 8}\l^2{\p V_g\over \p \l}
-{V_g\over 3}\b
\label{e11}\ee
In this case ${\p \b\over \p T}=0$ and (\ref{e10}) can be rewritten as
\be
{\p \b\over \p \l}={\b\over \l}-\left(4+{9\l^2\over 2\b}{\p \log V_g\over \p
\l}\right)\left(1-{\b^2\over 9\l^2}\right)
\label{e12}\ee
These equations are similar to those in the probe (quenched) limit.

Note also that although the equations of motion are linear in $x$,   the
$\b$-system (\ref{e8},\ref{e9}) is non-linear in $x$, with the understanding
that $A'^2\over e^{2A}$ is to be eliminated using \eqref{e6}.

\subsection{The UV Fixed point} \label{SecBetasUV}

Near the boundary where  $\l,T\to 0$,
we expect that $\gamma \to 0$,
so that equation (\ref{e6}) becomes
\be
12\left(1-{\b^2\over 9\l^2}\right){A'^2\over e^{2A}}=V_g-x
V_f(\l,0)+\cdots~~~\to ~~~{A'^2\over e^{2A}}={V_g-x V_f(\l,0)\over
12\left(1-{\b^2\over 9\l^2}\right)}+\cdots
\label{e14}\ee
According to the discussion of Sec.~\ref{SecBottomup}, the potentials are
expected to be regular at $\l=0$,
\be
V_g=V_0+V_1\l+V_2\l^2+{\cal O}(\l^3)\sp x V_f=W_{0}+W_{1}\l+W_{2}\l^2+{\cal
O}(\l^3)+(Z_0+Z_1\l+Z_2\l^2)T^2+\cdots
\ee
$$
h=h_0+h_1\l +h_2\l^2 + {\cal O}(\l^3) + {\cal O}(T^2)
$$
We also take the following Ansatz for the beta functions:
\be \label{bgUV}
\b=-b_0\l^2+b_1\l^3+\cdots +{\cal O}(T^2)
+\cdots\sp \g=(\g_0+\g_1\l+\g_2\l^2)T+\cdots
\ee
Inserting these into the equations (\ref{e8},\ref{e9}), we find
\be
b_0 = \frac{9}{8} \frac{V_1-W_1}{V_0-W_0} \sp b_1 =
\frac{207}{256}\frac{(V_2-W_2)^2}{(V_0-W_0)^2} - \frac{9}{4}
\frac{V_1-W_1}{V_0-W_0}
\ee
\be \label{egamma0}
\g_0^2 + 4 \g_0 -\frac{24 Z_0}{h_0 W_0 (V_0-W_0)} = 0
\ee
$$
\g_1 = \frac{12
Z_0}{(\g_0+2)h_0W_0(V_0-W_0)}\left(\frac{Z_1}{Z_0}-\frac{W_1}{W_0}-\frac{h_1}{
h_0}-\frac{V_1-W_1}{V_0-W_0}\right)
$$
To leading order the solutions around the UV fixed point are
\be
{1\over \l}={1\over \l_0}+b_0A+\cdots\sp T=T_0e^{\g_0 A}
\ee
In order for $T$ to have the proper UV dimension we must have $\g_0=-1$.
In the massless case, $T$ is dominated by the vev, and we have to choose
$\g_0=-3$. These solve Eq.~\eqref{egamma0} if
\be
 \frac{24 Z_0}{h_0 W_0 (V_0-W_0)} = -3 \ .
\ee

As we shall point out below, the combination $\gamma(\l,T)/T$ is mapped
to the anomalous dimension of the quark mass in QCD~\eqref{QCDgamma}.
Remarkably,
the solution~\eqref{bgUV} is consistent with QCD perturbation theory:
$\gamma(\l,T)/T$ has a series expansion in $\l$.
The leading term is fixed according to the UV dimension, and
the correction terms are identified with the anomalous dimension.

\subsection{Confining IR asymptotics}

In the IR,  a confining asymptotic has the property that  ${A'^2\over e^{2A}}\to
\infty$.
There are two possibilities for the tachyon:

1. If $T=0$, $\g=0$ and then $V_g\to V_g-x V_f(T=0)$.
In this case
\be
{A'^2\over e^{2A}}={V_g-x V_f(\l,0)\over 12\left(1-{\b^2\over
9\l^2}\right)}+\cdots
\ee

2. $T\to\infty$ in the IR,
$V_f\to 0$ exponentially, and we obtain the pure YM case in the IR.
Here
\be
{A'^2\over e^{2A}}={V_g\over 12\left(1-{\b^2\over 9\l^2}\right)}+\cdots
\ee
If we assume that $V_f(\l,0)$ does not grow faster than $V_g(\l)$ in the IR, the
solution in both cases is the same as in pure YM ($x=0$), \cite{ihqcd},
\be
\beta\simeq -{3\l\over 2}\left[1+{3\over 8}{1\over \log \l}+\cdots\right] \ .
\label{bym}\ee
This non-perturbative $\beta$-function indicates that the scaling dimension of
the dual operator $tr(F^2)$ in the IR is $\Delta={5\over 2}$ read from its
linear term. However this does not imply there is a scaling regime in the IR, as
the metric is far from AdS.

The tachyon equation becomes
\be
\,-{\partial\log  V_f \over \partial T}+\f\left(V_g-{x
V_f\sqrt{1\!+\!\f\g^2{A'^2\over e^{2A}}}}\right){\g\over 6}
+\left(4 +\b~{\partial \,\log V_f\over \partial \lambda}\,+{ \b\over 2}
{\partial \log \f \over \partial \lambda}\right)
{\f^2V_g^2~\g^3\over 144\left(1\!-\!{\b^2\over 9\l^2}\right)^2}
\label{ee9}
\ee
$$
+\left[\g{\p\gamma\over \p T}+\b{\p \g\over \p {\l}}+
\left(2\left(1\!-\!{\b^2\over 9\l^2}\right)+{\b }{\partial \log (\f V_f) \over
\partial \lambda}\right)\g+\left({1 \over 2}{\partial
\log \f \over \partial T}-{\partial \,\log V_f\over \partial
T}\right)\g^2\right]
$$
$$
 \times {\f V_g\over 12\left(1-{\b^2\over 9\l^2}\right)}=0
$$
In the IR,  for a class of potentials $\g\to -\infty$ and
$\sqrt{1+\f\g^2{A'^2\over e^{2A}}}\to -\sqrt{{\f
V_g\over 12\left(1-{\b^2\over 9\l^2}\right)}}~\g$.
We expect that  $V_f\sqrt{1+\f\g^2{A'^2\over e^{2A}}}\to 0$ as $V_f$ vanishes
exponentially while $\g$ increases polynomially with increasing $T$.
We also expect $\f V_g $ to be approximately constant and $\left(1-{\b^2\over
9\l^2}\right)\to {3\over 4}$.
We also expect ${\partial \log \f \over \partial T}\to 0$ and $\b{\p \g\over \p
{\l}}$ to be subleading.

Therefore the leading terms in equation (\ref{ee9}) are expected to be
\be
\left(4 +\b~{\partial \,\log V_f\over \partial \lambda}\,+{ \b\over 2} {\partial
\log \f \over \partial \lambda}\right)
{\f V_g~\g^3\over 9}
+\g{\p\gamma\over \p T}-{\partial \,\log V_f\over \partial
T}\g^2=0
\ee

If we define $\delta={1
\over \gamma}$, $\delta$ satisfies a linear equation
\be
{\p \delta\over \p T}+{\p \log V_f\over \p T}\delta={\f V_g\over 9}\left(4
+\b~{\partial \,\log (V_f\sqrt{\f})\over \partial
\lambda}\,\right)
\ee
with solution
\be
\delta ={C\over V_f}+{\f V_g\over 9V_f}\int^T_{T_*} V_f\left(4 +\b~{\partial
\,\log (V_f\sqrt{\f})\over \partial \lambda}\,\right) dT
\ee
with $T^*$ large enough so that we are in the IR regime.
For $\gamma$ we obtain
\be
\gamma={V_f\over {C}+{\f V_g\over 9}\int^T_{T_*} V_f\left(4 +\b~{\partial \,\log
(V_f\sqrt{\f})\over \partial \lambda}\,\right) dT}
\ee

Note that if $\f\sim \l^{-{4\over 3}}$, then $\b~{\partial \,\log
(\sqrt{\f})\over \partial \lambda}\simeq 1$. If we also assume  that $V_f$ has
factorized dependence, $V_f=V_{f\l}(\l)V_{fT}(T)$ we obtain

\be
\gamma={V_f\over {C}+{\f V_g\over 9}\left(5 +\b~{\partial \,\log (V_{f\l})\over
\partial \lambda}\,\right)\int^T_{T_*} V_f dT}
\ee
Note that from the whole one-family of solutions above, only one is diverging
for large $T$ and this is the only reliable solution.

For a tachyon potential of the form $V_{fT}=e^{-aT^2}$ the diverging solution is
\be
\g\simeq -{18a\over{\f V_g}\left(5 +\b~{\partial \,\log (V_{f\l})\over \partial
\lambda}\,\right)}T+\cdots.
\ee
 For the type of potentials we will be using $\f V_g/\sqrt{\log \l}$ approaches
a constant which we denote by $b$, and
$V_{f\l} \sim \l^2$
so that
\be \label{gammaasfin}
\g\simeq -{18a\over b\sqrt{\log\l}\left(2 -{9  \over 8}{1\over \log \l}
+\cdots\right)}T+\cdots\simeq- {9 a\over b\sqrt{\log\l}}T+\cdots
\ee

Note that if we assume a more elaborate tachyon potential of the form
$V_f=V_{f0}(\l)e^{-a(\l)T^2}$ with $a(\l)$ increasing with $\l$
in the IR then the asymptotic behavior of $\gamma$ changes and it vanishes
in the IR.
To capture this behavior from (\ref{ee9}) we take $\gamma\to 0$ in the IR
and $\sqrt{1+\f\g^2{A'^2\over e^{2A}}}\to 1$.
We also again expect that $\f V_g $ is approximately constant,
$\left(1-{\b^2\over
9\l^2}\right)\to {3\over 4}$, and ${\partial \log \f \over \partial T}\to 0$.
We also expect that the derivative terms $\g{\p\gamma\over \p T}$ and
$\b{\p \g\over \p{\l}}$ are subleading.

The two leading terms in (\ref{ee9}) are then expected to be
\be
-{\partial\log  V_f \over \partial T}+
{\f V_g~\b\over 9}{\partial \log (\f V_f) \over
\partial \lambda}\g =0
\ee
The solution  behaves for large T as
\be
\g\simeq {18a(\l)\over \b h V_g a'(\l)}{1\over T}+\cdots \simeq
-\frac{12}{b \frac{d \log a}{d \log \l} \sqrt{\log \l} }{1\over T} +\cdots
\ee
where $b$ is defined as above.

\subsection{Some simple $\beta$ functions}

We may engineer a $\beta$ function that interpolates between the one-loop
perturbative YM $\beta$-function, and the non-perturbative one in (\ref{bym}).
We could also have a $\gamma$ function interpolating between an operator with
dimension $\Delta_\mathrm{UV}$ in the UV, and $\Delta_\mathrm{IR}$ in the IR:
\be
\b(\l)=-b_0{\l^2\over 1+{2b_0\over 3}\l}\sp
\gamma=(\Delta_\mathrm{UV}-4)T{1+(\Delta_\mathrm{IR}-4)T^2\over
1+(\Delta_\mathrm{UV}-4)T^2}
\ee
The flow equations can be integrated to
\be
{1\over \l}-{2b_0\over 3}\log\l=b_0 A\sp {\log T\over
(\Delta_\mathrm{UV}-4)}+{\Delta_\mathrm{UV}-\Delta_\mathrm{IR}\over
2(\Delta_\mathrm{UV}-4)(\Delta_\mathrm{IR}-4)}\log\left(1+(\Delta_\mathrm{IR}
-4)T^2\right)=A
\ee
Such $\beta$ functions can be converted into potentials $V_g,V_f$, via the
equations (\ref{e8}), (\ref{e9}).

\begin{figure}
\centering
\includegraphics[width=0.33\textwidth]{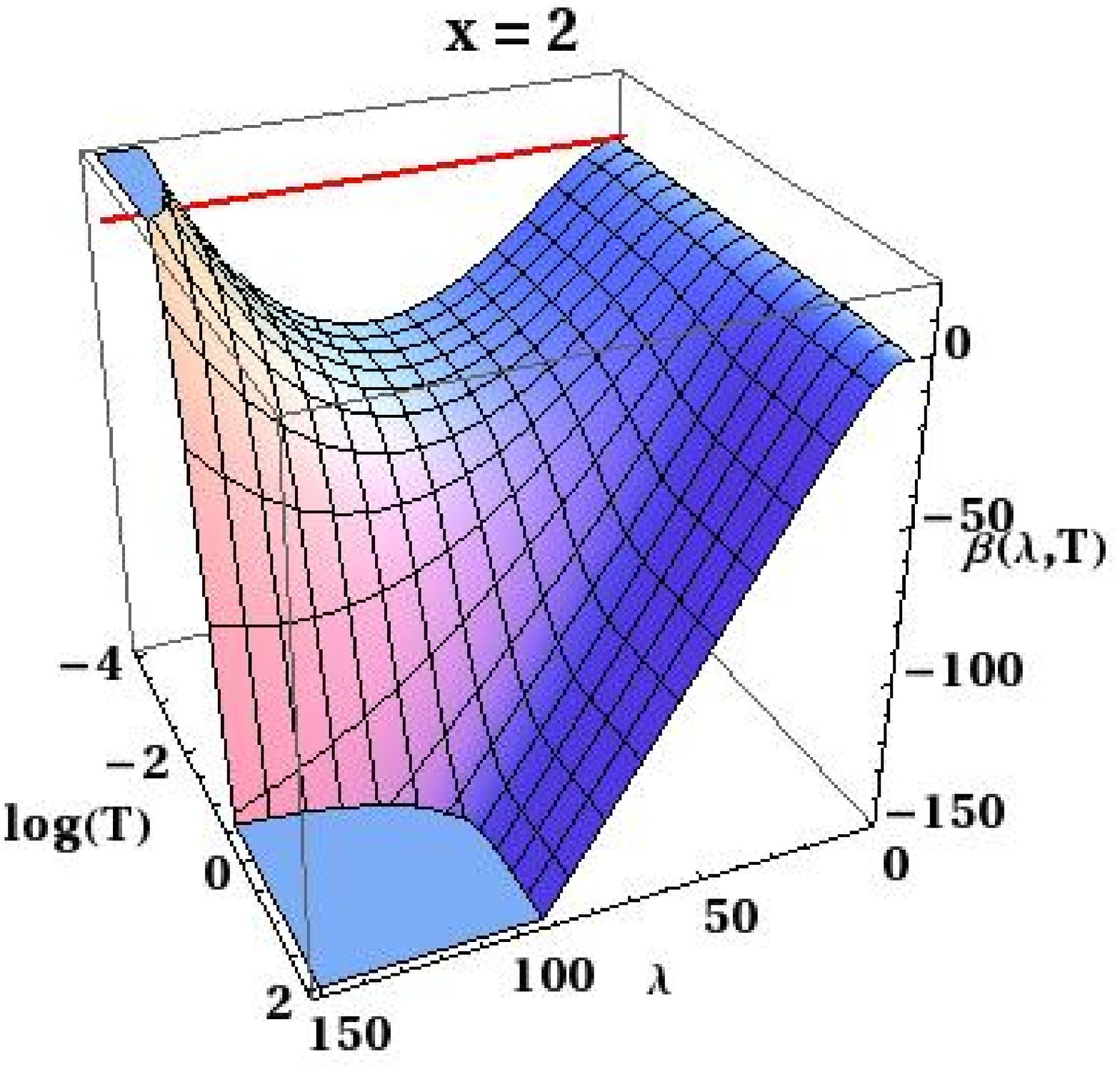}%
\includegraphics[width=0.33\textwidth]{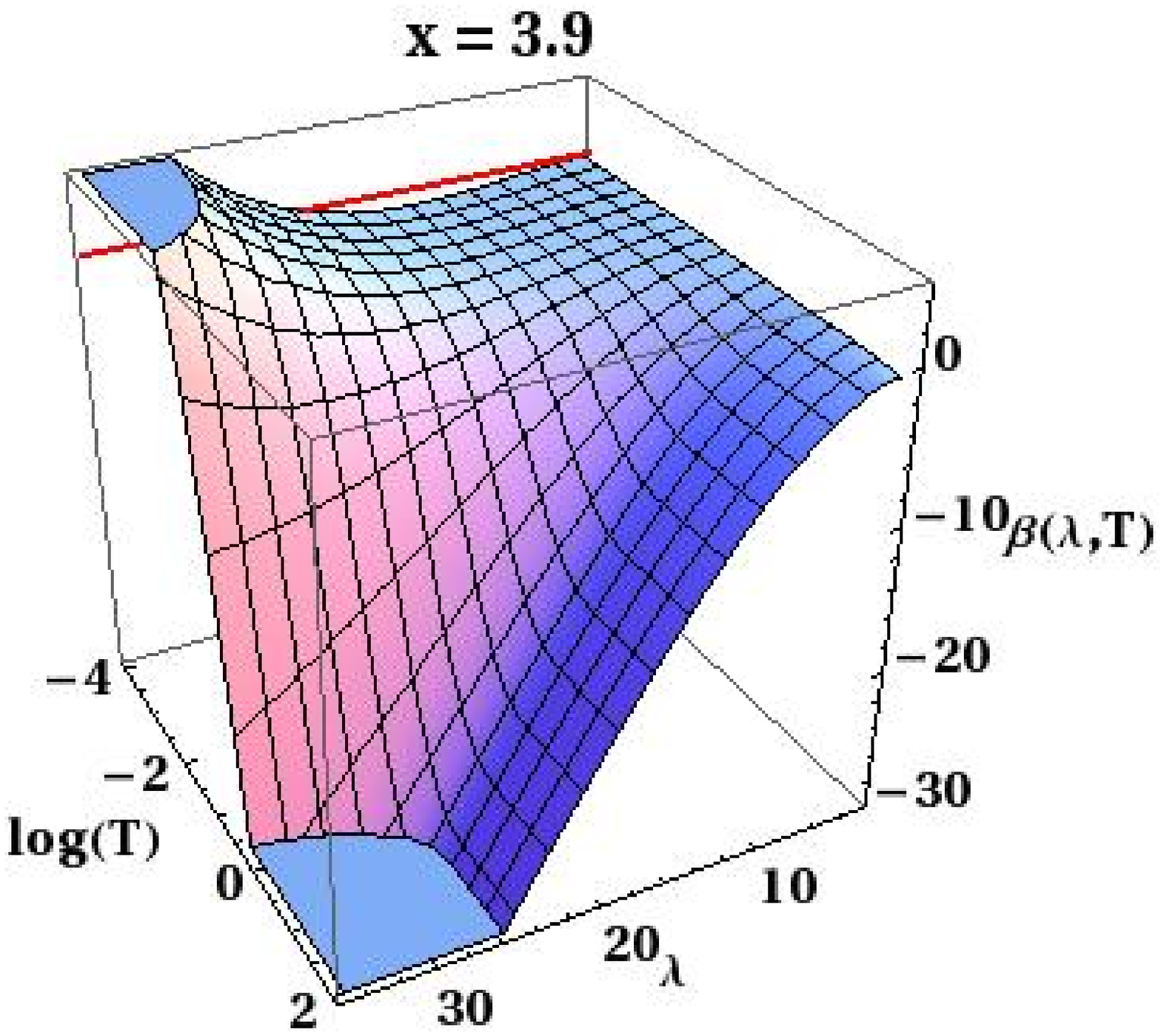}%
\includegraphics[width=0.33\textwidth]{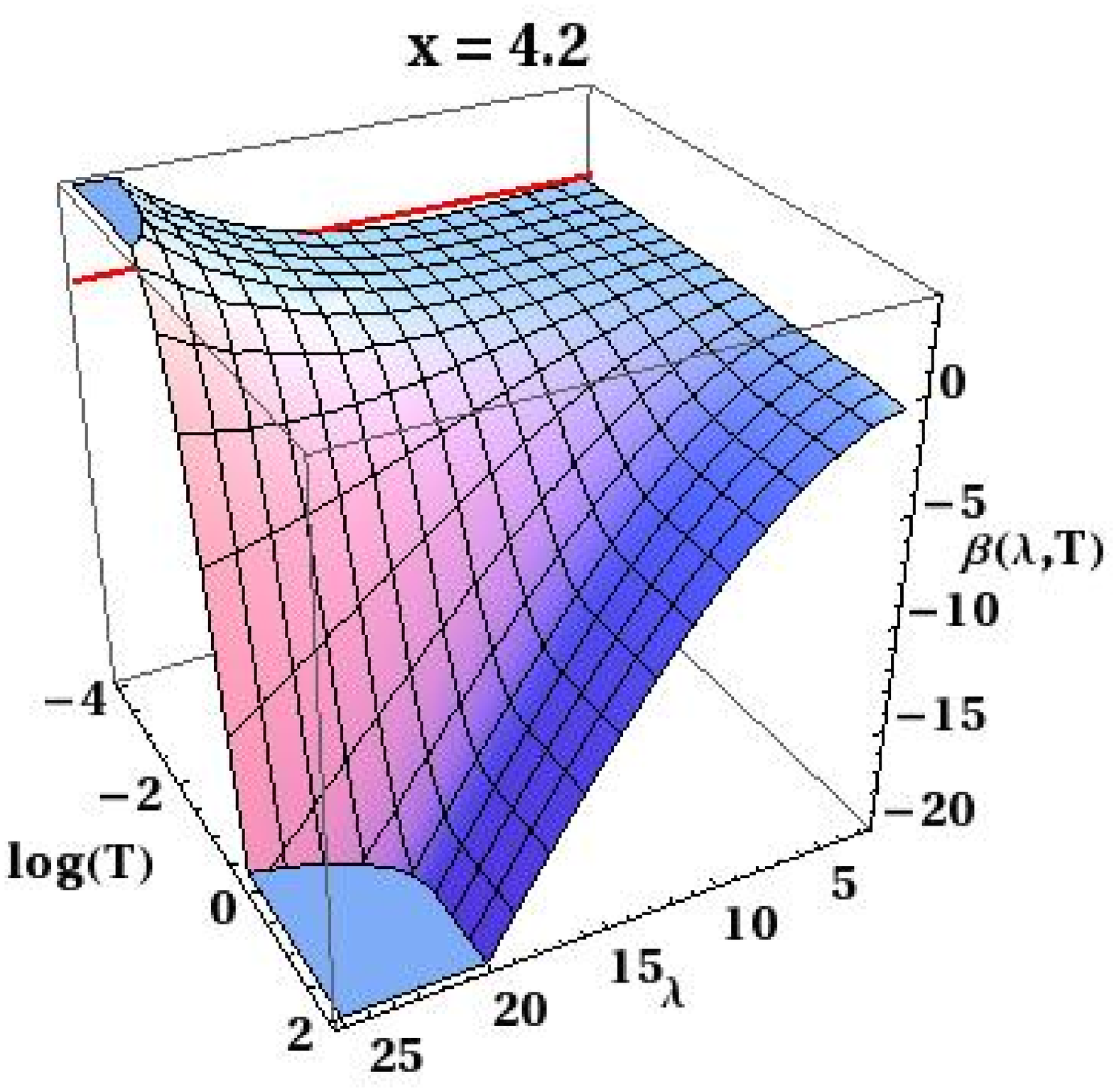}\vspace{-5mm}
\includegraphics[width=0.33\textwidth]{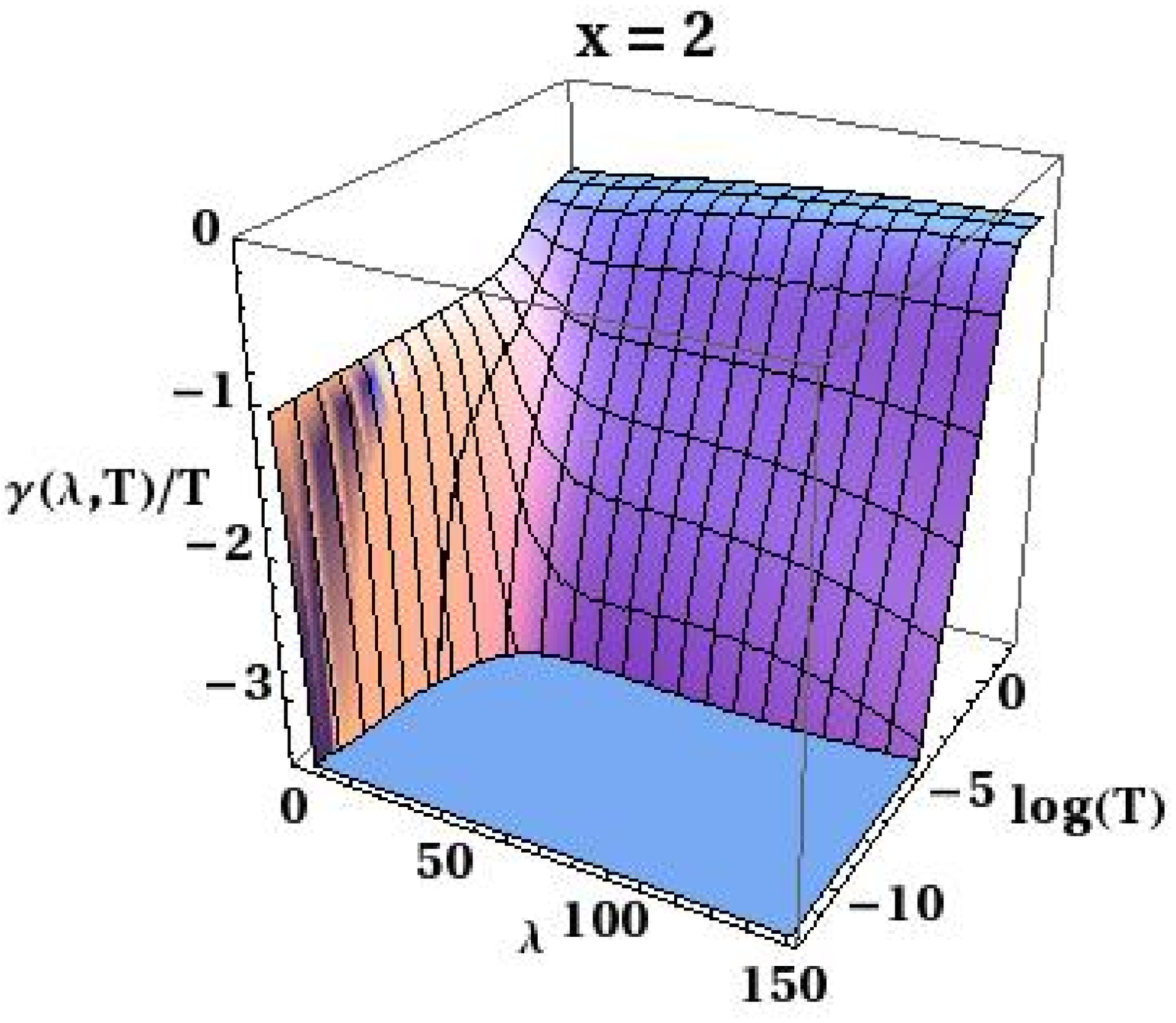}%
\includegraphics[width=0.33\textwidth]{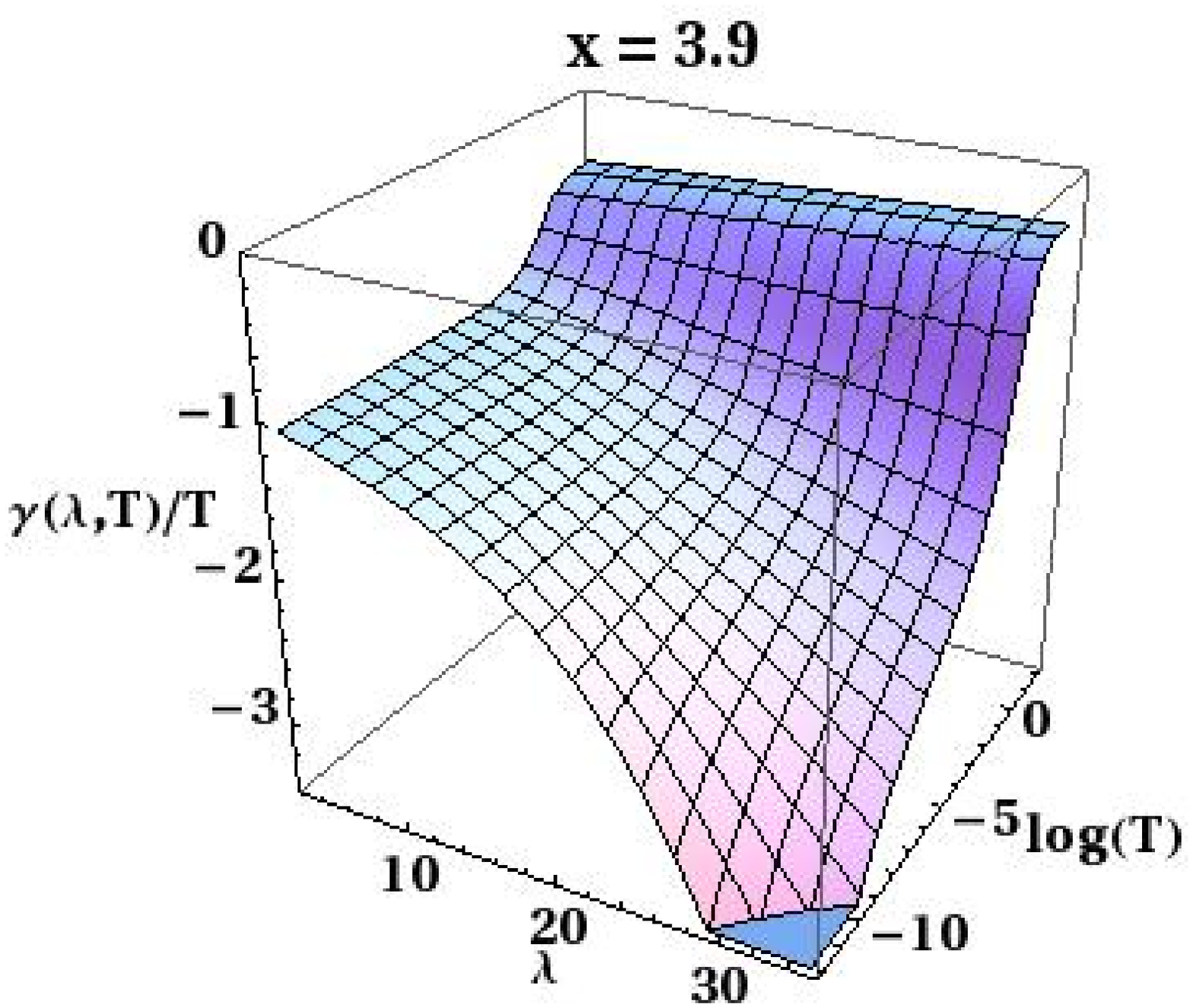}%
\includegraphics[width=0.33\textwidth]{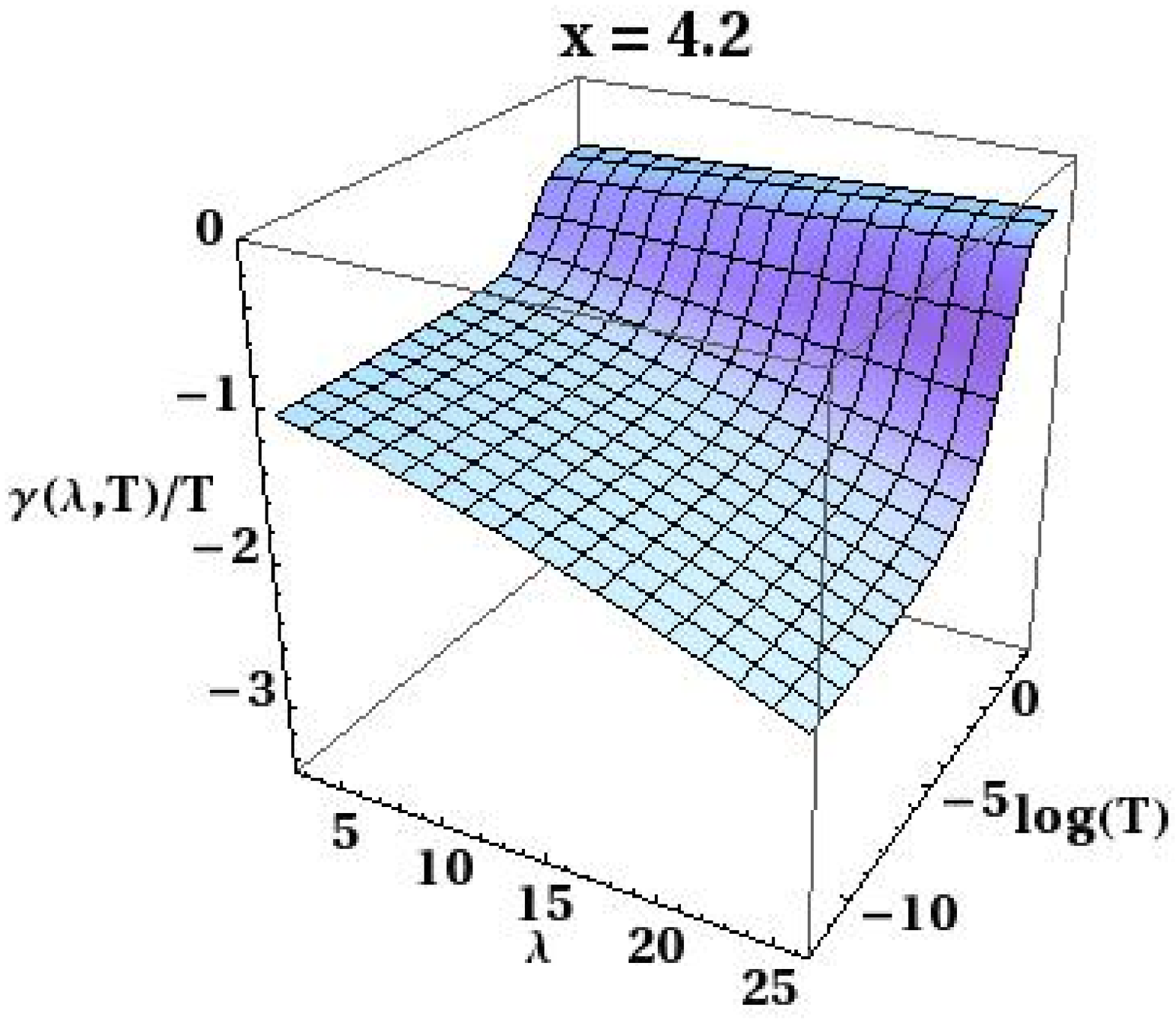}
\caption{The solutions for the $\beta$ (top row) and $\gamma$ (bottom row)
functions for $x=2$ (left), $x=3.9$ (middle), and $x=4.2$ (right). We added the
red lines on the top row at $\beta=0$ in order to show the location of the fixed
point. See the text for a detailed explanation.}
\label{figbetanum}
\end{figure}

\subsection{Numerical solutions for the $\beta$ and $\gamma$ functions}

The equations~\eqref{e6}, \eqref{e8}, and~\eqref{e9} for the $\beta$ and
$\gamma$ functions can be solved numerically for fixed potentials by combining
solutions evaluated along the RG flow, as detailed in Appendix~\ref{AppBeta}.
Fig.~\ref{figbetanum} shows the results for various values of $x$. We used the
potentials of scenario~I from Appendix~\ref{AppPotentials} and required that the
solutions flow to the good IR singularity, as explained in
Appendix~\ref{AppBeta}. Notice that the values $x=3.9$ and $4.2$ were chosen to
lie slightly above and below the edge of the conformal window, which is at $x_c
\simeq 3.9959$ for the potentials used in the plots.

The plots of the beta functions show a smooth transition as $T$ evolves from
small $\ll 1$ to large $\gtrsim 1$ values, reflecting the expectation of
Sec.~\ref{SecFP}. For small tachyon, the beta function has a fixed point
corresponding to the maximum of the effective potential $V_\mathrm{eff}$, which
moves to lower values of $\l$ as $x$ is increased, whereas for large $T$ the
$\l$-dependence of the beta function approaches the Yang-Mills form. The
$\gamma$ functions show a transition between the small and large $T$ regions as
well. For large $T$ and $\l$, the solutions agree with Eq.~\eqref{gammaasfin}.
For small $T$, the structure is richer. For $x=2$ the $\gamma$ tends to a
constant value as $T \to 0$ (except for very small $\l$) so that
$\gamma(\l,T)/T$, which is plotted in Fig.~\ref{figbetanum}, diverges. This
behavior is pushed for larger $\l$ as we increase $x$ to $3.9$, and has
disappeared for $x=4.2$, and the gamma function is instead linear in $T$ in
accordance with Eq.~\eqref{bgUV}. These changes are related to the tachyon
changing sign during the flow. If the tachyon has a zero, we hit the $T=0$ line
before reaching the UV singularity. Then $\gamma$ approaches
a constant value as $T \to 0$,
and  Eq.~\eqref{bgUV} does not apply. As will be discussed in detail below,
tachyon zeroes are indeed expected for $x<x_c$. Notice also that $\gamma/T$
tends to $-1$ as $\l \to 0$ with fixed $T$ at least for small $T$, reflecting
the expected value of $\gamma_0$ in Eq.~\eqref{bgUV}.

\section{The background solutions} \label{SecBG}

\subsection{Generic properties of the background}

 We start by discussing the symmetries and integration constants of the
equations of motion of V-QCD, \eqref{e1}-\eqref{e3}. We shall assume the
exponential Ansatz
\be
 V_f(\l,T) = V_{f0}(\l) \exp\left[-a(\l)T^2\right]
\ee
for the potential of the tachyon DBI action. Then the background EoMs have the
following symmetries
 \bea
 \mathrm{1.}\qquad    &&A  \to A + \log \Lambda \ ,\qquad    V_g \to
\Lambda^{-2} V_g  \ , \qquad V_{f0} \to \Lambda^{-2} V_{f0}  \\\nn
     &&T \to \Lambda T\ ,\qquad a  \to \Lambda^{-2} a \ ; \\
\mathrm{2.}\qquad && T \to \Lambda T\ ,\qquad a  \to \Lambda^{-2} a \ , \qquad h
\to \Lambda^{-2} h \ ; \\
 \label{transf}
\mathrm{3.}\qquad && r \to \Lambda(r-r_0)  \ , \qquad  A \to A - \log \Lambda \
.
 \eea
The first symmetry can be used to fix the value of the UV AdS  radius $\ell$,
and is usually associated with the units of energy in the boundary theory. The
second one will be used to fix the normalization of $h$ in the UV. The third one
is essentially different from the first two, since it does not involve the
potentials. It will therefore remain as a true symmetry of the background
solutions.

We may choose a set of independent equations of motion which contains one first
order and two second order differential equations. Therefore, their general
solution includes five integration constants. These can be identified as the
coefficients of the UV expansions of the fields as follows (assuming that the
solution has the standard UV singularity with $\l \to 0$, see
Appendix~\ref{AppUV}). The tachyon UV expansion has the usual free constants
related to the normalizable and non-normalizable solutions, identified as the
quark mass $m$ and the vacuum expectation value $\sigma$ of the $\bar q q$
operator, respectively.

 In close analogy, the solution for $\l$ involves two constants, identified as
the UV scale $\Lambda=\Lambda_\mathrm{UV}$ of the expansions, and another
constant $\hat A$ which we will define in Section~\ref{SecFE}, related to the
gluon condensate and the free energy of the system. The fifth integration
constant is simply the location of the UV singularity which can take to be $r=0$
by the translation symmetry of Eq.~\eqref{transf}.
We shall require that the system has a repulsive, ``good'' kind of IR
singularity, which fixes the values of the condensates $\sigma$ and $\hat A$ in
terms of $m$ and $\Lambda_\mathrm{UV}$.

In addition, we still have the scaling symmetry of Eq.~\eqref{transf}, which can
be used to vary the units of all constants, and reflects the corresponding scale
transformation of the dual field theory. Therefore, the single parameter which
characterizes all nontrivially linked physical backgrounds (for fixed potentials
and, in particular fixed $x=N_f/N_c$) is the ratio of the ``source''
coefficients $m/\Lambda_\mathrm{UV}$\footnote{Due to practical reasons we shall
often use a parameter $T_0$ linked to the tachyon behavior in the IR instead of
the quark mass to characterize different backgrounds.}.

For the choices of the functions $\f$, $V_g$, and $V_f$ of interest to us, the
tachyon typically decouples from the other fields asymptotically in the UV and
in the IR.
The UV and IR asymptotics are discussed in detail in Appendices~\ref{AppUV}
and~\ref{AppIR}, and we shall repeat only the main features of the physically
interesting possibilities here. The physically relevant UV asymptotics are
restricted. As pointed out in \cite{ihqcd}, the fields $A$ and $\l$ can be
expanded in $-1/\log r$ at the UV boundary $r=0$ in the probe limit.

 Similar expansions work also at finite $x=N_f/N_c$. The tachyon is required to
vanish linearly  $T(r) \simeq m r$ (or faster if $m=0$) in $r$ in the UV. Taking
$\e =-1/\log r \to 0$, the tachyon $T \simeq m \exp(-1/\e)$ vanishes
exponentially while $A$ and $\l$ have power-like behavior in $\e$. Since, in
addition, the functions $\f$ and $V_f$ must be regular in the UV (see
Sec.~\ref{SecBGUV} below), the tachyon can be set to zero in the leading order
action. We find that $A$ and $\l$ satisfy their probe limit equations of motion,
but with the dilaton potential $V_g$ replaced by $V_{\rm eff}(\l) = V_g(\l) - x
V_f(\l,T=0)$, which also verifies that the UV expansions of $\l$ and $A$ have
the same form as in the probe limit.

We shall only discuss cases where the tachyon indeed decouples asymptotically in
the IR. This is the case,
if we
take a tachyon potential having an exponential $T$ dependence, $V_f(\l,T)
\propto \exp(-a T^2)$, and the tachyon has a power-law (or faster) divergence as
$r \to \infty$. This is indeed what is suggested by tachyon condensation in
string theory.
 Therefore, the flavor part of the action is exponentially suppressed in the IR.
Consequently, the tachyon decouples asymptotically from $A$ and $\l$, and their
asymptotic expansions in the IR have exactly the same form as in the probe
limit, and are determined by the potential $V_g(\l)$.

In summary, even though all fields couple nontrivially for general values of the
coordinate, the probe limit description will be valid in the UV and IR. In
particular, this guarantees that the interpretation of the integration constants
is the same for finite $x$ as in the probe limit. The decoupling of the tachyon
in the IR leads to the system having similar ``good'' IR singularities as in the
probe case.

An important difference with respect to the probe limit discussion is that the
potentials which characterize the backgrounds in the UV and IR regions will be,
in general, qualitatively different. As discussed above in Section \ref{SecFP},
the potential $V_{\rm eff}(\l) = V_g(\l) - x V_f(\l,T=0)$ which controls the UV
behavior will be chosen such that it admits a fixed point (extremum of the
potential) at least for large values of $x$ as required by the Banks-Zaks
analysis.

The fixed point of $V_\mathrm{eff}$ will play an important role in the dynamics
in the intermediate region between UV and IR. In particular, for identically
vanishing tachyon,  the background simply flows from the UV fixed point at
$\l=0$ to the IR fixed point at finite $\l$. Adding, a tiny quark mass (or
chiral condensate), the solution in the UV region will not be changed
drastically. However, no matter how small the quark mass is,  the tachyon will
eventually become large and start coupling to $\l$ and $A$, which will drive the
system away from the IR fixed point.

 We can make two observations.
First, the special case of zero tachyon will be essentially different from all
other solutions. This solution has
vanishing quark mass and chiral condensate, and is therefore identified as the
solution corresponding to conserved chiral symmetry. We will discuss this
special case separately below.

Second, there is a possibility to have solutions which come very close to the IR
fixed point, but are eventually driven away by the increasing tachyon. Such
solutions will be identified with quasiconformal or ``walking'' dynamics of the
dual field theory, where the coupling constant remains approximately fixed over
a large range of energies. We shall see below how the phase structure of QCD in
the Veneziano limit, which includes a quasiconformal region, arises in our class
of models.

\subsection{Matching UV behavior with the QCD $\beta$-functions} \label{SecBGUV}

We shall now discuss the most important links of the potential functions to the
physics of the dual field theory.
We start by an analysis of the UV region,  where the behavior of the system can
be mapped to the QCD $\beta$-functions \cite{ihqcd} as already discussed in
Section~\ref{SecBetas}. In particular, in the probe limit  ($x \to 0$ limit),
the UV behavior is controlled by $V_g(\l)$, that has the expansion
\bea
 \label{Vgexp}
 V_g(\l) &=& V_0 + V_1 \l + V_2 \l^2 + \cdots
\eea
Here $V_0>0$ can be freely chosen and it fixes the AdS scale for $x=0$ as $V_0 =
12/\ell_0^2$. The other coefficients can be mapped to the Yang-Mills
$\beta$-function. At one-loop order we have \cite{ihqcd}
\be \label{YMVexp}
 V_g(\l)  = \frac{12}{\ell_0^2}\left[1 + \frac{8}{9}b_{0}^{\rm YM}\l
+ \cdots \right]
\ee
where $b_{0}̂^{\rm YM}$ is the one-loop coefficient of the $\beta$-function,
from which $V_1$ can be solved.

Moreover, as discussed above,
at finite $x$ the UV behavior is similar to the probe limit, but the role of
$V_g$ is taken by the potential $V_{\rm eff}(\l) = V_g(\l) - x V_f(\l,T=0)$.
Therefore, we take
\bea
 V_f(\l,T=0) &=& W_0 +W_1 \l +W_ 2 \l^2 + \cdots
\label{Vfexp}
\eea
and the relation to the QCD $\beta$-function reads
\be \label{QCDVexp}
V_{\rm eff}(\l) = V_g(\l)-xV_f(\l,0)  = \frac{12}{\ell^2}\left[1 +
\frac{8}{9}b_0\l
+ \cdots \right]
\ee
where $b_{0}$ is the leading coefficient of the $\beta$-function in the
Veneziano limit.

Similarly to $V_0$, the coefficient $W_0$ can be freely chosen, and  the other
coefficients can be solved from Eq.~\ref{QCDVexp}.
However, there are constraints: the AdS scale must remain positive for all
$0<x<11/2$, and $W_0$ should also be positive
(see Appendix~\ref{AppPotentials}). These boil down to
\be \label{ineqs}
 0 < W_0 < \frac{2}{11} V_0 \ .
\ee

In addition, as pointed out in Section~\ref{SecBetas}, we can map the UV
behavior of the tachyon action to the (anomalous) dimensions of the quark mass
and the chiral condensate of the dual field theory. For definiteness, we
parametrize $V_f(\l,T) = V_{f0}(\l)\exp(-a(\l)T^2)$ and assume that $\f$ depends
only on $\l$. Then the dimension of the quark mass constrains the UV behavior of
$\f(\l)$ and $a(\l)$. A detailed analysis is done in Appendix~\ref{TUV}.

Remarkably, assuming that the potential
functions have analytic expansions at $\l=0$ is
consistent with QCD perturbation theory. Indeed, requiring the dimension of the
quark mass to approach one in the deep UV fixes the leading terms as
\be
 \frac{\f(\l)}{a(\l)} = \frac{2 \ell^2}{3}\left(1+\f_1\l+\cdots \right) \ ,
\ee
and the next-to-leading coefficient $\f_1$ can be matched with the one-loop
anomalous dimension of the quark mass $\gamma_m(\l)$.

By using Eq.~\eqref{UVgammam} from Appendix, we obtain
\be
 - \gamma_0  = \frac{9}{8}\left[\frac{4}{3}\frac{8}{9}b_0 +
\frac{4}{3}\f_1\right]
\ee
where $\gamma_0$ is the leading coefficient of the anomalous dimension,  $-
d\log m/d\log\mu = \gamma_m(\l) = \gamma_0 \l +\cdots$. Notice that, for
example, the non-normalizable term in the tachyon solution~\eqref{TUVres} reads
after this identification
\be
 \frac{1}{\ell}T(r) = m r \left(-\log (r\Lambda)\right)^{-\gamma_0/b_0}\left[1 +
\mathcal{O}\left(\frac{1}{\log(r\Lambda)}\right)\right]
\ee
where the logarithmic correction is consistent with the one-loop solution for
the running quark mass in QCD.

\subsection{Condensate dimension and the edge of the conformal window}
\label{SecIRFPdim}

The most important new feature of the system discussed in this article is the
description of the phase diagram of QCD as a function of $x=N_f/N_c$ in the
Veneziano limit.
We shall now indicate which potentials lead to the desired structure. This
constraint to a large extent independent of the one discussed above that was set
by the UV expansions.

The phase structure is linked to
the dimension of the chiral condensate at the IR fixed point which is found at
the maximum of $V_\mathrm{eff} = V_g-xV_{f0}$ in the conformal window (in the
limit of small tachyon background).

We denote the value of the coupling at the fixed point by $\l_*$ so that
\be \label{lstardef}
 V_\mathrm{eff}'(\l) = V_g'(\l_*)-xV_{f0}'(\l_*)=0 \ .
\ee From 
the action (\ref{lagrang}) we can calculate the IR AdS scale
 \be
 {12\over \ell^2_\mathrm{IR}}=  V_\mathrm{eff}(\l_*)
 \ee
To calculate the dimension we need
the mass of $T$ in the IR fixed point, which can be extracted from the
action~\eqref{sf} and reads
\be
m^2_\mathrm{IR}=-{2a(\l_*)\over \f(\l_*)} \ ,
\ee
where we again parametrized 
$V_f(\l,T) = V_{f0}(\l)\exp(-a(\l)T^2)$. From this we obtain
\be \label{Tmasst}
\Delta_\mathrm{IR}(4-\Delta_\mathrm{IR})=-m^2_\mathrm{IR}\ell^2_\mathrm{IR}={24
a(\l_*)\over \f(\l_*)(V_\mathrm{eff}(\l_*)} \ .
\ee

The desired phase diagram is only obtained if the expression
$\Delta_\mathrm{IR}(4-\Delta_\mathrm{IR}) = -m_\mathrm{IR}^2 \ell_\mathrm{IR}^2$
has a certain dependence on $x$.
It must start at a value smaller than 4 at the end of the BZ region 
(which is guaranteed as
the standard UV boundary conditions fix it to $3$ there), grow as $x$ decreases,
and become 4 at some value of $x$ so that the BF bound
\cite{Breitenlohner:1982bm} is saturated \cite{Kaplan}. 
Indeed, the solutions near the edge of
the conformal window stabilize such that the critical $x_c$, defining the
location of the conformal phase transition for massless quarks, is determined by
\be \label{dimcond}
 \Delta_\mathrm{IR}(4-\Delta_\mathrm{IR})\Big|_{x=x_c} = 4 \ .
\ee
We shall discuss why this is the case in detail later on.
Note that in addition to the explicit dependence on $x$ in the factor $x
V_{f0}$, Eq.~\eqref{Tmasst} depends on $x$ through $\l_*$ due to the definition
of Eq.~\eqref{lstardef}.

We stress that the saturation of the BF bound means that for theories near the
critical $x_c$, $\Delta_\mathrm{IR} \to 2$, or in other words, the anomalous
dimension of the quark mass at the fixed point $\gamma_* \to 1$. That is, our
model reproduces the standard assumption for the energy dependence of the chiral
condensate near the edge of the conformal window. Recall that this is extremely
important for realizations of walking technicolor.

If the potentials are matched with the UV physics of QCD, the
expression~\eqref{Tmasst} can be further simplified. Fixing the tachyon mass in
the deep UV we obtain $a(0)/\f(0)=3/(2\ell^2)$. In the limit $T \to 0$, the beta
function $\beta(\l)=d\l/dA$ satisfies a first order differential equation which
depends on $V_\mathrm{eff}(\l)$, as can be immediately concluded by comparing
our action to the probe limit one \cite{ihqcd} for $T=0$. In terms of the phase
function $X$, defined by
\be \label{betadef}
 \frac{d\l}{dA} = \beta(\l) = 3\l X(\l) \ ,
\ee
we find \cite{ihqcd}
\be \label{Xde}
 \l \frac{dX(\l)}{d\l} = - \left( 8 X(\l)+ 3 \l \frac{d\log
V_\mathrm{eff}(\l)}{d\l}\right) \frac{1-X^2(\l)}{6 X(\l)} \ .
\ee
Solving this, we can parametrize the potential in terms of the
$\beta$-function:
\be
 V_\mathrm{eff}(\l) = V_{g}(\l)-x V_{f0}(\l) =
\frac{12}{\ell^2}\left(1-\frac{\beta(\l)^2}{9\l^2}\right)
\exp\left[-\frac{8}{9}\int_0^{\l}\frac{d\hat \l \beta(\hat \l)}{\hat
\l^2}\right] \ .
\ee
The IR dimensions now satisfy
\bea \label{Tmasst2}
\Delta_\mathrm{IR}(4-\Delta_\mathrm{IR}) &=& 24 {a(\l_*)\over
\f(\l_*)}\frac{\ell^2}{12}\left(1-\frac{\beta(\l_*)^2}{9
\l_*^2}\right)^{-1}\exp\left[\frac{8}{9}\int_0^{\l_*}\frac{d\l
\beta(\l)}{\l^2}\right]  \\\nn
 &=& 3 \frac{\f(0)}{\f(\l_*)}
\frac{a(\l_*)}{a(0)}\exp\left[\frac{8}{9}\int_0^{\l_*}\frac{d\l
\beta(\l)}{\l^2}\right] \ .
\eea
where in the second line we used the fact that $\beta(\l_*)=0$. If we do the
matching with perturbative QCD results in the UV, the $\beta$-function can be
identified as the QCD $\beta$-function in the Veneziano limit. Then the
exponential factor is small in the sense that it is roughly proportional to $b_0
\l_*$ where both the one-loop coefficient of the $\beta$-function $b_0$ and the
coupling at the fixed point $\l_*$ vanish in the BZ limit $x \to 11/2$.
Therefore we expect that $\Delta_\mathrm{IR}(4-\Delta_\mathrm{IR})$ depends on
$x$ mostly through the functions $\f$ and $a$. In the two-loop approximation of
the $\beta$-function we obtain
\be
 \frac{8}{9}\int_0^{\l_*}\frac{d\l \beta(\l)}{\l^2} = -\frac{4}{9} b_0 \l_* \ .
\ee

\begin{figure}
\centering
\includegraphics[width=0.5\textwidth]{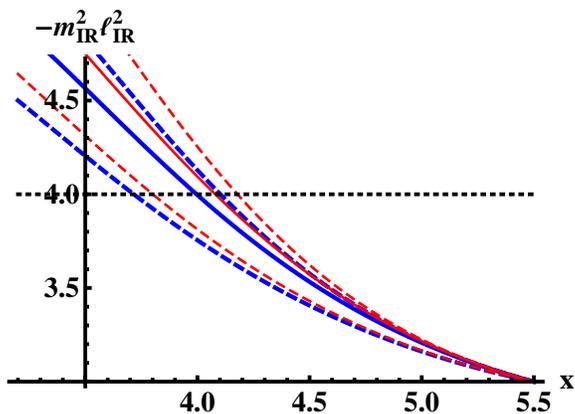}
\caption{The squared tachyon mass at the IR fixed point
$\Delta_\mathrm{IR}(4-\Delta_\mathrm{IR}) = -m_\mathrm{IR}^2 \ell_\mathrm{IR}^2$
as a function of $x$ for the two scenarios described in
Appendix~\protect\ref{AppPotentials}. Thick blue, and thin red curves are the results
for the scenarios I and II of Appendix~\protect\ref{AppPotentials}, respectively. The
dashed lines show the maximal changes as $W_0$ is varied from $0$ (upper curves)
to $24/11$ (lower curves).}
\label{figTmass}
\end{figure}

In Fig.~\ref{figTmass} we plot the dependence of
$\Delta_\mathrm{IR}(4-\Delta_\mathrm{IR})$ on $x$ for the two scenarios of
potential choices of Appendix~\ref{AppPotentials}. The solid lines give the
results for the value $W_0 = 12/11$ used in the Appendix, and the dashed lines
show the sensitivity of the result for the choice of $W_0$ as this parameter is
varied over its allowed range.
In all cases, $\Delta_\mathrm{IR}(4-\Delta_\mathrm{IR})$ intersects the value of
4, shown as the horizontal dotted line in the plot. This suggests that the class
of potentials that has the desired phase structure quite in general includes
those ones that are matched with the UV behavior of QCD. Moreover, the critical
value of $x$ is found within a quite narrow band
\be
3.7 \lesssim ~~x_c~~ \lesssim 4.2\;\;\;,
\label{cwi}\ee
and the largest source of uncertainty is the choice of $W_0$.\footnote{We have
checked this also for some additional potentials that are not discussed in this
article.} We stress that there is no strict bound on the allowed values of
$x_c$, but the numbers in Eq.~\eqref{cwi} rather give the expected magnitude for
the variation of $x_c$  within natural potential choices.

\begin{figure}
\centering
\includegraphics[width=0.5\textwidth]{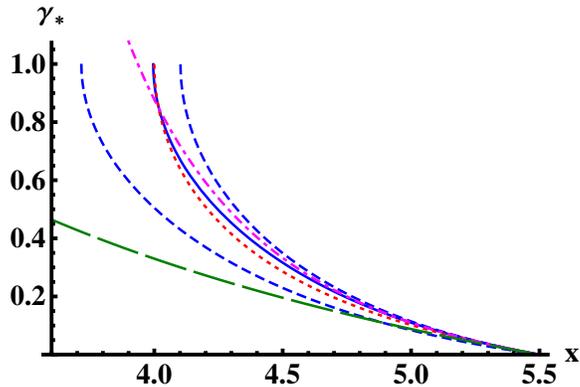}
\caption{The anomalous dimension of the quark mass at the IR fixed point
as a function of $x$ within the conformal window in various approaches.
The solid blue curve is our result
for the scenario I of Appendix~\protect\ref{AppPotentials}. The
dashed blue lines show the maximal change as $W_0$ is varied from $0$ 
(upper curve) to $24/11$ (lower curve).  
 The dotted red curve is the result from 
a Dyson-Schwinger analysis, the dot-dashed magenta curve is the prediction of 
two-loop perturbative QCD, and
the long-dashed green curve is based on an all-orders $\beta$-function.}
\label{figgammastar}
\end{figure}

It is also instructive to show the dependence of 
the anomalous dimension of the quark mass at the IR fixed point
on $x$ (see Fig.~\ref{figgammastar}). 
Within our model the anomalous dimension is
defined by $\gamma_*=\Delta_\mathrm{IR}-1$ where 
$\Delta_\mathrm{IR}$ is the smaller of the two roots. 
The result for the potentials of scenario I of Appendix~\ref{AppPotentials} 
is shown as the solid blue line in Fig.~\ref{figgammastar}, 
and as in Fig.~\ref{figTmass}, the dashed lines show 
the maximal variation as the parameter $W_0$ is varied over its allowed range.

Our result is very similar to the prediction obtained
by calculating the anomalous dimension from 
Dyson-Schwinger equations in the rainbow approximation \cite{Appelquist:1988yc},
and evaluating the result at the zero of the perturbative two-loop
$\beta$-function~\eqref{QCDbetaVen} (dotted red curve of 
Fig.~\ref{figgammastar}). As in our model,
the conformal window ends at the point where $\gamma_*$ reaches one 
and becomes complex in this approach.
The deviation from the simple perturbative estimate 
(dot-dashed magenta curve), which was obtained by 
using the two-loop anomalous dimension~\eqref{QCDgamma} instead, 
is also small. However, an all-orders 
$\beta$-function \cite{pica} (long-dashed green curve) predicts much
smaller values of $\gamma_*$ for low values of $x$.

\subsection{Constructing the background solutions} \label{secconstr}

 We will select concrete potentials for evaluating the background numerically.
These potentials must satisfy the constraints of the previous two subsections in
order to produce the desired phase structure of QCD in the Veneziano limit as
well as the perturbative UV physics. In addition, the system needs to have an
acceptable IR singularity where the tachyon diverges, which adds extra
requirements to the large $\l$ behavior of the potentials. However, as we shall
demonstrate briefly, different choices for the IR behavior do not change the
qualitative features of the background.

The construction of explicit potentials is detailed in
Appendix~\ref{AppPotentials}, where also the IR behavior is fixed by using the
generic  analysis of Appendix~\ref{AppIR}. For clarity, we repeat the final
result here. The potentials of the scenario I in Appendix~\ref{AppPotentials}
read
\bea
\label{Vgnumdef} V_g(\l) &=& 12 +  \frac{44}{9\pi^2} \l + \frac{4619}{3888\pi^4}
\frac{\l^2}{(1+\l/(8\pi^2))^{2/3}}\sqrt{1+\log(1+\l/(8\pi^2))} \\
 V_f(\l,T) &=& V_{f0}(\l) e^{-a(\l) T^2} \\
\label{V0numdef} V_{f0}(\l) &=& \frac{12}{11} +  \frac{4 (33-2 x)}{99 \pi ^2}\l
+  \frac{23473-2726 x+92 x^2}{42768 \pi ^4} \l^2 \\
 a(\l) &=& \frac{3}{22} (11 - x) \\
 \f(\l) &=&  \frac{1}{\left(1+\frac{115-16 x}{288 \pi ^2}\l\right)^{4/3}} \ .
\label{fnumdef}
\eea
We shall use this choice when calculating the background numerically, unless
stated otherwise.

Recall that only three of the equations of motion \eqref{e1}-\eqref{e3} are
independent. We choose a set of three second-order equations for the numerical
calculations, and treat the remaining first order equation as an extra
constraint. As usual, we choose boundary conditions that satisfy the constraint,
which is then automatically satisfied for all $r$.\footnote{In all numerical
calculations, we shall do the coordinate transformation from $r$ to $A$
discussed in appendix \ref{rA}, because after this,  the UV structure of the
solutions is reproduced more accurately. However, as this transformation is
straightforward, we shall continue to discuss the solution in the
``$r$-space''.}

When solving the equations numerically, it turns out that shooting from the IR
is numerically stable.\footnote{There is a potential instability related to the
constraint, as it is exactly satisfied only near  the IR cutoff due to numerical
effects. For general $r$ there is an error which may grow exponentially as we
solve the system towards the UV. However, we have the freedom of modifying the
system  of second order equations, which is used to calculate the solution, by
adding multiples of the constraint to some of the equations. In this way the
error can be made to decrease exponentially instead of growing towards the UV,
so that the instability is removed.}
We shall first discuss backgrounds where the tachyon has a nontrivial profile.
We fix the boundary conditions in the deep IR by using the asymptotic IR
expansions at the ``good'' IR singularity at $r=\infty$ (see
Appendix~\ref{goodIR}). For the functions given above,
the asymptotics becomes
\bea \label{IRexps}
 A(r) &=& -\frac{r^2}{R^2}+\frac{1}{2}\log\frac{r}{R}-\log R
+\frac{13}{8}+\log\left[\frac{27\ 6^{1/4}}{\sqrt{4619}}\right] -\frac{173\
R^2}{3456\ r^2} + \cdots \\
 \log \l(r) &=& \frac{3}{2}\frac{r^2}{R^2} -\frac{39}{16}+\log 8
\pi^2-\frac{151\ R^2}{2304\ r^2} + \cdots \nn \\ \nn
 T(r) &\sim& T_0 \exp\left[\frac{81\ 3^{5/6} (115-16 x)^{4/3} (11-x)}{812944\
2^{1/6}}\frac{r}{R}\right] \ .
\eea
Here we already used the translation symmetry of Eqs.~\eqref{transf} to set
$r_0$ in the formulas \eqref{IRresA} and \eqref{IRresl} to zero,\footnote{Notice
that with this choice of fixing the translation invariance, the UV boundary of
the final solution will not be at $r=0$. After obtaining the solution, we can
relax this condition and use translation invariance again to move the UV
singularity at $r=0$, if desired.} and we can further use the scaling symmetry
to fix $R=1$. After this, the solution (for fixed $x$) depends nontrivially on
only one free parameter $T_0$ in the IR, as is characteristic for a good IR
singularity.

To continue, we choose an IR cutoff $R_{\rm IR}$ such that the tachyon is large,
and therefore decoupled from the fields $A$ and $\l$. With the above choices of
potentials, a sufficient value turns out to be $T(R_{\rm IR})=70$: with this
choice, the tachyon has decoupled to a good precision, and $R_{\rm IR}/R$ is
large enough for the asymptotic expansions to work. We use the asymptotic
expansions to fix the values of $A$, $\l$, $T$, $A'$, and $T'$ at $r=R_{\rm
IR}$, and solve $\l'$ from the constraint (the first order EoM). The solution is
then obtained by numerically solving the set of second order equations of motion
toward the UV, until some of the UV asymptotics described in
Appendix~\ref{AppUV} are reached.

\begin{figure}
\centering
\includegraphics[width=0.4\textwidth]{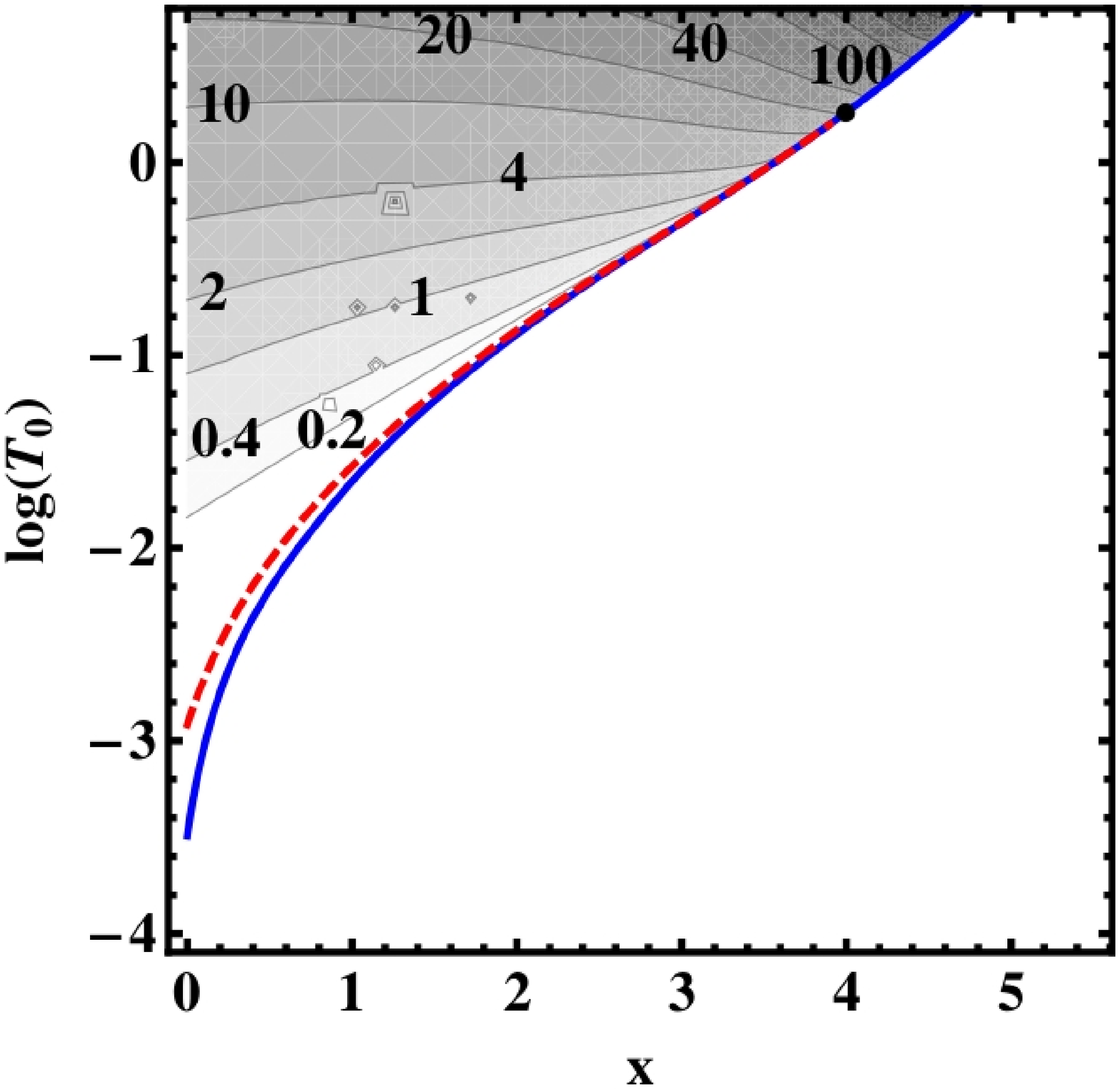}%
\hspace{1cm}
\includegraphics[width=0.4\textwidth]{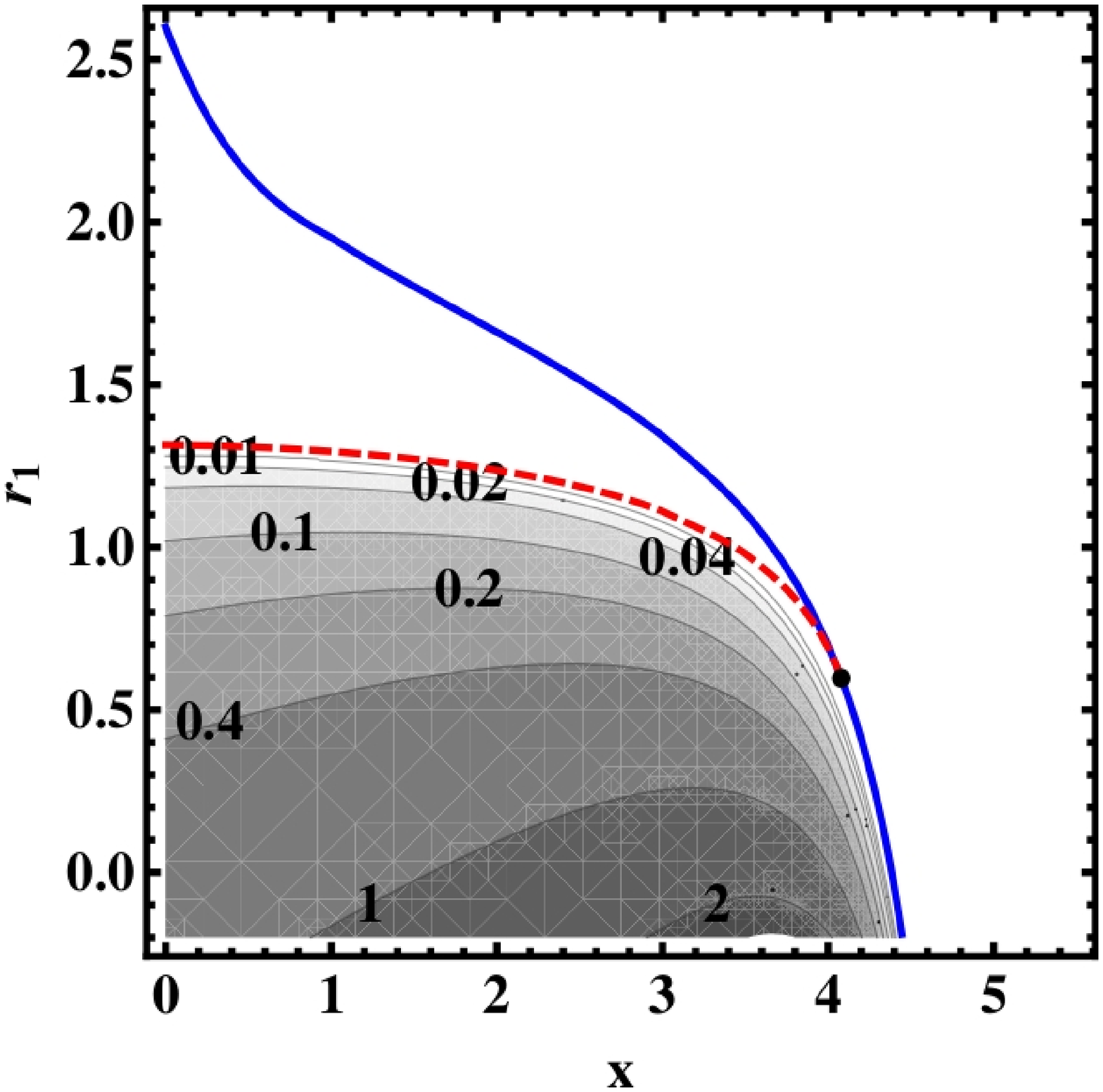}%
\caption{The UV behavior of the background solutions with good IR singularity
for the scenario I (left) and scenario II (right) defined in
Appendix~\protect\ref{AppPotentials}.
The thick blue curve represents a change in the UV behavior, the red dashed
curve has zero quark mass, and the contours give the quark mass. The black dot
where the zero mass curve terminates lies at the critical value $x=x_c$. For
scenario I (II) we have $x_c \simeq 3.9959$ ($x_c\simeq 4.0797$). See the text
for detailed explanation.}
\label{figuvbeh}
\end{figure}

The obtained UV behavior is depicted in Fig.~\ref{figuvbeh} (left) as a function
of the only remaining free parameters, $T_0$ and $x$. For comparison, we also
present the same plot for the scenario II of Appendix~\ref{AppPotentials} on the
right hand side, where $T_0$ is replaced by the parameter $r_1$ of
Eq.~\eqref{TasympscII}. For clarity, we shall only refer
to the variable $T_0$ in the
discussion below.
Because of the invariance of the Lagrangian under $T \to -T$, scanning over
positive $T_0$ is enough to catalog all possible solutions.  We shall explain
the notation and the results here, and discuss at qualitative level how the
structure arises, whereas the details are discussed in Appendix~\ref{AppBG}. We
shall also show concrete examples of the backgrounds as well as their $\beta$-
and $\gamma$-functions along the holographic RG flow below.

The solid blue line represents a change in the UV asymptotics. The standard UV
asymptotics of Appendix~\ref{TUV} is obtained left of the blue line, whereas
right of the blue line the solution ``bounces back''  at finite $\l$, as
discussed in Appendix~\ref{UVbounceback}. In the bounce back region
the $\beta$-function evaluated
along the RG flow becomes zero at finite $\l$, after which $\l$ starts growing
toward the UV, and the standard UV singularity is not reached. Left the blue
solid curve, the ``standard'' tachyon UV expansion of Appendix~\ref{TUV} defines
the quark mass and the chiral condensate, which are not defined right of, or on,
the blue line. The red dashed line has $m=0$. This line exists only for small
$x$ and terminates at a critical value $x_c \simeq 3.9959$ (in scenario I),
which matches with the definition of Eq.~\eqref{dimcond}. We stress that the
solution with zero quark mass does not exist for $x\geq x_c$.
Right of the red dashed line, the quark mass takes positive values and is
monotonic in $T_0$. The contours give the quark mass, obtained by fitting the
deep UV behavior of the tachyon solution to the expansion of
Appendix~\ref{AppUV}.\footnote{The mass here is given in IR units, i.e., we
actually plot $mR=m/\Lambda_\mathrm{IR}$, since we fixed the scale $R$ of the IR
expansions to unity in the numerics.} The backgrounds in the contoured region
will be identified as the physical ones, having lowest free energy.

\begin{figure}
\centering
\includegraphics[width=0.45\textwidth]{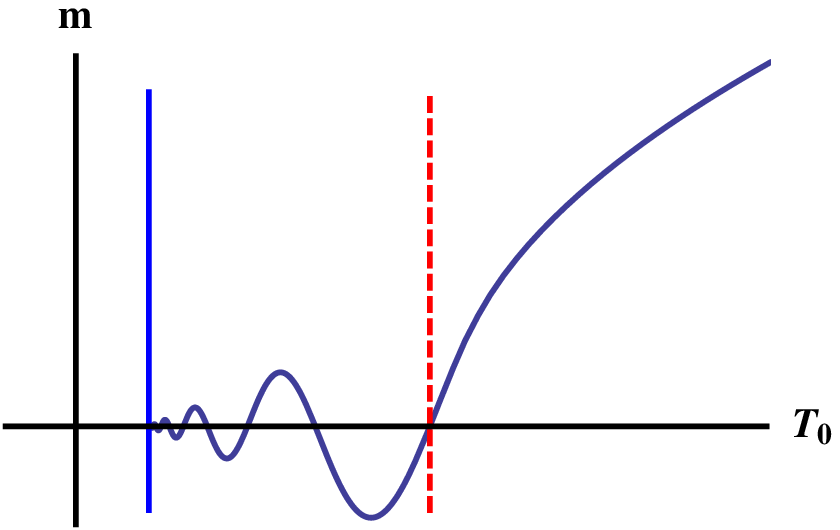}%
\hspace{5mm}
\includegraphics[width=0.45\textwidth]{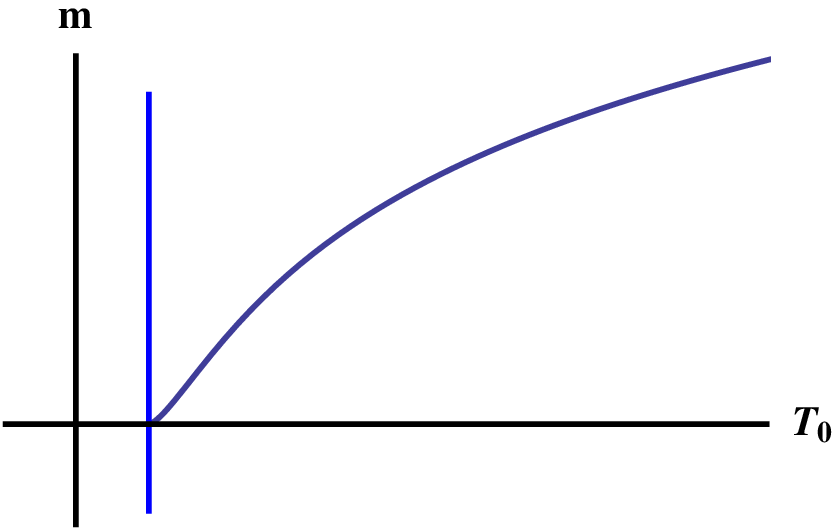}%
\caption{Left figure: Plot of the UV Mass parameter $m$,  as a function of the
IR $T_0$ scale, for $x<x_c$.
Right figure: Similar plot for $x\geq x_c$. The vertical solid blue and dashed
red lines show where corresponding lines are intersected in Fig.~\protect\ref{figuvbeh}.
While these plots are for model functions, similar plots on how the bare quark
mass $m$ depends on $T_0$, for different $x$ for the potentials of scenario I,
can be found in figure \protect\ref{figxscan} in Appendix~\protect\ref{AppBG}.}
\label{mT0}
\end{figure}

The value of the mass goes to zero when the solid blue (for $x>x_c$) or the red
dashed (for $x<x_c$) curves are approached from within the contoured region as
shown in Fig.~\ref{mT0}, which is not evident from Fig.~\ref{figuvbeh} due to
limited resolution. Between the blue solid and red dashed curves, the
quark mass is small, but depends on $x$ and $T_0$ in a complicated manner, as we
shall discuss below.

The right hand plot in Fig.~\ref{figuvbeh}, which was obtained after modifying
the
potentials in the IR, shows similar qualitative features as the left hand plot.
This is the case because the structure seen in Fig.~\ref{figuvbeh} arises from
the
 behavior of the solutions in the UV region and close to it, which is
analytically tractable and almost independent of the change of the
  potentials, if, e.g., we keep the quark mass fixed (see the discussion in
   Appendix~\ref{AppBG}). However, the mapping to the IR asymptotics
    (in particular to $T_0$ or $r_1$) is completely different in the two
     scenarios, which causes the differences between the plots.

Let us then discuss how the structure of Fig.~\ref{figuvbeh}
arises from the background solutions (see Appendix~\ref{AppBG} for a detailed
analysis). The main point is that the closer we are to the thick blue curve
(when approaching the curve from the right), the closer the background is to
reaching the IR fixed point, when the tachyon finally grows large and drives the
system away from it. Therefore the backgrounds near the blue curve will be
quasiconformal, or ``walking'', so that the coupling $\l$ is approximately
constant
over a large range of energies. The mass dependence can be then understood be
studying the tachyon EoM and in particular the tachyon mass at the IR fixed
point. Notice that as the red dashed curve of solutions with zero quark mass
ends on the blue curve as $x \to x_c$, quasiconformal behavior is expected in
this region.

To understand the behavior near the blue and red curves it is important to plot
the UV parameter, $m$,
versus the IR parameter $T_0$. We show this in figure  \ref{mT0}
 for $x<x_c$ (left) and $x \geq x_c$ (right).\footnote{As $m$ and $-m$ are
related by a chiral rotation by $\pi$, we expect that we can take $m\geq 0$. The
chiral rotation is reflected in the background solutions in the symmetry $T \to
-T$. Consequently, we can turn the negative mass solutions of Fig.~\ref{mT0} to
solutions with positive mass $|m|$ by changing $T_0 \to -T_0$.}
For $x \geq x_c$, there is a unique saddle point (regular classical
solution) for each value of the quark mass.

The situation for $x<x_c$ is more complex and reflects the fact
 that the chiral condensate operator violates the BF bound in
 the (potential) IR fixed point.\footnote{Notice that we have chosen potentials
where the fixed point exist even at arbitrary small positive values of $x$. For
another choice of potentials where the fixed point exist only up to a positive
limiting value $x_*$ (see Sec.~\ref{SecFP}), the structure is expected to be the
one described here at least for $x_*<x<x_c$, and the structure in the region $x
\leq x_*$ can be analyzed numerically.}
We see from the left of figure  \ref{mT0} that for each $m>0$,
 there is a finite number of regular classical solutions that
 we will label, by an integer $n=1,2,\ldots$, with $n=1$ being
 the rightmost solution in the figure (the one having the
  maximum $T_0$).

Understanding of the qualitative shape of the left figure \ref{mT0} is
given by the fact that for $x<x_c$,  $T\ll 1$ and $\l$ near the
fixed point value $\l_*$, the approximate solution of the tachyon is given by
$T\sim r^2\sin\left[k\log r+\phi\right]$. Note that the constant $k$ is fixed
for fixed $x$
 but the normalization and $\phi$ are determined by the boundary conditions and
IR regularity.
Therefore the tachyon starts at the boundary, evolves into the sinusoidal form
for a while, and then at end diverges.
Different solutions differ in the region in which they are sinusoidal,
 and it is this region that controls their number of zeros. This is explained in
more detail in Appendix~\ref{AppBG}.

For the solutions with high label $n$, tachyon changes sign several times before
diverging in the IR. As we move to the left toward the vertical blue line in
Fig.~\ref{mT0} (left), a new zero in the tachyon solution appears every time the
mass curve crosses the horizontal axis.
For $m=0$ we expect an infinite number of
regular
 solutions for all positive integers $n\geq 1$.
The presence of several such solutions reflect the violation
of the BF bound, and are reflecting the Efimov minima seen in other
contexts (see \cite{Kaplan,Jensen}). This agrees also with similar
 recent observations in \cite{kutasov}.

The hint of such multiple regular solutions/saddle points was
 seen already in \cite{ikp} that treated the flavor sector in
  the quenched approximation.
Indeed, a second regular solution was seen beyond the dominant one.
In that case a calculation of the spectrum of mesons in this
second solution indicated that this
saddle point was unstable, as the spectra were tachyonic.

It is interesting to point out that the presence of the Efimov-like
 tower of regular solutions is not tied uniquely to the existence
  of an IR fixed point solution
in the landscape of the bulk theory that violates the BF bound.
Even modifications of the potentials
that do not allow
 this IR fixed point may still have the Efimov tower. The qualitative
  reason is that once the bulk theory has a regime that is near
  critical, this is enough to trigger the presence of such multiple saddle
points.
  This is indeed the case in \cite{ikp} where in the quenched
  approximation a second solution exists even though there are no such IR fixed
point. We have checked
that exactly the same happens here in the probe limit $x\to 0$, where the fixed
point is absent, if the potentials of scenario~I are used. Moreover, for the
scenario~II the full Efimov tower remains even in the probe limit.

There is a more detailed analysis of the regular solutions in appendix
\ref{AppBG}.
The comparison of the free energies of the various regular solutions is made in
section \ref{SecFE}.

\subsection{Background solutions at vanishing quark mass}

To summarize the results of the above analysis, we identified the solutions of
the contoured region of Fig.~\ref{figuvbeh} as the physical ones. For $x<x_c$ we
found solutions for all $m \geq 0$, whereas for $x\geq 0$ we found that $m>0$ so
that the solution for $m=0$ was absent.

 We will discuss now how the background solutions  vary as we move around
Fig.~\ref{figuvbeh} (left).
  We will start with the massless case, where also an additional special
background exist (for all $x$), for which the tachyon is identically zero.

\begin{figure}
\centering
\includegraphics[width=0.38\textwidth]{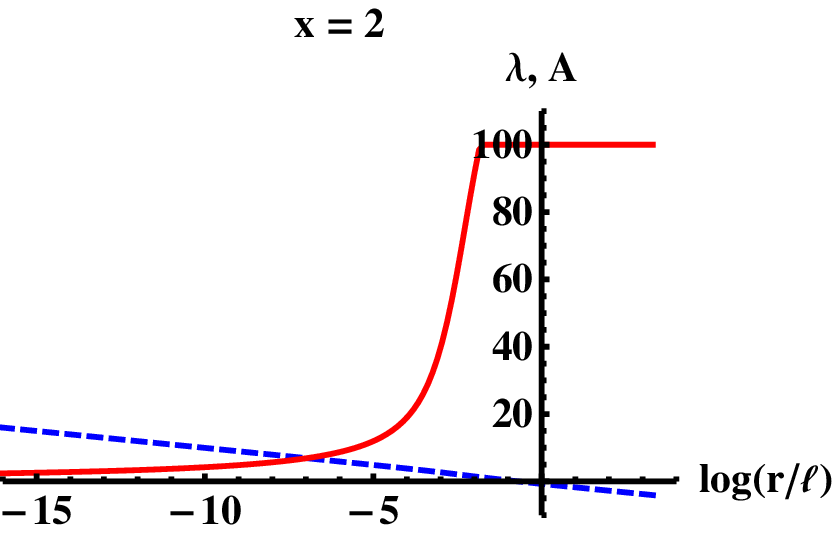}%
\hspace{1cm}
\includegraphics[width=0.38\textwidth]{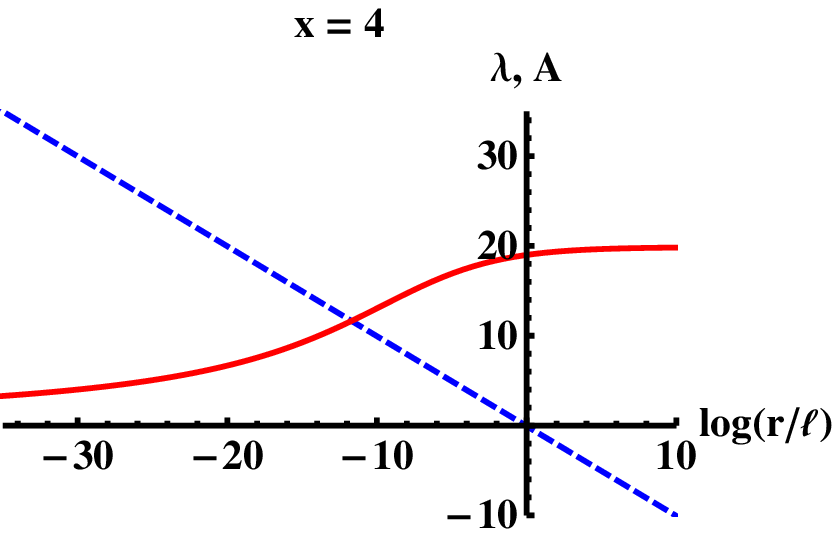}

\caption{The backgrounds with identically vanishing tachyon for $x=2$ (left) and
$x=4$ (right).
The red solid, and blue dashed curves are the values of $\l$, and $A$ as
functions of $\log(r/\ell)$, respectively.}
\label{figbackgroundT0}
\end{figure}

\subsubsection{Solution with identically vanishing tachyon} \label{SecBGT0}

Let us start by analyzing the special solution with vanishing tachyon, which is
not included in Fig.~\ref{figuvbeh}. In the absence of the tachyon, the system
is otherwise the one studied in \cite{ihqcd}, but with the dilaton potential
$V_g$ replaced by $V_\mathrm{eff}=V_g-xV_{f0}$. This potential is guaranteed to
have a maximum, corresponding to an IR fixed point, in the Banks-Zaks region
since we matched it with the QCD $\beta$-function. For the choice of
Eqs.~\eqref{Vgnumdef}-\eqref{fnumdef} the fixed point actually exists for all
$0<x<11/2$.\footnote{This is also the case for the potential associated to
scenario II, described in appendix \ref{AppPotentials}. We could also slightly
modify the potential so that the fixed point, disappears for $x<x_*<x_c$. We
have also analyzed such a case and find only minor differences from those
analyzed in detail  in this paper.}

 There is a single solution to the equations of motion that reaches the IR fixed
point, described in Sec.~\ref{TzeroIR} of Appendix~\ref{AppIR}. It is similar to
the backgrounds studied in \cite{Jarvinen:2009fe} where a $\beta$-function
inspired by supersymmetry was used. Similar backgrounds were also studied at
finite temperature in \cite{kajantieIRFP}.
We identify this special solution as the
background corresponding to the chiral symmetry conserving phase, as the vev
$\sigma$ vanishes. It is easy to construct the background numerically by
shooting from the vicinity of the IR fixed point, e.g., by using the expansions 
of Appendix~\ref{TzeroIR}.

  We plot the background for $x=2$ and $4$ in Fig.~\ref{figbackgroundT0}, where
we fixed the scale $R$ of the IR expansions~\eqref{lexpT0} and~\eqref{AexpT0} to
one. The geometry is expected to asymptote to AdS both in the UV and in the IR,
reflecting the flow from the IR fixed point to the standard UV one.
Actually $A$ is very closely linear in $\log r$ for all $r$, so that the
deviation from AdS is not visible in the plots. Similar observation was made in
\cite{Jarvinen:2009fe}. While the solution with vanishing tachyon exist for all
$x$, we will show in the next section that the other massless solution which
involves a nontrivial tachyon profile and therefore chiral symmetry breaking
(red dashed line in Fig.~\ref{figuvbeh}) has lower free energy whenever it
exist, i.e., for $x<x_c$. Therefore, this background which correspond to a field
theory flowing to an IR fixed point, is the physical one (for massless quarks)
only for $x\geq x_c$, and $x_c$ is indeed the edge of the conformal window.

\begin{figure}
\centering
\includegraphics[width=0.38\textwidth]{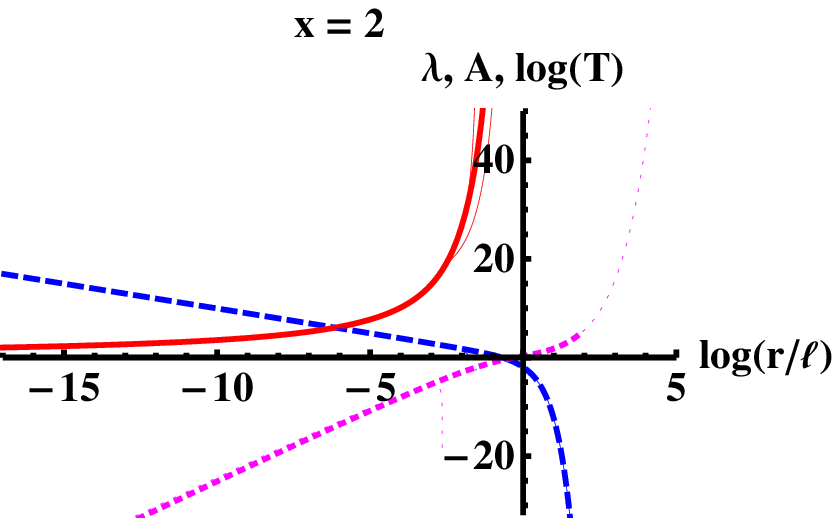}%
\hspace{1cm}
\includegraphics[width=0.38\textwidth]{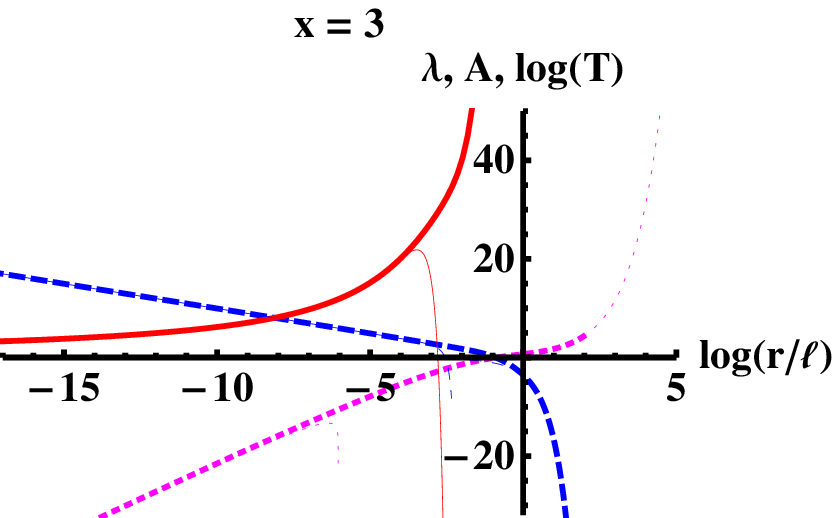}

\vspace{5mm}

\includegraphics[width=0.38\textwidth]{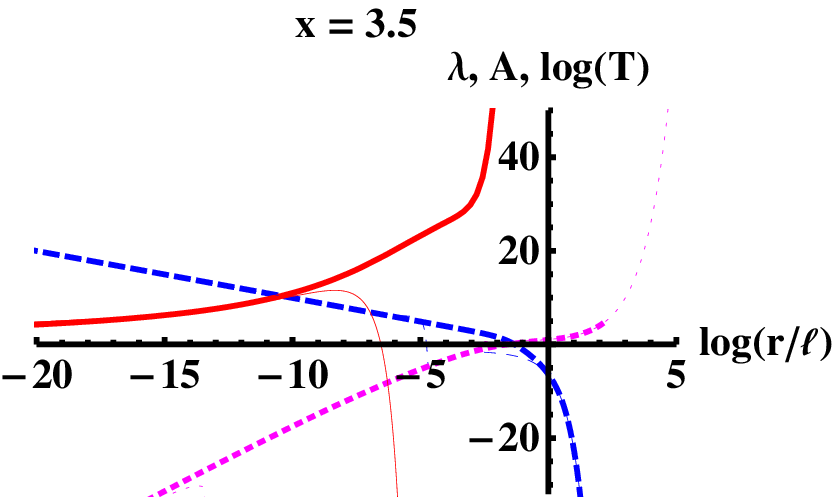}%
\hspace{1cm}
\includegraphics[width=0.38\textwidth]{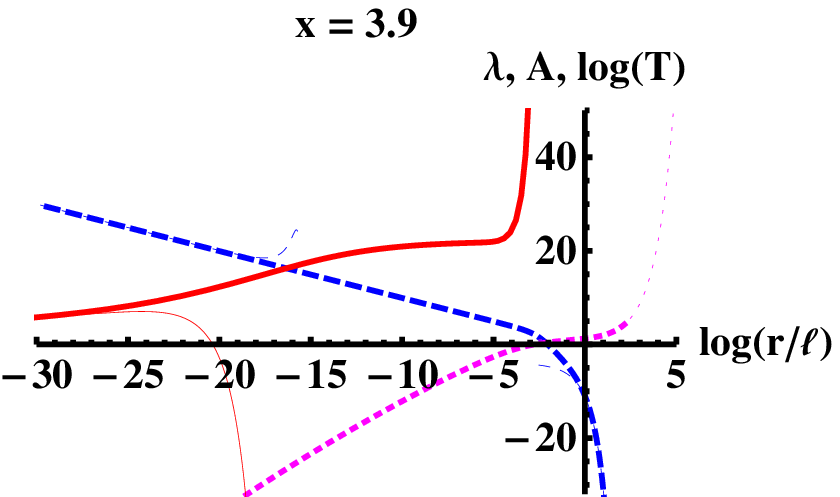}
\caption{The background for vanishing quark mass for various values of $x$ (see
the labels). The red solid, blue dashed, and magenta dotted curves are the
values of $\l$, $A$, and $\log T$ as functions of $\log(r/\ell)$, respectively.
The thin lines are the UV and IR expansions of the solutions.}
\label{figbackground}
\end{figure}

\subsubsection{Solutions having nontrivial tachyon dependence}

The backgrounds having  vanishing quark mass lie on the red dashed line in
Fig.~\ref{figuvbeh}. We plot the corresponding background as a function of $r$
for a few values of $x$ in Fig.~\ref{figbackground}. As we matched with the IR
expansions and chose their scale $R$ to be unity, the IR scale is approximately
fixed to $\mathcal{O}(1)$. The changing of the background as the critical value
$x_c\simeq 3.9959$ is approached,  is best visible in the solutions of $\l$ (the
solid red curves).

The dependence of the solutions on $x$ meets the expectations from field theory.
For $x=2$ the solution is ``running'': $\l$ has simple and smooth dependence on
$\log r$. As $x$ is increased to $3$, a small distortion appears which becomes
better visible for $x=3.5$. The solution of $\l$ is developing a plateau at $\l
\simeq 25$, as it approaches a fixed point. Indeed for $x=3.9$ the coupling
constant $\l$ ``walks'': it takes an approximately constant value as $r$ changes
by a few orders of magnitude. Such walking backgrounds have been studied in the
context of IHQCD by using a model beta function in the probe limit in
\cite{kajantie}.

 We also note that $A$ (dashed blue curves) depends linearly on $\log r$ up to
the IR region so that approximately $A \simeq -\log(r/\ell)$, and the metric is
thus very close to the AdS one, even over the quasiconformal region where $\l$
walks. The tachyon (dotted magenta curves) is small and decoupled from the
evolution of $A$ and $\l$ in the UV and in the walking region.\footnote{It is
difficult to obtain solutions with the value of mass close enough to zero to
produce the (quadratically vanishing) tachyon dependence of the massless
solutions when shooting from the IR due to limitations from numerical precision.
Therefore, we matched the background solution at $\log r \sim -8$ with a tachyon
solution that was obtained by shooting from the UV and assuming decoupling, and
plotted a combination of these to obtain the truly massless tachyon profile. A
similar procedure was required for Fig.~\ref{figgammas} below.} It becomes
$\mathcal{O}(1)$ (the curve crosses zero) only after the coupling has already
started to diverge. This agrees with our expectation that the UV behavior, the
behavior in the walking region, and in particular the phase structure is
basically independent of the choices of IR behaviors of the potentials and the
form of the tachyon action for large $T$.

We also show the UV and IR expansions of the various fields, given in
Eqs.~\eqref{UVexpsapp},~\eqref{TUVres} of Appendix~\ref{AppUV}, and in
Eqs.\ref{IRexps}, respectively, as thin lines where possible. These lines are
often poorly visible since they overlap with the background. In the running
region ($x=2$) the full solution can be obtained to a good approximation by
interpolating between the expansions. As $x$ increases, the region of validity
of the UV expansions is pushed to smaller $\log r$, and neither UV nor IR
expansions work in the walking region that grows as $x \to x_c$.

\begin{figure}
\centering
\includegraphics[width=0.35\textwidth]{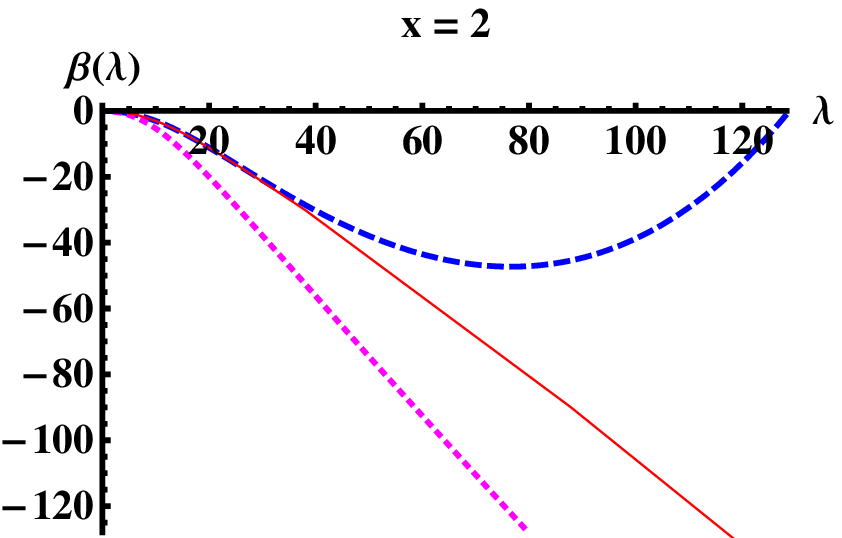}%
\hspace{1cm}
\includegraphics[width=0.35\textwidth]{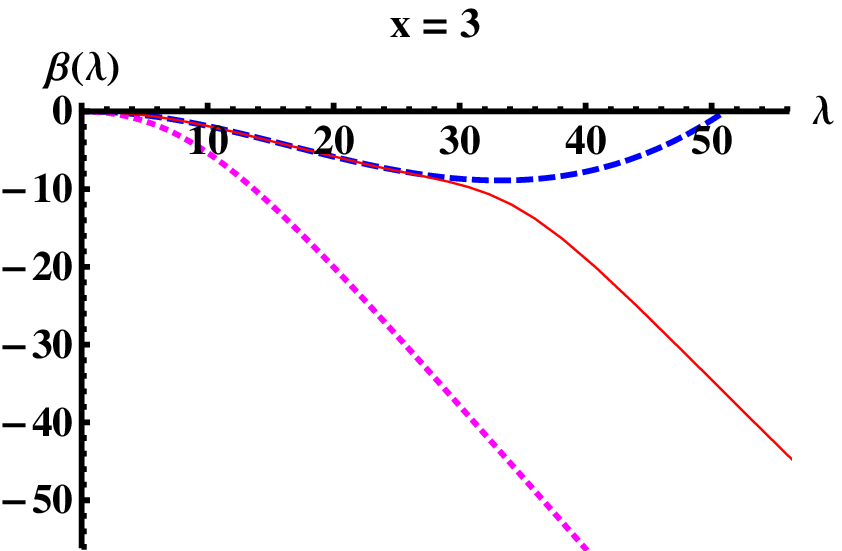}

\vspace{5mm}

\includegraphics[width=0.35\textwidth]{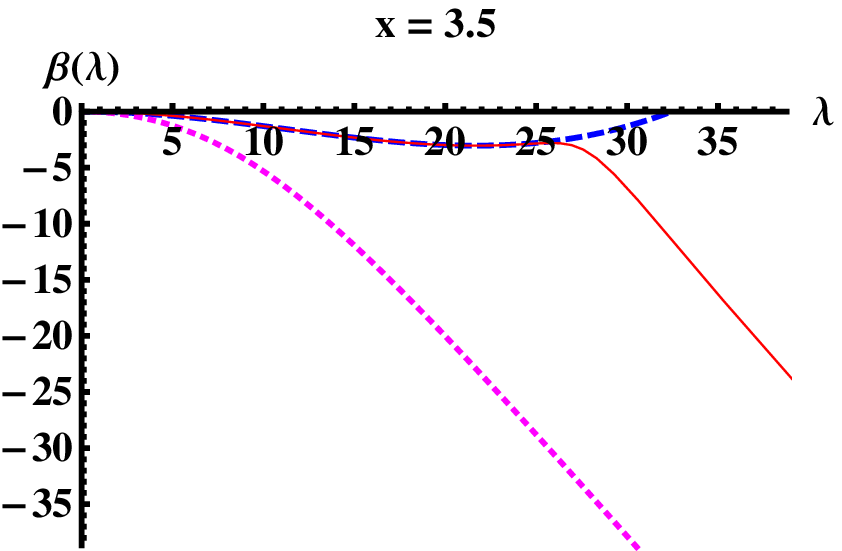}%
\hspace{1cm}
\includegraphics[width=0.35\textwidth]{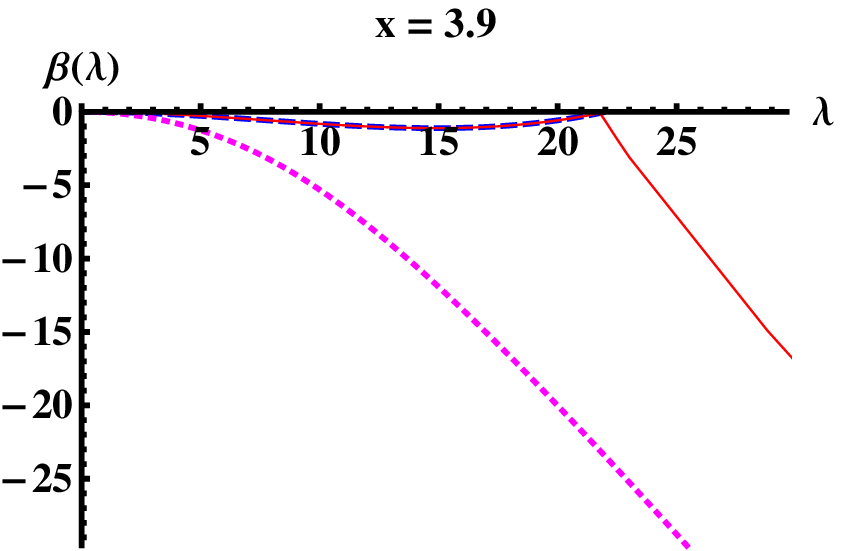}
\caption{The $\beta$-functions for vanishing quark mass for various values of
$x$ (see the labels). The red solid, blue dashed, and magenta dotted curves are
the $\beta$-functions corresponding to the full numerical solution ($d\l/dA$)
along the RG flow, the potential $V_\mathrm{eff}=V_g-xV_{f0}$, and the potential
$V_g$, respectively.}
\label{figbetas}
\end{figure}

\begin{figure}
\centering
\includegraphics[width=0.5\textwidth]{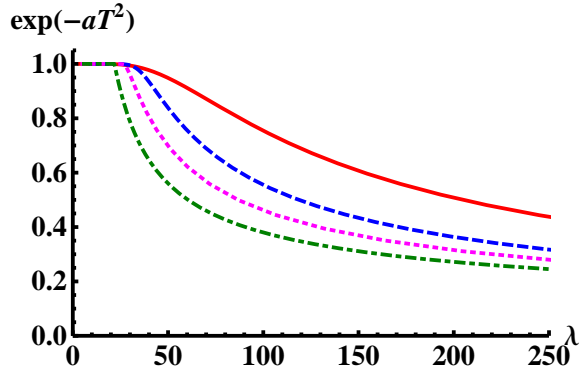}%
\caption{The factor $\exp(-a T^2)$ plotted against $\l$ along the RG flow for
various values of $x$. The red solid, blue dashed, magenta dotted, and green
dotdashed curves have $x = 2$, 3, $3.5$, and $3.9$, respectively.}
\label{figexpfactor}
\end{figure}

It is also illustrative to discuss the behavior of the system in terms of the
$\beta$-functions. For this we recall that in the absence of the tachyon,
potentials and beta functions are linked according to Eqs.~\eqref{betadef},
and~\eqref{Xde}.
By using this fact we can quantitatively estimate the validity of the tachyon
decoupling in the UV and in the IR that was discussed above.
First, recall that $V_f(\l,T)$ vanishes exponentially for large $T$.
Therefore, in accordance with the discussion of Sec.~\ref{SecFP}, the behavior
of $\l$ and $A$ is described by $V_g(\l)-xV_{f0}(\l)$ ($V_g(l)$) in the UV (IR)
where $T \to 0$ ($T \to \infty$).
Eq.~\eqref{Xde} gives directly the approximate beta function in the UV, whereas
in the IR we must replace $V_\mathrm{eff}(\l) \to V_g(\l)$ as in the Yang-Mills
case \cite{ihqcd}.

 We show in Fig.~\ref{figbetas} the $\beta$-functions corresponding to the UV
(dashed blue curves) and IR (dotted magenta curves) potentials, obtained by
solving Eq.~\ref{Xde}, and compare them to $d\l/dA$ evaluated along the RG flow
of the numerical solution (red solid curves) for various values of $x$. First,
notice that the $x$ dependence of the effective $\beta$-function $d\l/dA$ is as
expected from Fig.~\ref{figbackground}: for $x=2$ it is qualitatively similar to
the Yang-Mills $\beta$-function, and as $x$ approaches $x_c$ we find a typical
quasiconformal behavior where the fixed point is almost reached at a finite
value of the coupling.

As $\l \to 0$ the effective $\beta$-function $d\l/dA$ matches very well with the
expectation from the tachyon decoupling (the dashed blue curves).
Similarly, toward the IR ($\l \to \infty$) the asymptotics of the red curves are
similar to the magenta ones, which were obtained by taking $\l \to \infty$.
However the convergence towards the decoupling limit (blue curves in the UV, and
magenta ones in the IR) is slower in the IR than in the UV. The main reason for
this is understood by studying Fig.~\ref{figbackground}. Since we plot the
$\beta$-functions as functions of the coupling $\l$, and $\l$ diverges in the IR
much faster than the tachyon, the tachyon decoupling as $\l \to \infty$ takes
place slowly. This is confirmed in Fig.~\ref{figexpfactor} by plotting the
factor $\exp(-a T^2)$ of the tachyon potential, which controls the tachyon
decoupling, as a function of $\l$. Indeed this factor approaches the constant
value of one quickly in the UV, whereas the convergence to zero in the IR is
slower.

Notice the clear similarity in the $\l$ dependencies of this factor and the
$\beta$-function $d\l/dA$ in Fig.~\ref{figbetas}.

\begin{figure}
\centering
\includegraphics[width=0.4\textwidth]{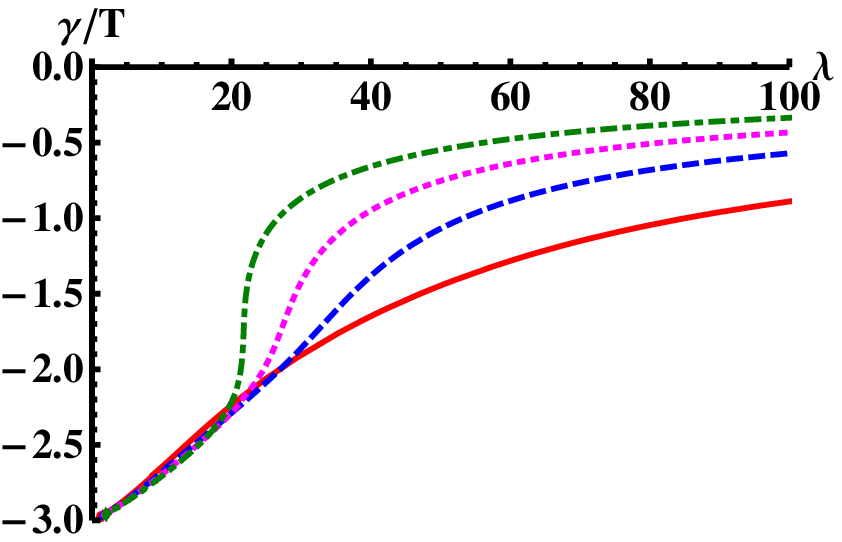}%
\hspace{1cm}
\includegraphics[width=0.4\textwidth]{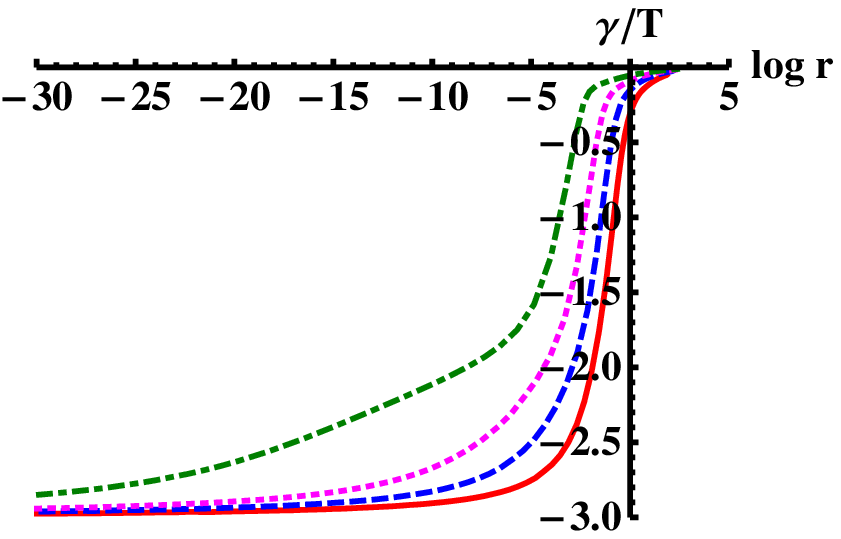}
\caption{The gamma functions along the RG flow for vanishing quark mass as a
function of $\l$ (left) and $\log r$ (right) for various values of $x$. The red
solid, blue dashed, magenta dotted, and green dotdashed curves have $x = 2$, 3,
$3.5$, and $3.9$, respectively.}
\label{figgammas}
\end{figure}

Finally we plot the effective gamma function
\be
 \frac{\gamma}{T} = \frac{1}{T} \frac{d T}{dA} = \frac{d\log T}{dA}
\ee
along the RG flow against $\l$ and $\log r$ in Fig.~\ref{figgammas}. When the
quark mass is zero, $\gamma/T$ approaches $-3$ in the deep UV (see
Appendix~\ref{TUV}). In the UV region $\gamma/T$ is approximately independent of
$x$ and increases with $\l$ until it reaches $-2$ near the value $\l\simeq \l_c$
where the fixed point develops as $x \to x_c$.
This is in line with the discussion of the preceding sections.
In particular, when plotted as a function of $\log r$ we see that $\gamma/T$ is
close to $-2$ in the walking region, corresponding to the saturation of the BF
bound.  We have checked that this behavior gets more pronounced as we choose
values of $x$ even closer to $x_c$ so that a plateau near the value $-2$
develops in the left hand plot of Fig.~\ref{figgammas}.

\subsection{Backgrounds at generic quark masses}

\begin{figure}
\centering
\includegraphics[width=0.4\textwidth]{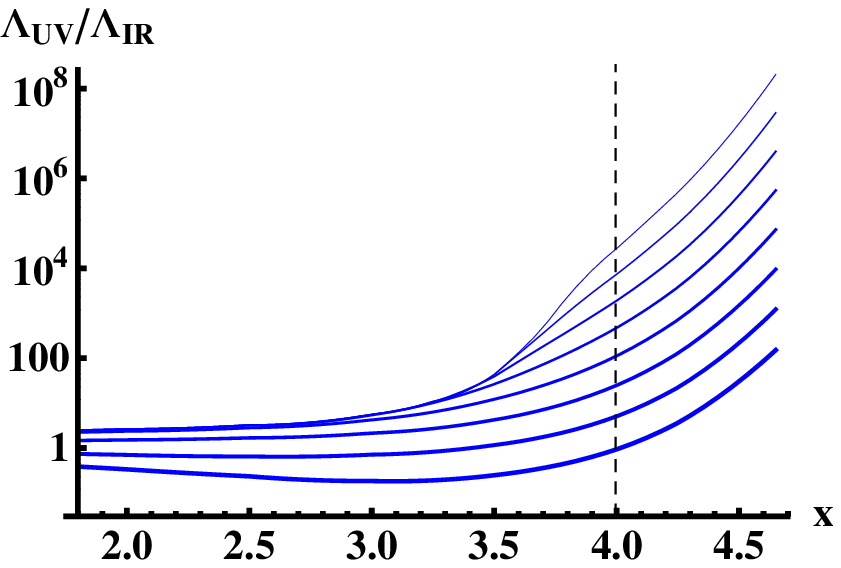}%
\hspace{1cm}
\includegraphics[width=0.4\textwidth]{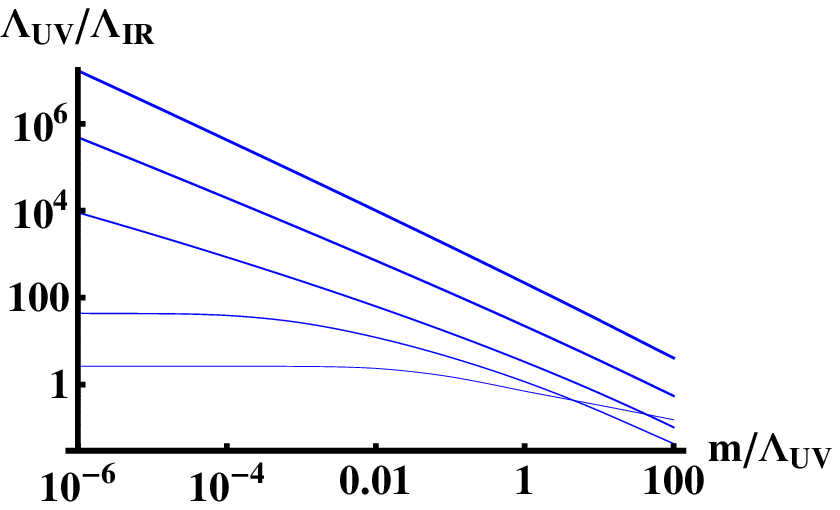}
\caption{Left: The ratio of the UV ($\Lambda_\mathrm{UV}$) and IR
($\Lambda_\mathrm{IR}$) scales as a function of $x$ for $m/\Lambda_\mathrm{UV} =
10^{-6},10^{-5},\ldots,10$ from thin to thick lines (top to bottom). The
vertical dashed line is at critical $x_c\simeq 3.9959$. Right: The ratio of the
UV and IR scales as a function of  $m/\Lambda_\mathrm{UV}$ for $x  = 2$, $3.5$,
$3.9$, and $4.25$ from thin to thick lines (bottom to top).}
\label{figscalesep}
\end{figure}

 We now discuss the solutions of Fig.~\ref{figuvbeh} for generic quark masses.
As pointed out above, the massive solutions (the contoured region left of the
blue and dashed red lines in Fig.~\ref{figuvbeh}) exist for all values of $x$.
Except for the modified UV asymptotics of the tachyon, no new classes of
qualitatively different backgrounds with respect to the massless case are found.

However, adding a mass introduces a new scale to the system, which affects the
background in a different way depending on its size and the value of
$x$.\footnote{See \cite{dietrich} for an analysis within a different framework.}
This effect may be illustrated by studying the ratio of the UV and IR scales of
the background, which measures how close to the IR fixed point the system comes.
 We define the scales in term of the UV and IR expansions, i.e.,
$\Lambda_\mathrm{UV}=\Lambda$ in Eqs.~\eqref{UVexpsapp} and
$\Lambda_\mathrm{IR}=1/R$ in Eqs.~\eqref{IRexps}, which we have fixed to one.

The dependence of $\Lambda_\mathrm{UV}/\Lambda_\mathrm{IR}$ on the quark mass in
UV units $m/\Lambda_\mathrm{UV}$ and on $x$ is depicted in
Fig.~\ref{figscalesep}, and meets the expectations from field theory. The
left-hand plot shows the ratio as a function of $x$ for various choices for the
quark mass. We see that there is a qualitative difference between the regions
with $x<x_c$ and $x>x_c$. For $x<x_c$, chiral symmetry breaks spontaneously even
for $m=0$, and there is some range of small masses where the background is
essentially independent of $m$. This is best seen on the right hand plot, where
$\Lambda_\mathrm{UV}/\Lambda_\mathrm{IR}$ levels for small masses for values of
$x$ below the critical line (lowest curves).

When the mass grows large enough (essentially larger than the scale of the
spontaneous symmetry breaking), it starts to fix the IR scale directly, and
$\Lambda_\mathrm{UV}/\Lambda_\mathrm{IR}$ decreases. For $x>x_c$ there is no
spontaneous chiral symmetry breaking, and the IR scale is determined smoothly by
the value of the mass. From the log-log plot on the right we see that the
dependence of $\Lambda_\mathrm{UV}/\Lambda_\mathrm{IR}$ on $m$ is a power-law.
Naively one could expect that the IR scale is directly given by the quark mass,
$m \sim \Lambda_\mathrm{IR}$, so that $\Lambda_\mathrm{UV}/\Lambda_\mathrm{IR}
\propto (m/\Lambda_\mathrm{UV})^{-1}$ (which is also the result one gets by
approximating $T(r)/\ell = m r$ and using the arguments of
Appendix~\ref{AppBG}). The nontrivial energy dependence of the quark mass
modifies the power from $-1$ to smaller values.

\section{The free energy} \label{SecFE}

 We now analyze the free energy for zero quark mass. In this case we identified
two distinct solutions, one with identically vanishing tachyon and the other
with nontrivial tachyon background. We shall show that the latter one is
energetically favorable in the region where it exists ($x<x_c$). We start with
the generic definition of free energy for our action.

The free energy is given by the on-shell Euclidean action plus counterterms
(which we will not need in this article). From (\ref{lagrang}) the Euclidean
action takes the form:
\bea
{\cal S}&=&-(M^3\,N^2)\int d^5x
\bigg[\sqrt{g}\left(R-{4\over3}{(\partial\lambda)^2\over\lambda^2}
+V_g(\lambda)\right) \nn\\
&&-x
\,V_f(\lambda, T)\sqrt{\det\left(g_{ab}+\f(\lambda, T) \partial_a T\,\partial_b
T\right)}\bigg]\ .
\label{euclact}
\eea
By using the Einstein equations (\ref{einsteq}) we can eliminate $R$, which
leads to
\bea
{\cal S}_{\rm os}&=&-(M^3\,N^2)\int d^5x\Bigg[
-{2\over3}\,e^{5A}\,V_g(\lambda)+{2\over3}\,x\,e^{5A}\,V_f(\lambda,T)\,\sqrt{
1+e^{-2A}\,\f(\lambda,T)\,T'^2}\\ \nn
&&-{x\over3}{e^{3A}\,\f(\lambda,T)\,T'^2\over \sqrt
{1+e^{-2A}\,\f(\lambda,T)\,T'^2}}\Bigg]\,.
\label{euos1}
\eea
Next, one can solve $V_g$ from (\ref{e1}) and (\ref{e2}):
\bea
e^{5A}\,V_g(\lambda)&=&(3A'\,e^{3A})'+{x
\over2}\,e^{5A}\,V_f(\lambda,T)\,\sqrt{1+e^{-2A}\,\f(\lambda,T)\,T'^2} \nn\\
&&+ {x \over2}\,e^{5A}{V_f(\lambda,T)\over
\sqrt{1+e^{-2A}\,\f(\lambda,T)\,T'^2}}\,,
\eea
and inserting this into (\ref{euos1}) the on-shell action can be integrated:
\beq
{\cal S}_{\rm os}=2\,M^3\,N^2 \int
d^5x\,(A'\,e^{3A})'=2\,M^3\,N^2\,V_4\left[A'\,e^{3A}\right]^{r_0}_\epsilon\,.
\label{euclos}
\eeq
Here $\epsilon$ and $r_0$ are the UV and IR cutoffs, respectively. In all the
backgrounds which we consider here the contribution from the IR vanishes, so we
will drop that term.
Following \cite{ihqcd} the Gibbons-Hawking boundary term is given by
\beq
{\cal S}_{GH}=2\,M^3\,N^2 \int_{\partial M} d^4x\,\sqrt{
h}\,K\,=8\,M^3\,N^2\,V_4\,A'(\epsilon)\,e^{3A(\epsilon)}\,,
\eeq
where we have used that \cite{ihqcd} $K=4e^{-A}A'$. We obtain the following
expression for the free energy of the system:
\beq
{\cal E}=6\,M^3\,N^2\,V_4\,A'(\epsilon)\,e^{3A(\epsilon)} \ .
\label{freeng}
\eeq

\subsection{The free energy difference of the $m=0$ backgrounds}

The free energy calculated above is obviously divergent as $\epsilon \to 0$, and
needs to be regularized. However, different solutions with the same UV boundary
conditions (the quark mass $m$ and the UV scale $\Lambda_\mathrm{UV}$) have
the same divergent terms as well as counterterms and differ only through a
finite term which can be extracted from the UV expansions. This is the case in
particular for the two backgrounds having zero quark mass, one with vanishing
tachyon and the other with a nontrivial tachyon solution. we now discuss in
detail how the free energy difference between these two backgrounds can be
obtained, by expanding all quantities as series at the UV singularity $r=0$.
Since the leading UV free energy behaves as $1/\epsilon^4$, corrections ${\cal
O}(r^4)$ to the behavior of $A$ and $\l$  will possibly contribute to the finite
terms.

As discussed above, in the UV the tachyon decouples from the equations of motion
for $A$ and $\l$. For $m=0$, the leading corrections to these equations due to
the tachyon are suppressed by $T^2$ or $e^{-2A} T'^2$, i.e., by ${\cal O}(r^6)$.
Therefore the coupling to  tachyon does not affect the free energy directly in
the massless case, and we can set it to zero.

Let us take
\be \label{Videf}
 V_{\rm eff}(\l)=V_g(\l)-x V_f(\l,0)=V_g(\l)-x
V_{f0}(\l)=\frac{12}{\ell^2}\left[1 + V_1 \l +V_2 \l^2+\cdots \right] \ .
\ee

The finite contribution to the free energy can be studied by writing
\bea \label{Alexps1}
 A(r) &=& A_0(r) +r^4 A_1(r) + {\cal O}(r^6) \\ \nn
 \l(r) &=& \l_0(r) + r^4 \l_1(r) + {\cal O}(r^6)
\eea
in close analogy to Eqs.~\eqref{kirb-bo}, where $A_i(r)$ and $\l_i(r)$ have now
series expansions in $1/\log r$ at $r \to 0$. The expansions of for $A_0$ and
$\l_0$ are given in Eqs.~\eqref{UVexpsapp}. Inserting these as well as suitable
Ans\"atze for $A_1$ and $\l_1$ in the EoMs of $A$ and $\l$, and  expanding up to
${\cal O}(r^4)$ we find
\bea \label{Alexps2}
 A_1(r) &=& \hat A\left[ 1 -\frac{19}{12\log(r \Lambda)} + \cdots \right]\\ \nn
 \l_1(r) &=& \frac{\hat A}{V_1}\left[-5 +\frac{445 V_1^2-320 V_2}{36 V_1^2
\log(r \Lambda)} + \cdots\right]
\eea
where $\Lambda$ is the same UV scale that appears in the expansions of $A_0$ and
$\l_0$, and $\hat A$ is a free parameter. It is recognized as an integration
constant of the EoMs that did not appear in the leading order
($\mathcal{O}(r^0)$) UV expansions. Two solutions having the same $\Lambda$ and
$m=0$, but different IR behavior, will have different $\hat A$ as its value is
not fixed by the EoMs asymptotically in the UV. Inserting the results for $A_1$
and $\l_1$ in the expression (\ref{freeng}) for the free energy, we find
\be \label{DEres}
 \Delta {\cal E} = 6 M^3 N^2 V_4 \ell^3 \Delta \hat A \ ,
\ee
where $\Delta {\cal E}$ and $\Delta \hat A$ are the differences between the two
solutions in free energy and the constant $\hat A$, respectively.

It is useful to also study the corresponding variation in the $\beta$-function.
We will actually use the phase variable defined by
\be
 X(r) = \frac{\l'(r)}{3 \l(r) A(r)}  = \frac{\beta}{3\l} \ .
\ee
We may again write
\be \label{Xexpdef}
 X(r) = X_0(r) + r^4 X_1(r) + {\cal O}(r^6) \ .
\ee
Substituting the expansions (\ref{Alexps1}), (\ref{Alexps2}) in the definition
of $X$, we identify
\be \label{X1UVexp}
 X_1(r) = \hat A\left[-\frac{15}{2} \log(r \Lambda) + \frac{
 5  [80 V_1^2 -
    64 V_2- \log[-\log(r \Lambda)] (23 V_1^2 - 64 V_2) ]}{24 V_1^2} + \cdots
\right] \ .
\ee
By using the logarithmic expansion from above, the result can be expressed in
terms of $\l$:
\be \label{Xexpres}
 X(\l) = X_0(\l) + \frac{20 \hat A}{3 V_1 \Lambda^4}\exp\left[  \frac{\log(9
V_1/8) (23 V_1^2 - 64 V_2)}{9 V_1^2}\right] e^{-\frac{32}{9 V_1 \l}}
\l^{\frac{14}{9}-\frac{64 V_2}{9 V_1^2}}\left[1+{\cal O}\left(\l\right)\right]
\ee

Finally, we recall that when the tachyon has decoupled, $X$ satisfies
\be \label{Xde2}
 \l X'(\l) = \left[8 +
    \frac{3 \l}{
     X(\l)} \frac{d}{d \l}
      \log V_{\rm eff}(\l)\right] \frac{X(\l)^2 -
    1}{6} \ .
\ee
Indeed it is easy to check that the result~(\ref{Xexpres}) is consistent with
this equation, and that $\hat A$ is the integration constant which parametrizes
all solutions of the differential equation for given potential $V$.

\begin{figure}
\begin{center}
\includegraphics[width=0.5\textwidth]{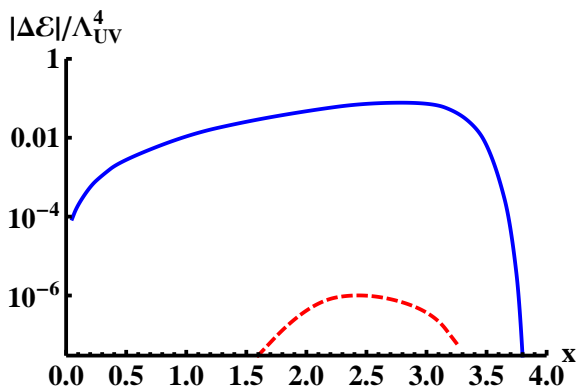}%
\end{center}
\caption{The free energy difference between the chiral symmetry conserving
($T\equiv 0$) and breaking ($T \to \infty$ as $r \to \infty$) solutions. Solid
blue curve is the difference for the chiral symmetry breaking solution having a
monotonic tachyon with no zeroes, and the dashed red curve is the difference for
the solution having one zero.}
\label{figDE}
\end{figure}

The remaining task is to extract the coefficients $\hat A$ from the chiral
symmetry breaking (with nontrivial tachyon profile) and conserving backgrounds
(with identically vanishing tachyon) which have zero quark mass, and then use
the formula~\eqref{DEres} to calculate the free energy difference. Extraction of
the coefficients is done by studying the variation $X_1$ of the phase function,
and details are given in Appendix~\ref{AppExtrDE}.

We find that the chiral symmetry breaking solution is the energetically
favorable one in the region where it exists, i.e., for $0<x<x_c$. We plot the
free energy difference (setting $M^3N^2V_4=1$) in Fig.~\ref{figDE} (solid blue
curve).

Notice that $\Delta {\cal E}$ approaches zero both as $x \to 0$ and as $x \to
x_c$. In the case $x\to 0$, we do expect that the effect vanishes as indeed the
number of flavors, which controls the backreaction of the tachyon,
vanishes. One expects linear dependence $\Delta {\cal E} \propto x$: since the
background configurations behave smoothly as $x \to 0$, the energy difference
arises due to the explicit $x$ dependence in the action and due to its linear
effect on the background. The case $x\to x_c$ will be discussed in the next
section. For $x_c<x<5.5$ only one solution with zero quark mass exists (the one
with $T \equiv 0$), as discussed in Sec.~\ref{SecBG}, so there is no need for
comparison.

There are also solutions with zero quark mass where the tachyon has one or more
zeroes (see Appendix~\ref{AppBG}). We have verified that these solutions have
larger free energies than the one without a tachyon zero. The red dashed line in
Fig.~\ref{figDE} is the free energy difference between the background with
vanishing tachyon and the one with a nontrivial tachyon solution having one
zero. The energy difference between the $T\equiv0$ solution and the solutions
having more than one tachyon zeroes is even smaller.

Therefore,  for $x<x_c$ and $m=0$,  the standard tachyon solution has the lowest
free energy and the chirally symmetric one,   $T=0$, has the largest  free
energy. All other undulating solutions have free energies that are between these
two (and closest to the $T=0$ solution).

 For $|m|>0$, the standard tachyon solution has the lowest free energy and the
non-standard ones, have higher free energy.

\section{Scaling below the conformal window} \label{secBKT}

\subsection{BKT scaling of the chiral condensate}

We shall now argue that the chiral condensate $\propto \sigma$ obeys the BKT
\cite{Kosterlitz:1974sm} or Miransky \cite{Miransky} scaling behavior,
\be
\sigma \propto \exp\left(-\frac{c}{\sqrt{x_c-x}}\right) \ .
\ee
as $x$ approaches the critical value $x_c$ where the solution ceases to exist.
This argument will be supported with numerical results in section \ref{numres}.
The ratio of the UV and the IR energy scales will show similar scaling. This
behavior is known to arise both in Dyson-Schwinger \cite{Miransky,Holdom2} and
holographic approaches \cite{Kaplan,BKT,kutasov,kajantie}.
Indeed, our analysis has many
similarities with both Dyson-Schwinger and earlier holographic approaches, and
in particular with the recent study \cite{kutasov} of a related model.

We shall not give a precise proof but only sketch how the scaling arises. It is
enough to study the region near the UV where the tachyon is small $T(r) \ll 1$,
so the tachyon decouples from the EoMs of $\l$ and $A$. We will neglect the
logarithmic corrections to the tachyon (see Appendix~\ref{TUV}) which play no
role in the scaling argument.

In the deep UV, where the coupling is small $\l \ll 1$, the tachyon behaves
as
\be \label{TdeepUVapp}
 T(r) \sim \sigma r^3 \ .
\ee
As $r$ increases $T$ stays small, and $\l$ starts to approach the fixed point
value $\l=\l_*$ which maximizes $V_g(\l)-xV_{f0}(\l)$. The behavior of $\l$ and
$A$ in this region is given by the $T=0$ asymptotics of Sec.~\ref{TzeroIR}:
\bea \label{lirfp}
 \l&=&\l_* + {\cal O}\left[ \left(\frac{r}{r_\mathrm{UV}}\right)^{-\delta}
\right] \\
 A&=& -\log (r-r_0) + A_0 + {\cal O}\left[
\left(\frac{r}{r_\mathrm{UV}}\right)^{-2 \delta} \right]
\eea
where $\delta$ is a positive parameter defined in Eq.~\eqref{deltadef}.
This approximation is valid for some intermediate region $r_\mathrm{UV} \ll r
\lesssim r_\mathrm{IR}$, where at the scale $r_\mathrm{IR}$ the tachyon finally
becomes ${\cal O}(1)$, and drives the system away from the IR fixed point. We
choose $x$ near the critical value, $0<x_c-x \ll 1$. The tachyon IR mass was
calculated above in Sec.~\ref{SecBG}:
\be
\Delta_\mathrm{IR}(4-\Delta_\mathrm{IR})=-m^2_\mathrm{IR}\ell^2_\mathrm{IR}
=G(\l_*,x) \ ,
\ee
where
\be
 G(\l,x) \equiv {24 a(\l)\over \f(\l)(V_{g}(\l)-x V_{f0}(\l))} \ .
\ee

Keeping formally $x$ fixed while varying $\l_*$, the right hand side will become
equal to four as $\l_*$ reaches a critical value $\l_c$: $G(\l_c,x)=4$. The
scaling behavior will appear as $\l_* \to \l_c$ from above, and then also $x \to
x_c$. We  expand around this point:
\be
\Delta_\mathrm{IR}(4-\Delta_\mathrm{IR})= G(\l_*,x)  = 4 + \frac{\p}{\p
\l}G(\l_c,x) (\l_*-\l_c) + \cdots  \equiv 4 + \kappa (\l_*-\l_c) + \cdots \ .
\ee
For $0<\l_*-\l_c\ll 1$, we obtain
\be
 \Delta_\mathrm{IR} \simeq 2 \pm i \sqrt{\kappa (\l_*-\l_c)}\;.
\ee
The tachyon solution becomes
\be \label{Tirfp}
 T(r) \simeq C_\mathrm{fp} \left(\frac{r}{r_\mathrm{UV}}\right)^2
\sin\left(\sqrt{\kappa (\l_*-\l_c)} \log\frac{r}{r_\mathrm{UV}} + \phi\right) \
.
\ee
When $\l$ moves away from the fixed point,  while moving towards the UV,  it
will at some point become smaller than $\l_c$ so that the asymptotic
solution~\eqref{Tirfp} fails. However, as we shall see, the leading scaling
originates from the region where $\l_*-\l \ll \l_*-\l_c$, and we need the
solution for smaller $\l$ only to make contact with the deep UV behavior and the
definition of $\sigma$. Even for $\l \lesssim \l_c$,  $\kappa (\l_c-\l)$ is
small for almost the whole range of $r$ with  $r_\mathrm{UV} < r$. Therefore,
within the this region,  it is sufficient to  neglect $\kappa (\l_c-\l)$ and
approximate $T(r) \propto r^2$. Further, we introduce an intermediate scale
$\hat r$ where $\l_* - \l \sim \l_* - \l_c$, and write the approximation as
\begin{align} \label{Tintermapp}
 T(r) &\simeq \hat C \left(\frac{r}{r_\mathrm{UV}}\right)^2\ ;&
&r_\mathrm{UV} \ll r \lesssim \hat r \ , \ \ \l\lesssim \l_c \\
\label{Tirfpfin}
 T(r) &\simeq C_\mathrm{fp} \left(\frac{r}{r_\mathrm{UV}}\right)^2
&\!\!\!\!\!\sin\left(\sqrt{\kappa (\l_*-\l_c)} \log\frac{r}{\hat r} + \hat
\phi\right) \ ; \quad \ &\hat r \lesssim r \ll r_\mathrm{IR} \ , \ \ \l_c
\lesssim \l < \l_*
\end{align}
where we use $\hat r$ instead of $r_\mathrm{UV}$ as the reference value of the
logarithm for later convenience.

For $r \gtrsim r_\mathrm{IR}$
there is no obvious way to write a good approximation for the tachyon solution.
However, as we shall see, such an approximation is not necessary for finding the
scaling behavior.

The scaling behavior can be found by matching the tachyon solutions in the
different regions. First, we require that the solutions of
Eq.~\eqref{TdeepUVapp} and~\eqref{Tintermapp} join approximately continuously at
$r \simeq r_\mathrm{UV}$. This gives
\be
 \hat C \sim \sigma r_\mathrm{UV}^3 \ .
\ee
Notice that $\sigma$ is not a free parameter here, but it will later be fixed by
the matching procedure.
Further requiring approximate continuity at $r \simeq \hat r$ we
find\footnote{This matching involves a subtlety which does not affect the
scaling (see Appendix~\ref{AppBKT}).}
\be \label{CfpUV}
 C_\mathrm{fp} \sim \sigma r_\mathrm{UV}^3 \quad \ ; \qquad \hat \phi \sim 1 \ .
\ee
The remaining task is to match with the unknown IR behavior at $r \simeq
r_\mathrm{IR}$. First, $r_\mathrm{IR}$ was defined as the scale where the
tachyon becomes ${\cal O}(1)$ and drives the system away from the fixed point,
so $T(r_\mathrm{IR}) \sim 1$, which fixes $r_\mathrm{IR}$ in terms of $\sigma$:
\be \label{Cfpfix}
 C_\mathrm{fp} \sim \sigma r_\mathrm{UV}^3 \sim
\left(\frac{r_\mathrm{UV}}{r_\mathrm{IR}}\right)^2 \ .
\ee
Finally, we need to match the solution to $T'(r_\mathrm{IR})$, which is ${\cal
O}(1/r_\mathrm{IR})$, since the tachyon EoM is apparently regular in this
region. Notice that we must indeed fix this number to have a solution that
asymptotes to the ``good'' singularity in the IR. The good IR asymptotics have
one free parameter, the normalization of the tachyon in the IR. This is however
already determined by requiring $T(r_\mathrm{IR}) \sim 1$. Therefore, the
argument of the sine function in \eqref{Tirfpfin} is basically fixed to a given
${\cal O}(1)$ number at $r=r_\mathrm{IR}$, which gives the desired BKT scaling:
\be \label{sinargmatch}
\sqrt{\kappa (\l_*-\l_c)} \log\frac{r_\mathrm{IR}}{\hat r} =  {\cal O}(1)
\ee
so that
\be
\frac{r_\mathrm{IR}}{\hat r} \sim \exp\left(\frac{K}{\sqrt{\l_*-\l_c}}\right) \
,
\ee
with $K$  positive. Finally, we notice that the connection between $\hat r$ and
$r_\mathrm{UV}$ can be obtained from \eqref{lirfp} by using the definition of
$\hat r$. This results in a power law:
\be
 \frac{\hat r}{r_\mathrm{UV}} \sim \left(\l_*-\l_c\right)^{-\frac{1}{\delta}}
\ee
which can be neglected as a subleading correction to the exponential scaling.
Taking this into account
\be
 \frac{r_\mathrm{IR}}{r_\mathrm{UV}} \sim
\exp\left(\frac{K}{\sqrt{\l_*-\l_c}}\right) \sim \exp\left(\frac{\hat
K}{\sqrt{x_c-x}}\right) \ .
\ee
The inverse of this scaling result is expected to hold for any ratio of IR and
UV energy scales independently of their precise definitions.
By using Eq.~\eqref{Cfpfix} we find the scaling results for $\sigma$:
\be
 \sigma \sim \frac{1}{r_\mathrm{UV}^3}  \exp\left(-\frac{2
K}{\sqrt{\l_*-\l_c}}\right) \sim \frac{1}{r_\mathrm{UV}^3}  \exp\left(-\frac{2
\hat K}{\sqrt{x_c-x}}\right) \ .
\ee
Notice that $x_c$ and $\l_c$ were defined by $G(\l_*(x_c),x_c)=4$ and
$G(\l_c,x)=4$, respectively, so that $\l_*=\l_c$ at $x=x_c$. Since $G$ is smooth
in this region, $\l_*-\l_c$ could readily be replaced by $x_c-x$ in the results
above (after rescaling $K$).

In the above discussion we neglected several subtleties.  These issues are
analyzed in Appendix~\ref{AppBKT}. In particular, after more careful analysis,
we are able to find explicit results for $K$ and $\hat K$:
\bea \label{Krestext}
 K = \frac{\pi}{\sqrt{\kappa}} = \frac{\pi}{\sqrt{ \frac{\p }{\p \l}G(\l_c,x)}}
\ ; \qquad \hat K = \frac{\pi}{\sqrt{- \frac{d }{d x}G(\l_*(x),x)\big|_{x=x_c}}}
\ .
\eea

As a final remark, we stress that the above sketch was to a large extent
independent of the details of the model. In particular, we did not need any
information on the nonlinear terms in the tachyon EoM and on how the IR boundary
conditions are fixed.

\subsection{Scaling of the free energy} \label{SecFEBKT}

Let us then study the scaling of the free energy difference $\Delta {\cal E}$
between the solutions without and with chiral symmetry breaking (at zero quarks
mass), which was studied numerically in Sec.~\ref{SecFE}. Linearizing the EoMs
for $A$ and $\l$ by writing
\bea \label{AlexpsBKT}
 A(r) &=& A_0(r) +r^4 A_1(r) + {\cal O}(r^6) \\\nn
 \l(r) &=& \l_0(r) + r^4 \l_1(r) + {\cal O}(r^6) \ ,
\eea
we related  $\Delta {\cal E}$ with the variation of the leading coefficients of
$A_1$ and $\l_1$ when expressed as a series in $\log r$ at $r=0$. The source of
this variation is the difference in the tachyon solution, and it can be analyzed
in the limit $x \to x_c$. The tachyon contributes corrections $\mathcal{O}(T^2)$
and $\mathcal( e^{2A}(T')^2)$ to the EoMs of $A$ and $\l$. For zero quark mass
these contributions are ${\cal O}(r^6)$ in the deep UV and thus decoupled from
the variations of Eqs.~\eqref{AlexpsBKT}. However, as we have pointed out, in
the walking region $r_{\rm UV} \lesssim r \lesssim r_{\rm IR}$ the tachyon is
$\mathcal{O}(r^2)$, and therefore the tachyon contributions ${\cal O}(r^4)$ do
couple to $A_1$ and $\l_1$.\footnote{The next-to-leading terms of $A$ and $\l$
are still ${\cal O}(r^4)$ in this region since there are no sources which could
change this.} Consequently, we expect that, e.g.,
\be
 r^4 \Delta A_1(r) \sim T(r)^2 \ ; \qquad r_{\rm UV} \lesssim r \lesssim r_{\rm
IR} \ ,
\ee
where $\Delta A_1(r)$ denotes the difference in $A_1$ between the solutions with
intact and broken chiral symmetry.
We expect that to a good enough approximation, the size of $A_1$ equals the
coefficient $\hat A$ defined in Eq.~\eqref{Alexps2} even in the walking region
where the UV series does not converge.\footnote{This can be verified by
inserting the Ans\"atze~\eqref{AlexpsBKT} to the EoMs and studying the solutions
similarly as done for $X_1$ in Appendix~\ref{AppExtrDE}.} Therefore, we obtain
\be
 \Delta \hat A \sim \sigma^2 r_{\rm UV}^2
\ee
and finally the scaling result for the free energy difference reads
\be \label{DEscaling}
 \frac{\Delta {\cal E}}{ M^3 N^2 V_4 } \sim \Delta \hat A  \sim \sigma^2r_{\rm
UV}^2 \sim r_{\rm UV}^{-4} \exp\left(-\frac{4 K}{\sqrt{\l_*-\l_c}}\right) \sim
r_{\rm UV}^{-4}  \exp\left(-\frac{4 \hat K}{\sqrt{x_c-x}}\right) \ .
\ee

\begin{figure}
\begin{center}
\includegraphics[width=0.5\textwidth]{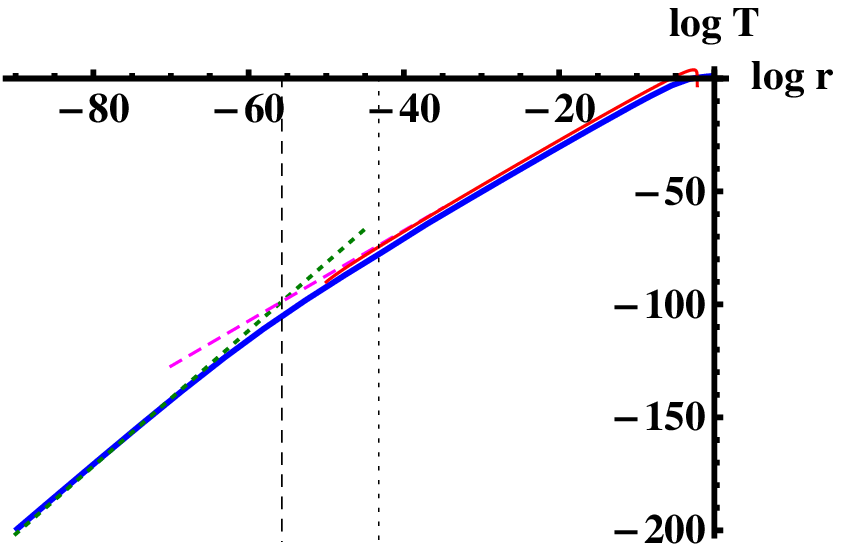}%
\includegraphics[width=0.5\textwidth]{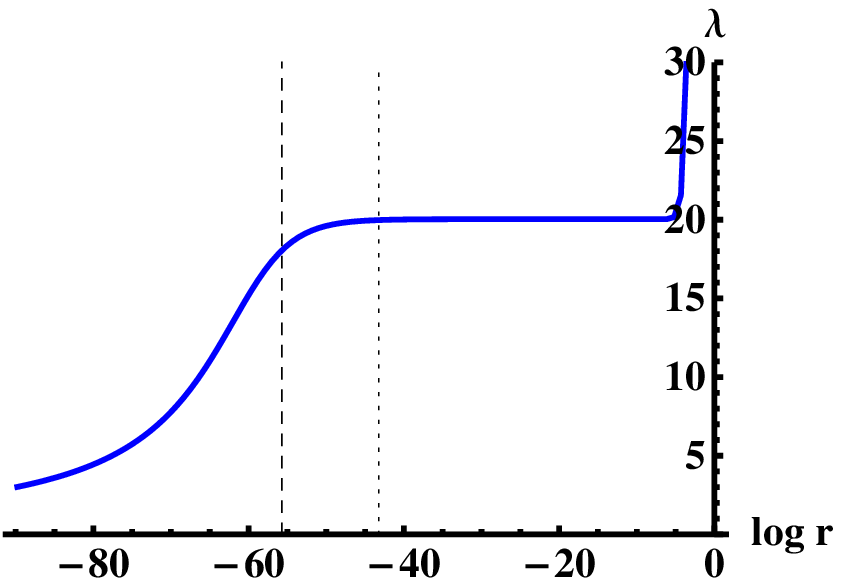}%
\end{center}
\caption{The tachyon $\log T$ (left) and the coupling $\l$ (right) as functions
of $\log r$ for an extreme walking background with $x=3.992$. The thin lines on
the left hand plot are the approximations used to derive the BKT scaling (see
the text for explanation), and the vertical dashed (dotted) lines mark $\log
r_\mathrm{UV}$ ($\log \hat r$).}
\label{figextremesol}
\end{figure}

\subsection{Comparison with numerical results\label{numres}}

We now compare the analytic results above to numerical solutions. In
Fig.~\ref{figextremesol} we plot the tachyon (left) and the coupling $\l$
(right) as functions of $\log r$ in an extreme walking case with $x=3.992$ such
that $x_c-x \simeq 0.004$. The numerical tachyon solution with zero quark mass
(blue thick curve on the left) was obtained by gluing together the various
solutions described in Appendix~\ref{AppExtrsigma}. We compare the solution to
the analytic approximations of Eqs.~\eqref{TdeepUVapp}, \eqref{Tintermapp} and
\eqref{Tirfpfin}, shown as thin green dotted, magenta dashed, and solid red
curves, respectively. The parameters of these curves were chosen such that
$\sigma$ has the extracted value (see Appendix~\ref{AppExtrsigma}),
$r_\mathrm{UV} = 1/\Lambda_\mathrm{UV}$ with $\Lambda_\mathrm{UV}$ obtained by
fitting $\l$ to its UV expansion, and $\hat r$ is the value where $\l$ reaches
$\l_c$. The parameters $\l_c$, $\l_*$ and $\kappa$ were calculated directly from
the potentials, and $\hat \phi$ was given an arbitrary small value. The
agreement between the approximation and the full numerical solution is
remarkably good.

\begin{figure}
\begin{center}
\includegraphics[width=0.5\textwidth]{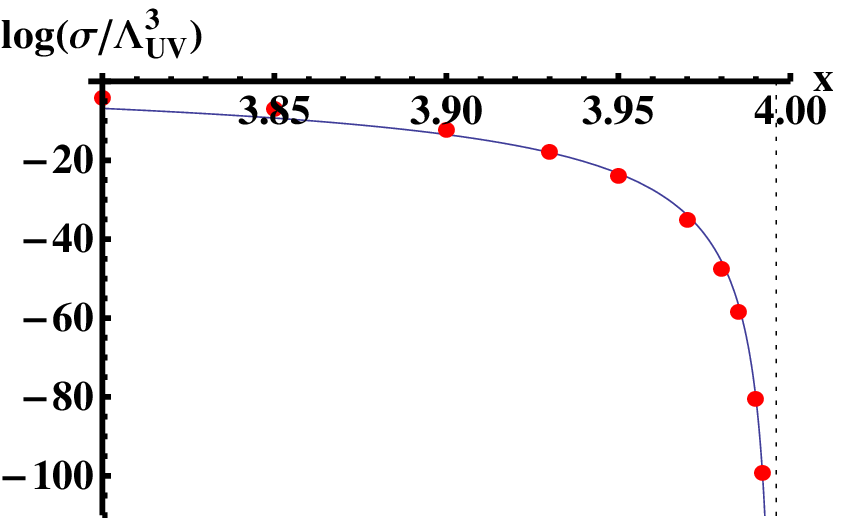}%
\includegraphics[width=0.5\textwidth]{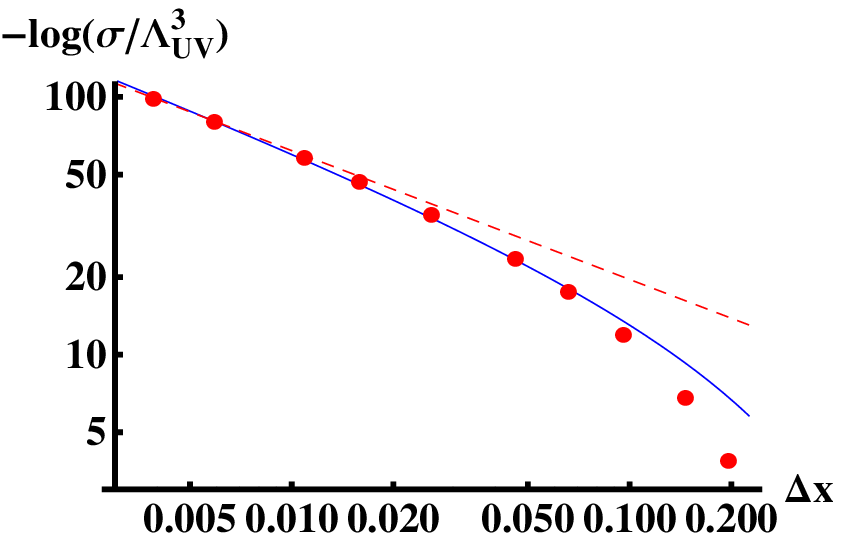}%
\end{center}
\caption{Left: $\log(\sigma/\Lambda^3)$ as a function of $x$ (dots), compared to
a BKT scaling fit (solid line). The vertical dotted line lies at $x=x_c$. Right:
the same curve on log-log scale, using $\Delta x = x_c-x$. To guide the eye we
added the straight dashed line corresponding to the BKT scaling fit without a
constant term, that was fixed to go through the data point with smallest $\Delta
x$.}
\label{figlogsigmavsx}
\end{figure}

\begin{figure}
\begin{center}
\includegraphics[width=0.5\textwidth]{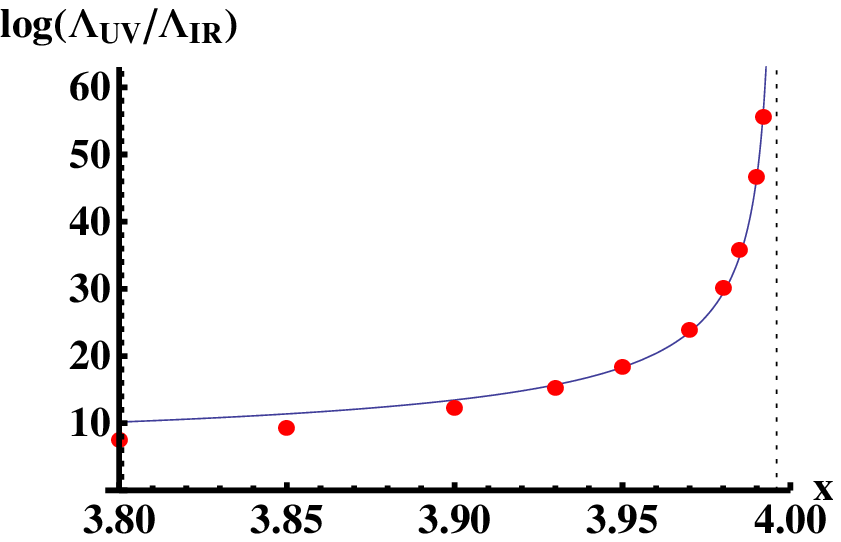}%
\includegraphics[width=0.5\textwidth]{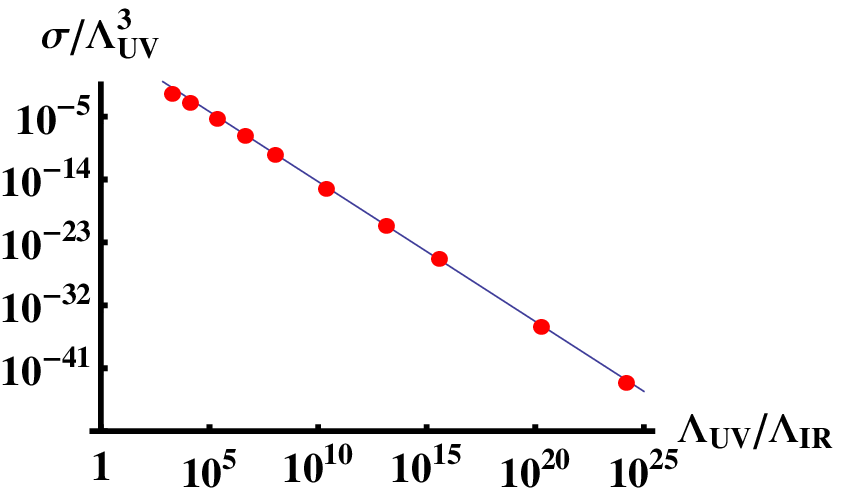}%
\end{center}
\caption{Left: $\log(\Lambda_\mathrm{UV}/\Lambda_\mathrm{IR})$ as a function of
$x$ (dots), compared to a BKT scaling fit (solid line). Right:
$\sigma/\Lambda^3$ plotted against $\Lambda_\mathrm{UV}/\Lambda_\mathrm{IR}$ on
log-log scale. The line is a fit with the expected power law.}
\label{figloglambda}
\end{figure}

We have also compared the expected scaling to the values of chiral condensate
that were extracted from the background (see Sec.~\ref{SecBG}) as detailed in
Appendix~\ref{AppExtrsigma}.
The results for various values of $x$ are the dots in Fig.~\ref{figlogsigmavsx}.
The solid line is a fit to the BKT scaling behavior, given by
\be
 \log  \frac{\sigma}{\Lambda^3}  = 8.6-\frac{6.8}{\sqrt{x_c-x}} \ ,
\ee
which describes the data well. The analytic result of Eq.~\eqref{Krestext} gives
for the potentials used
\be
 2 \hat K = 6.10
\ee
which agrees with the fitted result $6.8$ within the precision of the fit. We
have checked that using only a few of the data points with the highest $x$
brings the fitted valued of $\hat K$ closer to the analytical one.

In Fig.~\ref{figloglambda} (left) we compare the ratio of the scales
$\Lambda_\mathrm{UV} =\Lambda$ and $\Lambda_\mathrm{IR}=1/R$ as defined by the
UV and IR expansions of the background, respectively, to a BKT scaling fit with
$\hat K \simeq 3.4$. We also checked  that the scaling of Eq.~\eqref{Cfpfix},
\be
 \frac{\sigma}{\Lambda_\mathrm{UV}^3} \sim
\left(\frac{r_\mathrm{UV}}{r_\mathrm{IR}}\right)^2 \sim
\left(\frac{\Lambda_\mathrm{IR}}{\Lambda_\mathrm{UV}}\right)^2 \ ,
\ee
is satisfied to a high precision on Fig.~\ref{figloglambda} (right).

\section{Conclusions}
\label{SecConcl}

We analyzed a novel class of holographic models (V-QCD), which reproduces the
main features of QCD in the Veneziano limit of large $N_f$ and $N_c$ with
$x=N_f/N_c$ fixed.

V-QCD is on one hand based on a successful holographic model of Yang-Mills (YM)
theory, and termed improved holographic QCD (IHQCD). IHQCD contains a dilaton
coupled to five-dimensional gravity background. Its characteristic feature is
the holographic renormalization group flow of the YM coupling constant,
identified as the exponential of the dilaton,
as a function of the energy scale, identified roughly
as the inverse of the bulk coordinate.
On the other hand, the model builds on earlier work on including matter in
holographic models via flavor branes in the quenched approximation, i.e.,
neglecting the backreaction of the brane on the dilaton and the background
metric. In particular, we use the tachyon Dirac-Born-Infeld (DBI) action
originally introduced by Sen.

Putting together these two frameworks, the dynamics in the Veneziano limit is
modeled by a system of a dilaton and a tachyon coupled to five-dimensional
gravity. The dilaton action is fixed as in IHQCD. For the tachyon we consider a
generalized DBI action, where   the dilaton dependence is parametrized in terms
of a few potentials, which are a priori unknown.
We may then constrain the unknown potentials, among other methods, by requiring
that the UV physics implements that of (perturbative) QCD and that the solutions
are regular in the IR.

The essential observation for uncovering the dynamics of the system, is the
identification of the effective potential. It involves terms from both the
dilaton and DBI actions, and takes the role of the dilaton potential of IHQCD.
For large $x$, the perturbative Bank-Zaks IR fixed point can, and must be
implemented trough the effective potential. Even further away from the
Banks-Zaks region, the solutions can continue to flow to the fixed point in the
IR. Interestingly, the fixed point can also be ``screened'' by the tachyon
dynamics, such that the theory comes very close to it and the coupling almost
freezes, but eventually starts running again in the deep IR. This kind of
backgrounds are termed as the quasiconformal or ``walking'' ones.

In this article we did a detailed analysis of the backgrounds and the
zero-temperature phase structure of V-QCD. Our main results are as follows:
\begin{itemize}

 \item We generalize the holographic RG flow of IHQCD to include the evolutions
of both the dilaton and the tachyon, which are controlled by the holographic
$\beta$- and $\gamma$-functions, respectively. Remarkably, for potentials that
are analytic in the UV, the interplay of the dilaton and the tachyon
automatically results in the anomalous dimension of the quark mass having
physically reasonable UV asymptotics, i.e., a power series in the 't Hooft
coupling.

 \item If we require that the UV expansions of the potentials capture the
essential QCD physics, and choose potentials that join smoothly with their UV
expansions, the V-QCD phase diagram is always the physically relevant one. That
is, for zero quark mass we find a phase transition at $x=x_c$. The conformal
window, where the backgrounds have an IR fixed point, extends from $x=x_c$ up to
the maximal value $x=11/2$ where asymptotic freedom is lost. Below $x=x_c$,
chiral symmetry is broken, and the theory has similar behavior in the deep IR
as the QCD we have observed in Nature.

 \item The edge of the conformal window is stabilized such that the dimension of
the quark mass at the IR fixed point approaches two (the anomalous dimension
approaches one) as the edge is approached. Under reasonable assumptions for the
potentials, we find values of $x_c$ within a narrow band around the value
$x_c=4$. High value of the anomalous dimension is of importance for applications
to (walking) technicolor.

 \item Below $x_c$ but close to the edge of the conformal window, we find
quasi-conformal or walking backgrounds.

 \item Backgrounds for (any) nonzero quark mass exist. In the conformal window
the quark mass triggers chiral symmetry breaking, and below $x_c$ introducing
the quark mass affects the scales of
the theory in a physically reasonable manner.

 \item We verify that the standard vacua which give the above physical phase
structure also have the lowest free energies. They have monotonically increasing
or identically vanishing tachyon profiles, when the chiral symmetry is broken or
intact, respectively. For $x<x_c$ we also found a tower of unstable ``Efimov''
vacua, for which the tachyon can change its sign arbitrary many times.

 \item Finally, as $x$ approaches $x_c$ from below for the backgrounds with zero
quark mass, we show that the chiral condensate, as well as the length of energy
range where the background stays close to the fixed point, obey the
characteristic Miransky or BKT scaling law.

\end{itemize}

\section{Acknowledgments}

We thank D.~Arean and C.~Krishnan for contributions at an early stage of this
project.
We also thank D.~D.~Dietrich, U.~G\"ursoy, P.~Hoyer, I.~Iatrakis, N.~Jokela,
K.~Kajantie, E.~Keski-Vakkuri, D. Kutasov,
M.~Lippert, R.~Meyer, T.~Morita,  V.~Niarchos, A.~Paredes, A.~Parnachev
and T.~Ryttov
for discussions and comments.
This work was in part supported by the European grants
FP7-REGPOT-2008-1: CreteHEPCosmo-228644 and  PERG07-GA-2010-268246.

\newpage
\appendix
\renewcommand{\theequation}{\thesection.\arabic{equation}}
\addcontentsline{toc}{section}{APPENDIX\label{app}}
\section*{APPENDIX}

\section{General solutions for $\beta$ and $\gamma$ } \label{AppBeta}

It is not difficult to construct numerically $\beta$ and $\gamma$ which solve
the partial differential equations \eqref{e8} and \eqref{e9}. We can use the
fact that they were derived from a set of ordinary differential equations. More
precisely, the equation have the form
\bea \label{PDEform1}
 \beta \frac{\d \beta}{\d \l } + \gamma \frac{\d \beta}{\d T } &=& {\cal
F}_\beta\left(\beta,\gamma\right) \\ \label{PDEform2}
 \beta \frac{\d \gamma}{\d \l } + \gamma \frac{\d \gamma}{\d T } &=& {\cal
F}_\gamma\left(\beta,\gamma\right)
\eea
where  ${\cal F}_\beta$ and ${\cal F}_\gamma$ are independent of the partial
derivatives of the $\beta$- and $\gamma$-functions (and the dependence on
potentials was left implicit).

Now we can apply a standard method for solving first order partial differential
equations (PDEs), which in this
case applies even for a pair of PDEs. We first look for curves along which the
system reduces to ordinary DEs. Not surprisingly, such curves coincide with the
holographic RG flow. That is, we can implicitly define such family of curves, by
requiring that each curve $(\l(A),T(A))$, where $A$ parametrizes the curve,
satisfies
\be \label{flowdef}
 \l'(A) = \beta(\l(A),T(A)) \quad \ ; \qquad T'(A) = \gamma(\l(A),T(A)) \ .
\ee
Then we notice that the differential operators in \eqref{PDEform1} and
\eqref{PDEform2} become derivatives along the curves:
\be
 \frac{d}{dA}\beta(\l(A),T(A)) = {\cal F}_\beta\left(\beta(A),\gamma(A)\right)
\quad \ ; \qquad \frac{d}{dA}\gamma(\l(A),T(A)) = {\cal
F}_\gamma\left(\beta(A),\gamma(A)\right) \ .
\ee
Further, the equations along the curves are essentially in one-to-one
correspondence with the system \eqref{e1}-\eqref{e3}. Indeed, the original
system of DEs can be formally recovered be eliminating the $\beta$ and $\gamma$
functions by using Eqs.~\eqref{flowdef}. Therefore, any solution of the original
system satisfies the PDEs along the curve $(\l(r),T(r))$ that it defines.

As the final step of the method, we notice that the PDEs only depended on the
derivatives of $\beta$ and $\gamma$ along the curves. The derivatives
perpendicular to the curves can be freely chosen. Therefore, \emph{any}
(continuously parametrized) family of curves which satisfies~\eqref{flowdef}
will define a solution to the PDEs in some region of the $(\l,T)$-plane. That
is, the general solution to the PDEs is given by the planes that the solutions
of Eqs.~\eqref{e1}-\eqref{e3} draw in $(\beta,\l,T)$ and $(\gamma,\l,T)$-spaces
as the boundary conditions to the equations are varied in an arbitrary manner.

Remarkably, the amount of degrees of freedom of the solution matches with the
expectation for a system of two first order PDEs, for which the solution should
depend on two arbitrary functions. Since $\beta$, $\gamma$, $\l$, and $T$ are
all invariant under the symmetry of \eqref{transf}, the number of integration
constants in the system \eqref{e1}-\eqref{e3}  which are relevant for the
solution of PDEs is three.

A generic one-dimensional family of these parameters, and consequently a generic
solution to the PDEs, can be defined by giving the dependence of two of these in
terms of the third one
which makes two arbitrary functions.

We have demonstrated how the boundary conditions of the PDEs are mapped to those
of the original set of EoMs. Now the analysis of the solutions of the EoMs (see
Appendices~\ref{AppUV} and~\ref{AppIR}) suggests at least two natural ways to
choose special one-parameter families of the boundary conditions, and to define
special solutions to the PDEs.
\begin{enumerate}

 \item Require that the solutions to EoMs end in the ``good'' IR singularity of
Sec.~\ref{specIR}
(with varying quark mass).

\item Require the more generic IR singularity of Sec.~\ref{Xeq1as} and keep the
quark mass fixed.

\end{enumerate}
(In the second case, the ``good'' IR singularity is expected to arise as some
limit of the more generic IR behavior so that it appears at a boundary of the
region in $(\l,T)$-space where the $\beta$ and $\gamma$ functions are defined.)

\section{A coordinate transformation\label{rA}}

It turns out that it is convenient to solve the system numerically using $A$  as
a coordinate instead of $r$. Therefore, we present the system after this
transformation:
\bea
12-\frac{6 \dot q}{q} +\frac{4 \dot \lambda ^2}{3 \lambda ^2} &=& q^2
V_g(\lambda )-q^2 x\ V_f(\lambda ,T) \sqrt{1+\frac{\f(\lambda ,T)\ \dot
T^2}{q^2}} \\ \nn
12-\frac{4 \dot \lambda ^2}{3 \lambda ^2} &=& q^2 V_g(\lambda )-\frac{q^2 x\
V_f(\lambda ,T)}{\sqrt{1+ q^{-2}\f(\lambda ,T)\ \dot T^2}} \\ \nn
\frac{\ddot \lambda }{\lambda }+\frac{4 \dot \lambda }{\lambda }-\frac{\dot q
}{q }\frac{\dot \l }{\l }-\frac{\dot \lambda ^2}{\lambda ^2}&=&\frac{3}{8}\ q^2
\lambda\, \left(\!- \frac{dV_g}{d\l}+x  \, \sqrt{1\!+\!q^{-2}\f\ \dot T^2}\
\frac{\partial V_g}{\partial \lambda}+\frac{ x\  \dot T^2 }{2 q^2\,
\sqrt{1\!+\!q^{-2}\f \ \dot T^2}}\ \frac{\partial \f}{\partial\lambda}\
V_f\!\right) \\ \nn
\ddot T+ 4 \dot T-\frac{ \dot q }{q} \dot T &=& -\frac{\f\, \dot
T^3}{q^2}\left[4 + \dot \l \ \frac{\p}{\p\l} \log(\sqrt{\f}\ V_f) \right] + \dot
T^2 \frac{\p}{\p T}\log\frac{V_f}{\sqrt{\f}} \nonumber\\ \nn
&&-\dot T\dot  \lambda \frac{\p}{\p\l} \log\left( \f V_f\right) +\frac{q^2 }{\f
}\frac{\p}{\p T}\log V_f
\eea
where the dots are derivatives with respect to $A$ and we defined
\be
 q(A) = e^A \frac{dr}{dA} \ .
\ee

\section{Examples of explicit potential choices} \label{AppPotentials}

 We will construct explicit examples of potentials, which give physically
reasonable backgrounds.
We consider an Ansatz for $V_f$ of the form
\be
 V_f(\l,T) =V_{f0}(\l) e^{-a(\l)T^2} \ ,
\ee
and assume that $\f$ does not depend on $T$.
First, we expect in the UV
\bea
 \label{Vgexpapp}
 V_g(\l) &=& V_0 + V_1 \l + V_2 \l^2 + \cdots \\
 V_f(\l,T=0) &=& V_{f0}(\l) = W_0 +W_1 \l +W_ 2 \l^2 + \cdots
 \label{Vfexpapp}
\eea
The potentials must also produce the good kind of IR singularity discussed in
Appendix~\ref{goodIR} with $P=1/2$ \cite{ihqcd}, which constrains the
asymptotics of $V_g$ to
\be
 V_g(\l) \sim \l^{4/3}\sqrt{\log\l} \ ; \qquad \l \to \infty
\ee
(assuming that $V_{f0}$ plays no role in the IR for the chiral symmetry breaking
solution as tachyon and/or $\l$ diverge so that $\exp(-a(\l)T^2)$ tends to
zero).

In addition, the IR behavior of $V_{f0}$ should be chosen such that the fixed
point (maximum) of $V_g(\l)-xV_{f0}(\l)$, which is guaranteed to be present at
large $x \to 11/2$ if we fix the small $\l$ series of the potential
appropriately, continues to exists up to sufficiently low $x$. This is most
easily achieved if $V_{f0}$ diverges faster than $V_g$ as $\l \to \infty$, so
that the fixed point exists for all $x$. Another possibility, mentioned in
Sec.~\ref{SecFP}, is that the fixed point exist only for $x \geq x_*$, where
$x_*>0$ is relatively small. We have checked that choosing such potentials does
not change any results at qualitative level, and do not discuss this choice
further here.
We also require that the potentials are analytic at $\l=0$. A simple Ansatz that
meets these requirements and involves free coefficients up to two-loop order is
\bea
 V_g(\l) &=& V_0 + V_1\l + V_2
\frac{\l^2}{(1+\l/\l_0)^{2/3}}\sqrt{1+\log(1+\l/\l_0)} \\
 V_{f0}(\l) &=& W_0 + W_1\l + W_2 \l^2 \ .
\eea
Notice that the AdS radius
\be
 \ell = \sqrt{\frac{12}{V_0 - x W_0}}
\ee
must be well defined for all $x$ up to $x=11/2$, which sets an upper bound for
$W_0$ for given $V_0$. We also expect $V_f$ to be positive at small
$\l$.\footnote{Negative $W_0$ is also problematic since, at least for small
$|W_0|$, it generates a zero of $V_f$, which causes the tachyon solution to
become singular.}

Therefore, we take
\be \label{ineqs2}
 0 \le W_0 \le \frac{2}{11} V_0 \ .
\ee
However, as discussed in Appendix~\ref{TUV}, the sum of the anomalous dimensions
of the quark mass and the chiral condensate is not equal to 4 if $W_0=0$ (then
$\delta=1$ in Appendix~\ref{TUV}). Therefore, we discard this option. Notice
also that if we saturate the upper limit with $W_0 = \frac{2}{11} V_0$, the AdS
radius
diverges in the Banks-Zaks limit $x \to 11/2$ unless we choose an $x$-dependent
$V_0$.

In addition, we can fix the coefficients $V_i$, $W_i$ by mapping to the field
theory $\beta$-functions as
\bea \label{YMVexpapp}
 V_g(\l)  &=& \frac{12}{\ell_0^2}\left[1 + \frac{8}{9}b_{0}^{\rm YM}\l +\frac{23
(b_0^{\rm YM})^2-36b_1^{\rm YM} }{81} \l^2 + \cdots \right] \\
 V_g(\l)-xV_{f0}(\l)  &=&  \frac{12}{\ell^2}\left[1 + \frac{8}{9}b_0\l +
\frac{23 b_0^2-36b_1 }{81} \l^2 + \cdots \right] \ .
\eea
Here the (scheme independent) QCD $\beta$-function in the Veneziano limit with
vanishing quark masses up to two-loop order are

\be
 b_0 = \frac{2}{3} \frac{11 - 2 x}{(4  \pi)^2} \ ; \qquad b_1  = -\frac{2}{3}
\frac{34 - 13 x}{(4  \pi)^4}
\ee
and $b_i^{\rm YM} = b_i|_{x=0}$. Setting $V_0 =12$ and $W_0=12/11$, which lies
in the middle of the allowed region of Eq.~\eqref{ineqs2}, we obtain
\begin{align}
 &V_1  = \frac{44}{9\pi^2}  \ ; \qquad &V_2 &= \frac{4619}{3888\pi^4}\ ; \\
 &W_1  =  \frac{4 (33-2 x)}{99 \pi ^2} \ ; \qquad &W_2 &=  \frac{23473-2726 x+92
x^2}{42768 \pi ^4}
\end{align}
Further, we choose $\l_0=8\pi^2$ to prevent the higher order terms in the UV
expansion of the potentials from growing unnaturally large.

In addition, we need to choose the functions $a(\l)$ and $\f(\l)$ appropriately.
As discussed in Section~\ref{SecBG} and in Appendix~\ref{TUV}, we must have
\be
 \frac{\f(\l)}{a(\l)} = \frac{2 \ell^2}{3}\left(1+\f_1\l+\cdots \right)
\ee
Here the coefficient $\f_1$ can be matched with the one-loop anomalous dimension
of the quark mass, which reads in the Veneziano limit
\be
 \gamma_m(\l) = \frac{3}{(4\pi)^2} \l + \cdots
\ee
By matching with Eq.~\eqref{UVgammam} from Appendix, we obtain
\be
  -\frac{3}{(4\pi)^2} = \frac{9}{8}\left[\frac{4}{3}\frac{8}{9}b_0 +
\frac{4}{3}\f_1\right]
\ee
from which
\be
 \f_1 = - \frac{115-16 x}{216 \pi ^2} \ .
\ee

The IR behavior of the functions $\f(\l)$ and $a(\l)$ is linked to the tachyon
behavior in the IR for the solution which breaks chiral symmetry. They are
discussed in detail in Appendix~\ref{goodIR}, where essentially only two
different cases, which are consistent with the good IR singularity, are found.
These possibilities are produced by the following choices.

\begin{itemize}
 \item[I]  We can choose the function $a(\l)$ to be constant, and the function
$\f(\l)$ to have power-law IR asymptotics:
        \be
         \f(\l) = \frac{1}{(1-\frac{3 \f_1}{4}\l)^{4/3}} \ ; \qquad a(\l) = a_0
= \frac{3}{2\ell^2} = \frac{3}{22} (11 - x) \ ,
        \ee
        which corresponds to the special case of $\rho=4/3$ and $\sigma=0$ in
Appendix~\ref{goodIR}. In this case the tachyon diverges exponentially,
        \be
           T(r) \sim T_0 \exp\left[\frac{81\ 3^{5/6} (115-16 x)^{4/3}
(11-x)}{812944\ 2^{1/6}}\frac{r}{R}\right]
        \ee
        as $r\to\infty$. Here $R$ is the IR scale of the solutions which is
defined by Eqs.~\eqref{IRresA}, \eqref{IRresl} and~\eqref{Rrel}. $T_0$ is the
only free parameter.

\item[II]   The other choice is practically a generalization of the first one.
It has a simple form of $\f(\l)$, but more complicated $a(\l)$:
        \be
         \f(\l) = \frac{1}{(1+\l/\l_0)^{4/3}} \ ; \qquad a(\l) = \frac{3}{22}
(11 - x)\frac{1-\f_1\l +\f_2\l^2}{(1+\l/\l_0)^{4/3}} \ ,
        \ee
        so that $\rho=4/3$ and $\sigma=2/3$ in Appendix~\ref{goodIR}.
The extra term proportional to $\f_2$ was added to make $\sigma$ positive. We
choose its coefficient to be small, $\f_2=1/\l_0^2$. The tachyon behaves as
        \be \label{TasympscII}
          T(r) \sim \frac{27\ 2^{3/4} 3^{1/4} }{\sqrt{4619}}
\sqrt{\frac{r-r_1}{R}}
        \ee
        for large $r$. Here $r_1$ is a free parameter.

\end{itemize}

We have checked that both scenarios lead to qualitatively similar results. In
the numerical calculations we use, for definiteness, the choice I, unless stated
otherwise.

\section{UV behavior} \label{AppUV}

In this Appendix we shall discuss the UV behavior of the system
\eqref{e1}-\eqref{e3} in general. This analysis should be compared to that
carried out in \cite{ihqcd,gkmn} in the absence of the tachyon backreaction.
Recall that apart from the two degrees of freedom of the transformation
\eqref{transf}, the solutions contain three integration constants. In the
discussion below, the degrees of freedom refer to these three ``nontrivial''
constants.

We shall not discuss the most general behavior of the solutions, but make some
physically motivated assumptions. In particular, we restrict ourselves to the
potentials $V_g$ and $V_f$ which are bounded as $\l \to 0$, and which are smooth
at any finite $\l$. In general we are interested two types of potentials: ones
that start from a constant value at $\l=0$, are monotonic as $\l$ increases, and
approach $+\infty$ as $\l \to \infty$, and ones that start at a constant value
at $\l=0$, increase until they reach a maximum at some $\l=\l_*$, and thereafter
monotonically decrease to $-\infty$ as $\l \to \infty$. This should be kept in
mind while reading the analysis below, as some of the arguments below may fail
for more generic potentials, even though no assumptions are listed explicitly.
Further, we mostly restrict to effective $\beta$-functions $\beta_{\rm eff} =
d\l/dA$ which are negative.

\subsection{Generic behavior}

\subsubsection{Singularity at $\l = 0$}

Near the standard UV singularity, the tachyon must behave schematically as
$T(r)\sim m r + \sigma r^3$ whereas the other fields have a logarithmic
dependence on $r$. Therefore the tachyon decouples asymptotically as $r \to 0$,
and we may analyze its UV behavior by first solving $A$ and $\l$ with $T=0$ and
then by analyzing the tachyon EoM for this background.

\paragraph{Asymptotic behavior of $A$ and $\l$}

We  take
\be \label{Videfapp}
 V_{\rm eff}(\l)=V_g(\l)-x V_f(\l,0)=\frac{12}{\ell^2}\left[1 + V_1 \l +V_2
\l^2+\cdots \right] \ .
\ee
Then the (leading) UV expansions of $A$ and $\l$ can be written as
\bea \label{UVexpsapp}
A(r) &=& -\log\frac{r}{\ell} + \frac{4}{9 \log(r \Lambda)}  \\
&&+ \frac{
  \frac{1}{162} \left[95  - \frac{64 V_2}{V_1^2}\right] +
   \frac{1}{81} \log\left[-\log(r \Lambda)\right] \left[-23 + \frac{64
V_2}{V_1^2}\right]}{
  \log(r \Lambda)^2} +{\cal O}\left(\frac{1}{\log(r\Lambda)^3}\right) \nn \\\nn
  V_1 \l(r)&=&-\frac{8}{9 \log(r \Lambda)} + \frac{
   \log\left[-\log(r \Lambda)\right] \left[\frac{46}{81} - \frac{128 V_2}{81
V_1^2}\right]}{\log[r \Lambda]^2}+{\cal
O}\left(\frac{1}{\log(r\Lambda)^3}\right) \ .
\eea
Notice that they contain no free parameters (in addition to $\Lambda$). In fact,
after using the equations of motion there is one degree of freedom left in the
coefficients of the above expansion, but as it turns out, this freedom can be
eliminated by rescaling $\Lambda$. We have removed this extra parameter by
requiring that the coefficient of the $1/(\log r\Lambda)^2$ term in the
expansion of $\l$ vanishes.

\paragraph{Tachyon UV asymptotics} \label{TUV}

 We take
\be
 V_f(\l,T) = e^{-a(\l)T^2}V_f(\l)
\ee
and parametrize
\bea \label{Vhexps}
 V_{\rm eff}(\l) &=& V_g(\l)-x V_f(\l,0)=\frac{12}{\ell^2}\left[1 + V_1 \l +V_2
\l^2+\cdots \right] \\\nn
 x V_f(\l) &=& \l^\delta\left[W_0 + W_1 \l +W_2 \l^2+\cdots \right] \nn\\
 \frac{\f(\l)}{a(\l)} &=& \frac{2 \ell^2}{3}\left[1 + \f_1 \l +\f_2 \l^2+\cdots
\right]
\eea
where $\delta$ is a nonnegative integer. Here the leading coefficient of $\f/a$
was already fixed in order to have the correct UV mass of the tachyon
\cite{ckp}. We  further assume that
\be
 \f(\l) = \l^\xi(1 + {\cal O}(\l)) \ .
\ee
It is enough to study the linear terms in the tachyon EoM, which read
\bea
 &&T''(r) + \Bigg[-3-\frac{\delta +\xi}{\log(r\Lambda)}
+ {\cal O}\left(\frac{1}{\log(r\Lambda)^2}\right)\Bigg]\frac{T'(r)}{r}\\ \nn
&&+ \Bigg[3+\frac{8 (\f_1\!+V_1\!)}{3 V_1 \log(r\Lambda)}
  + {\cal O}\left(\frac{1}{\log(r\Lambda)^2}\right)
\Bigg] \frac{T(r)}{r^2} = 0 \ .
\eea From 
this one could expect that the solution has the form
\be
T(r) \sim m r\left (1+\mathcal{O}\left(1/\log r\right)\right) + \sigma r^3\left
(1+\mathcal{O}\left(1/\log r\right)\right) \ ,
\ee
i.e., the logarithmically suppressed corrections to the EoM show up as
logarithmically suppressed corrections to the functions. However, this is not
the case: an Ansatz for the solution which assumes this kind of corrections
fails. The correct asymptotics reads
\bea \label{TUVres}
 \frac{1}{\ell}T(r) &=& m r
(-\log(r\Lambda))^{\frac{4}{3}-\frac{\delta}{2}-\frac{\xi}{2}+\frac{4 \f_1}{3
V_1}} \left[1+\frac{C_1 + C_2 \log(-\log(r\Lambda))}{\log(r\Lambda)}+ {\cal
O}\left(\frac{1}{\log(r\Lambda)^2}\right)\right] \\ \nn
&&+\sigma r^3
(-\log(r\Lambda))^{-\frac{4}{3}+\frac{3\delta}{2}+\frac{3\xi}{2}-\frac{4 \f_1}{3
V_1}} \left[1+\frac{D_1 + D_2 \log(-\log(r\Lambda))}{\log(r\Lambda)}+ {\cal
O}\left(\frac{1}{\log(r\Lambda)^2}\right)\right]
\eea
where $C_i$ and $D_i$ are known functions of $\delta$, $\xi$, $\f_1$, $V_1$,
$V_2$, $W_1$, and $W_2$, which we suppressed for brevity.

The ``surprising'' logarithmic power corrections in Eq.~\eqref{TUVres} can
actually be identified as the nontrivial running of the quark mass and the
condensate in the UV, which arises as their anomalous dimensions are different
from zero. To make this explicit, we calculate the gamma function $T'/A'$ in the
UV. For $m\ne 0$ it is dominated by the linear tachyon solution:
\be \label{UVgammam}
 \frac{\gamma}{T} = \frac{T'}{TA'} = -1  -\frac{\frac{4}{3}-\frac{\delta+\xi}{2}
+ \frac{4 \f_1}{3 V_1} }{\log(r\Lambda)} + {\cal
O}\left(\frac{1}{\log(r\Lambda)^2}\right)
\ee
whereas for $m=0$ we find
\be \label{UVgammacond}
 \frac{\gamma}{T} = -3 +\frac{\frac{4}{3}-\frac{3 (\delta+\xi)}{2} + \frac{4
\f_1}{3 V_1} }{\log(r\Lambda)} + {\cal O}\left(\frac{1}{\log(r\Lambda)^2}\right)
\ .
\ee
The next-to-leading terms in~\eqref{UVgammam} and~\eqref{UVgammacond} are mapped
to the one-loop anomalous dimensions of the quark mass and the chiral condensate
in QCD, respectively. Since they should add up to zero, we must have
$\delta+\xi=0$. The easiest way to satisfy this is to take  $\delta=0=\xi$. In
particular, the expression $h(\l)=\l^{-4/3}$ with $\xi=-4/3$, which was found in
the probe limit \cite{ihqcd}, does not work, since $\delta$ was required to be
an integer to ensure that the $\beta$-functions have power series with integer
powers at $\l=0$.

Finally, it is easy to verify that the UV expansion presented here match with
those of Sec.~\ref{SecBetasUV}, which were derived by using the holographic beta
functions.

\subsubsection{A bounce back at finite $\l$} \label{UVbounceback}

A bounce back may take place when the potential $V_{\rm
eff}(\l)=V_g(\l)-xV_f(\l,T=0)$ has a maximum at some $\l=\l_*$ signaling the
presence of an infrared fixed point, and $V'(\l)<0$ for $\l>\l_*$. If the
tachyon is sufficiently small, the effective $\beta$-function $d\l/dA$ hits
zero, and becomes positive when the system is evolved toward the UV. Therefore
the coupling has a finite minimum ($>\l_*$) and the above ``standard'' UV
singularity at $\l=0$ is not reached. All fields are analytic at the point where
$\l'=0$. The bounce back behavior is found in the white regions of
Fig.~\ref{figuvbeh}. Examples of the $\beta$ and $\gamma$-functions evaluated
along the RG flow for the bounce back scenario are shown as the dotted curves in
Fig.~\ref{figxscan} in Appendix~\ref{AppBG}.

\subsection{Special case: UV fixed point at finite $\l$} \label{AppUVFP}

We have identified one special UV singularity, which is found as a limiting case
between the two first generic behaviors discussed above (the blue curve of
Fig.~\ref{figuvbeh}). In this case the asymptotic solution is expected to depend
on two integration constants.  The solution terminates as the $\beta$-function
$d\l/dA$ approaches zero at the maximum $\l_*$ of the effective potential
$V_{\rm eff}(\l)=V_g(\l)-xV_f(\l,T=0)$. The singularity is found at a fixed
value of $r=r_*$ where $A$ diverges and $\l$ approaches $\l_*$ from above.
Examples of the $\beta$ and $\gamma$-functions evaluated along the RG flow with
this UV fixed point are shown as the thick blue curves of the middle and right
columns of Fig.~\ref{figxscan} in Appendix~\ref{AppBG}.

Depending on the value of the tachyon mass at the fixed point (see
Sec.~\ref{SecIRFPdim}), the asymptotics may be written in two different forms.
Let us recall the definition of the dimension $\Delta$ at the fixed point:
\be \label{Tmassapp}
\Delta(4-\Delta)={24 a(\l_*)\over \f(\l_*) V_{\rm eff}(\l_*)} \ .
\ee

When $x>x_c$ the right hand side of the definition is smaller than 4 so that
there are two real roots $\Delta =\Delta_\pm$. The geometry approaches the AdS
one near the fixed point,
\bea
  A(r)  &=& -\log r + \log \ell_* + A_0 (\Lambda r)^{2\Delta_-} + \cdots\\
 \l(r) &=& \l_* + \l_0 (\Lambda r)^{2\Delta_-} + \cdots\\
 T(r) &=&  T_0 (\Lambda r)^{\Delta_-} + \cdots \ ,
\eea
where  $\Delta_-$ is the smaller root, $\ell_*^2 =12/V_{\rm eff}(\l_*)$, we
chose $r_*=0$, and the constants $A_0$, $\l_0$, and $T_0$ satisfy two
constraints which can be solved from the EoMs \eqref{e1}-\eqref{e3}.
There are two free parameters which can be taken to be the coefficients of the
tachyon solutions with the dimensions $\Delta_\pm$. The solution associated to
$\Delta_+$ will appear at the next-to-leading order only if we choose the
coefficient $T_0$ of the above solution to vanish.

When $x<x_c$ we have two complex roots $\Delta_\pm \equiv 2 \pm i k$. Now the
asymptotics reads
\bea
  A(r)  &=& -\log r + \log \ell_* +(\Lambda r)^{4}\left[A_1+ \hat A_1
\sin\left(2 k \log(\Lambda r) +\phi_A \right)\right] + \cdots \\\
 \l(r) &=& \l_* + (\Lambda r)^{4}\left[\l_1+ \hat \l_1 \sin\left(2 k
\log(\Lambda  r) +\phi_\l \right)\right] + \cdots \\\
 T(r) &=&  T_0 (\Lambda r)^{2} \sin\left(k \log(\Lambda  r) +\phi_T \right) +
\cdots \ .
\eea
The coefficients in the next-to-leading terms for $\l$ and $A$ can
be solved from the EoMs  \eqref{e1}-\eqref{e3} by inserting the tachyon
asymptotics. The free parameters are $T_0$ and $\phi_T$ in this case.

\section{IR behavior}  \label{AppIR}

In this Appendix we discuss the IR behavior of the system \eqref{e1}-\eqref{e3}
in general. As above, we restrict to certain quite simple potentials and to
cases where the $\beta$-function $d\l/dA$ takes negative values in the vicinity
of the IR singularity. The IR structure is much richer than the UV one mostly
because there are much less obvious constraints on the potentials.

We shall discuss here both generic and special singularities. Of these the most
interesting ones for us will be the the special ones that depend only on one 
free parameter (excluding trivial reparametrization symmetries). 
They can be identified as the ``good'', or fully repulsive, IR singularities.
Examples of such singularities are identified below in Sec.~\ref{SecTIR}
partially based on the analysis of Sec.~\ref{goodIR}.

\subsection{Generic cases}

There are two generic IR ``singularities'' which in fact do not involve
divergences of any of the fields. Therefore, they also appear independently of
the details of the potentials to a large extent.

\subsubsection{Divergence of the derivative of the tachyon}

A typical, generic IR behavior is similar to what was found in the probe limit
in \cite{ihqcd}, where the tachyon goes to a constant value but its derivative
diverges. Indeed, the Ansatz
\bea \label{Tsqrtexp}
 T(r) &=& T_* + T_1\sqrt{r_*-r}+T_2(r_*-r)+\cdots \\
 A(r) &=& A_* + A_1(r_*-r) + A_2(r_*-r)^{3/2}+\cdots \\
 \l(r) &=&  \l_* + \l_1(r_*-r) + \l_2(r_*-r)^{3/2}+\cdots
\eea
solves the equations of motion \eqref{e1}-\eqref{e3} quite in general. We do not
present the rather complicated constraint equations which follow for the
constants in the expansions, but it is not difficult to check that the solution
has three independent integration constants and is thus indeed generic. Since
none of the fields diverge at $r=r_*$, it is natural to take all the potentials
to be analytic at the point of expansion, and the solution is expected to exist
to a large extent independently of the choices for them.

Notice also that there is no real singularity at $r=r_*$: one can make a
coordinate transformation such that all fields are analytic in the vicinity of
this point. A natural choice that realizes this is to use $T$ as the coordinate.
Of course, this leads to all fields being double-valued functions of $r$, with
the two branches having the same absolute value of $T_1$ but opposite signs. If
this is allowed, it is not hard to find analytic solutions which, for example,
start at a UV singularity at $\l=0$, bounce back at a point where the tachyon
derivative diverges, and return to another singularity at $\l=0$.

\subsubsection{A bounce back as $d\l/dA \to 0$} \label{bounceb}

There is also a three-dimensional space of solutions where the coupling reaches
a maximum value, and then starts to decrease with decreasing $A$ so that the
$\beta$-function $d\l/dA$ becomes positive. All fields are analytic in $r$ at
the point where the $\beta$-function is zero. We have not checked how the
solutions continue to evolve in the region of positive $\beta$-function.

\subsection{Special IR singularities}

In addition to the generic IR behaviors discussed above, we have identified
several true IR singularities, where the fields $A$ and $\l$ diverge.
Note  that above,  the divergence of the tachyon derivative was proportional to
the parameter $T_1$, which could take both positive and negative values.
Solutions with arbitrary small $|T_1|$ also exist, as well as the limit $T_1 \to
0$ where the solution, which in general ends in the divergence of $A$ and $\l$
rather than $T$. Therefore we expect that spaces of generic solutions with
positive and negative $T_1$, respectively, will be separated by a subspace of
solutions with singularity in the IR. The behavior of the system in this case
depends strongly on the asymptotics of the potentials. However, the tachyon
often decouples asymptotically from $\l$ and $A$, in particular if the tachyon
diverges in the IR, which is the expected behavior for physically relevant
singularities \cite{ckp,ikp}. Here we shall assume that the decoupling takes
place, since completely general classification of the singularities seems
daunting.

As the tachyon decouples, the classification of singularities for $A$ and $\l$
follows earlier studies \cite{ihqcd,gkmn}.
We shall review the results here for clarity. Assuming that the tachyon tends to
$T_0$ in the IR, the effective potential that drives the metric and the coupling
in the IR is
\be
 V_{\rm IR}(\l) = V_g(\l)- x V_f(\l,T_0)
\ee
where $T_0$ can be infinite in which case $V_{\rm IR}(\l) = V_g(\l)$.
We parametrize
\be
 V_{\rm IR}(\l) =  \l^{2Q}\left(\log\l\right)^P\left(V_0+\frac{V_1}{\log\l}
+\frac{V_2}{(\log\l)^2} + \cdots \right)
\ee
as $\l \to \infty$. The equations of motion are Eqs.~\eqref{e1} and~\eqref{e2}
with $T(r) \equiv T_0$ so that $T'(r)=0$. There are two types of singularities
(see \cite{gkmn}): ``generic'' ones where
\be
 X = \frac{1}{3\l}\beta = \frac{1}{3\l}\frac{d\l}{dA} \to -1
\ee
in the IR, and ``special'' ones where $X \to - \frac{3}{4} Q$.

\subsubsection{Generic metric singularity} \label{Xeq1as}

The generic singularities exist for $Q \le 4/3$ and depend on one free
parameter. The system is solved by the Ansatz
\bea
 A &=& \frac{1}{3} \log \delta r -\log R + A_1 \delta r^{(8-6Q)/3} (-\log \delta
r)^P + \cdots \\
 \l &=& \frac{1}{\delta r}\left(\l_0+\l_1 \delta r^{(8-6Q)/3} (-\log \delta r)^P
+ \cdots \right)
\eea
where
\be
 \delta r = (r_*-r)/R
\ee
is the ``conformally invariant'' distance from the singularity, $\l_0$ is the
free parameter, and the dropped terms are suppressed by $1/\log\delta r$.
Plugging this in the equations of motion,
\bea
 A_1 &=& \frac{3 V_0 \l_0^{2 Q}}{88-114 Q+36 Q^2}\\
 \l_1 &=& -\frac{27 (2 Q-1) V_0 \l_0^{2 Q+1}}{16 \left(44-57 Q+18 Q^2\right)} \
.
\eea
For $Q=4/3$ the singularity exists if $P<1$. In this case the asymptotics reads
\bea
 A &=& \frac{1}{3} \log \delta r -\log R + \frac{V_0 \l_0^{8/3}}{3 (1+P)}
(-\log \delta r)^{P+1} + \cdots \\
 \l &=& \frac{1}{\delta r}\left(\l_0 - \frac{5 V_0 \l_0^{11/3}}{8 (1+P)} (-\log
\delta r)^{P+1} + \cdots \right) \ .
\eea

\subsubsection{Special metric singularity} \label{specIR}

This kind of singularity exist for $0<Q<4/3$ so that the asymptotic value
$-4Q/3$ of $X$ lies between zero and one. We need to require that the potential
is asymptotically positive, $V_0>0$. The solution does not involve any
integration constants in addition to the ones linked to the reparametrization
symmetry, which suggest that when combined with a proper tachyon solution,
``good'' IR asymptotics can be identified.

If $2/3<Q<4/3$ we find a singularity at finite value $r_*$ of $r$:
\bea
 A &=& \frac{1}{9 Q^2/4-1} \log \delta r -\log R + \mathcal{O}\left(1/\log
\delta r\right) \\
 \l &=& -\frac{Q}{Q^2-4/9}\log \delta r + \frac{1}{2Q}\bigg[2 \log 2+(1-2 P)
\log 3 + (P-2) \log \left(9 Q^2-4\right) \\\nn
 &&+\log \left(16-9 Q^2\right)-P \log Q-P \log (-\log  \delta r)-\log V_0 \bigg]
+ \mathcal{O}\left(1/\log \delta r\right) \ ,
\eea
where again $\delta r = (r_*-r)/R$. Here  $R$ and $r_*$ are the integration
constants which reflect the reparametrization symmetry, but no other free
parameters appear.

If $0<Q<2/3$ similar formulas hold for $r \to \infty$:
\bea
 A &=& -\frac{1}{1-9 Q^2/4} \log \hat r-\log R + \mathcal{O}\left(1/\log \hat
r\right) \\
 \l &=& \frac{Q}{4/9-Q^2}\log \hat r + \frac{1}{2Q}\bigg[2 \log 2+(1-2 P) \log 3
+ (P-2) \log \left(4-9 Q^2\right) \\\nn
 &&+\log \left(16-9 Q^2\right)-P \log Q-P \log (\log  \hat r)-\log V_0 \bigg] +
\mathcal{O}\left(1/\log \hat r\right) \ ,
\eea
where now $\hat r = (r-r_0)/R$.

\label{goodIR}

If $Q=2/3$, and $P<1$, there is a singularity at $r=\infty$. The asymptotic
solution reads
\bea \label{IRresA}
 A &=& - \left(\frac{r-r_0}{R}\right)^{\alpha }+A_0-\frac{1}{2} \frac{P}{1-P}
\log\frac{ R} {r-r_0}+\frac{5}{6}+\frac{P}{4}+\frac{1}{2} P
\log\frac{3}{2}+\frac{2 V_1}{3 P V_0}\\
&&+\frac{ -52 P^2 V_0^2+4 P^3 V_0^2+27 P^4 V_0^2+64 V_1^2-64 P V_1^2+128 P V_0
V_2}{288 P (1+P) V_0^2} \left(\frac{R}{r-r_0}\right)^{\alpha } + \cdots\nn\\
\log \l &=& +\frac{3}{2} \left(\frac{r-r_0}{R}\right)^{\alpha }
-\frac{5}{4}-\frac{3 P}{8}-\frac{V_1}{P V_0}\\
&&+\frac{ -20 P^2 V_0^2-40 P^3 V_0^2+9 P^4 V_0^2-64 V_1^2+64 P V_1^2-128 P V_0
V_2}{192 P (1+P) V_0^2} \left(\frac{R}{r-r_0}\right)^{\alpha }+ \cdots\nn
\label{IRresl}
\eea
where
\bea
 \alpha &=&\frac{1}{1-P} \\
 R &=&\frac{2^P 3^{1-P} }{(1-P) e^{A_0} \sqrt{V_0}} \ .
\label{Rrel}
\eea

As pointed out in \cite{ihqcd}, this special case produces a good match with the
IR physics of QCD, in particular if we choose $P=1/2$. We will
confirm below that the potentials of the tachyon action can be chosen such
that the tachyon asymptotics also meets all requirements known to us, 
and the produced
singularity is of the ``good'' kind.
We will use these singularities in our analysis, 
and the above formulas will be
used to fix the IR boundary conditions for the numerical solutions. 

If $Q=2/3$, and $P>1$, we find a singularity at finite value $r=r_*$ of the
coordinate. The asymptotics transforms to
\bea
 A &=& - \left(\frac{R}{r_*-r}\right)^{ \bar \alpha }+A_0-\frac{1}{2}
\frac{P}{1-P}  \log\frac{ R} {r_*-r}+\frac{5}{6}+\frac{P}{4}+\frac{1}{2} P
\log\frac{3}{2}+\frac{2 V_1}{3 P V_0}\\
&&+\frac{ -52 P^2 V_0^2+4 P^3 V_0^2+27 P^4 V_0^2+64 V_1^2-64 P V_1^2+128 P V_0
V_2}{288 P (1+P) V_0^2} \left(\frac{r_*-r}{R}\right)^{\bar \alpha } +
\cdots\nn\\
\log \l &=& +\frac{3}{2} \left(\frac{R}{r_*-r}\right)^{\bar \alpha }
-\frac{5}{4}-\frac{3 P}{8}-\frac{V_1}{P V_0}\\
&&+\frac{ -20 P^2 V_0^2-40 P^3 V_0^2+9 P^4 V_0^2-64 V_1^2+64 P V_1^2-128 P V_0
V_2}{192 P (1+P) V_0^2} \left(\frac{r_*-r}{R}\right)^{\bar \alpha }+ \cdots\nn
\eea
where $\bar \alpha =1/(P-1)$ and $A_0$ is related to $R$ as in Eq.~\eqref{Rrel}.

Finally, for $Q=2/3$ and $P=1$ the metric factor $A$ diverges exponentially as
$r \to \infty$,
\bea
 A &=& - \exp\left(\frac{r-r_0}{R}\right)-\log R + \frac{r-r_0}{2R}+\frac{1}{2}
\log 6- \frac{1}{2}\log V_0+\frac{13}{12} +\frac{2 V_1}{3V_0}\\\nn
&&+\frac{128 V_2-21 V_0}{576 V_0}\exp\left(-\frac{r-r_0}{R}\right) + \cdots\\
\log \l &=& +\frac{3}{2}
\exp\left(\frac{r-r_0}{R}\right)-\frac{V_1}{V_0}-\frac{13}{8}  +\frac{ -51
V_0-128 V_2}{384 V_0}\exp\left(-\frac{r-r_0}{R}\right)+ \cdots \ .
\eea

\subsubsection{Tachyon behavior} \label{SecTIR}

To complete the analysis, one should insert each of the above asymptotics to the
tachyon EoM and check  what the tachyon asymptotics is for various choices of
the potentials $V_f$ and $\f$, and start looking for the ``good'' kind of
singularities. Once the potentials are fixed, one can check if a solution, which
is consistent with the assumption that the tachyon decouples, indeed exists. If
it does, it can depend on one or two additional parameters. For the good, fully
repulsive singularities  the number of free parameters (excluding those related
to the reparametrization symmetry) is equal to one, i.e., we must have a special
metric singularity combined with a one-parameter tachyon asymptotics.
In addition we should require that the tachyon diverges in the IR, since that
kind of solutions have bulk flavor anomalies similar to those of QCD
\cite{ckp,ikp}.

Here we shall restrict to
the special metric IR singularity with $Q=2/3$ and $P<1$, since it is expected
to include the most interesting cases due to additional constraints from
confinement and excitation spectra \cite{ihqcd}.
We parametrize
\be
 V_f(\l,T) = V_{f0}(\l)\exp\left(-a(\l)T^2\right) \ ; \qquad \f = \f(\l)
\ee
With this parametrization, the tachyon EoM reads
\be \label{Teomg}
 T'' + F_1 T' + F_2 T + F_3 T'^3 + F_4 T'^2 T +F_5 T' T^2 + F_6 T'^3 T^2 = 0
\ee
where
\begin{align}
 &F_1  = 3 A' + \l'\frac{d}{d\l} \log(\f(\l) V_{f0}(\l))  \ ; \qquad &F_2 &=
\frac{2 a(\l) e^{2A}}{\f(\l)}\ ; \\
 &F_3  = \f(\l) e^{-2A}\left[4 A' + \l' \frac{d}{d\l}\log(
\sqrt{\f(\l)}V_{f0}(\l))\right]   \ ; \qquad &F_4 &= 2 a(\l)\ ; \\
 &F_5  = -\l' \frac{da(\l)}{d\l}   \ ; \qquad &F_6 &= -e^{-2A} \f(\l) \l'
\frac{da(\l)}{d\l}
\end{align}
and the primes are derivatives with respect to $r$.
Notice that the last two terms vanish if $a(\l)$ is constant.

We  consider generic power-law asymptotics
\be
 \f(\l) \sim \f_0 \l^{-\rho} \ ; \qquad a(\l) \sim a_0 \l^\sigma \ ; \qquad
V_{f0}(\l) \sim W_0 \l^\tau
\ee
of the potentials at large $\l$, and introduce a shorthand notation for the
asymptotic behavior in \eqref{IRresA}-\eqref{IRresl}:
\bea
 A &=& - \left(\frac{r-r_0}{R}\right)^{\alpha
}+\frac{\a-1}{2}\log\frac{r-r_0}{R} +  A_c+
\mathcal{O}\left(\frac{r-r_0}{R}\right)^{-\a} \\
\log\l&=& \frac{3}{2} \left(\frac{r-r_0}{R}\right)^{\alpha } + \l_c+
\mathcal{O}\left(\frac{r-r_0}{R}\right)^{-\a}
\eea
where we set $r_0=0$. Then the leading behavior of the coefficients $F_i$ at
large $r$ is
\bea
 F_1&  \sim& -\frac{3 \alpha  (2+\rho-\tau  )  }{2
r}\left(\frac{r}{R}\right)^{\alpha } \\
 F_2 &\sim &\frac{2 a_0  e^{2 A_c+\rho  \lambda _c+\sigma  \lambda _c}}{\f_0}
\left(\frac{r}{R}\right)^{\alpha -1} \exp\left[\left(\frac{3}{2} \rho
+\frac{3}{2} \sigma-2\right)\left(\frac{r}{R}\right)^{\alpha} \right]  \\
 F_3&  \sim & -\frac{\f_ 0 \alpha  (16-6 \tau +3 \rho ) e^{-2 A_c-\rho  \lambda
_c}}{4 R} \exp \left[\left(2-\frac{3}{2} \rho
\right)\left(\frac{r}{R}\right)^{\alpha }\right]\\
F_4 & \sim & 2 a_ 0 e^{\sigma \lambda _c}\exp\left[\frac{3\sigma}{2}
\left(\frac{r}{R}\right)^{\alpha }\right] \\
 F_5 &  \sim & -\frac{3 a_ 0 \alpha  \sigma   e^{\sigma \lambda _c} }{2 r}
\left(\frac{r}{R}\right)^{\alpha }\exp \left[ \frac{3}{2}\sigma
\left(\frac{r}{R}\right)^{\alpha }\right] \\
F_6 &\sim& -\frac{3 a_ 0 \f_ 0 \alpha  \sigma  e^{-2 A_c-\rho  \lambda _c+\sigma
 \lambda _c}}{2 R}  \exp \left[\left(2+\frac{3}{2} \sigma-\frac{3}{2} \rho
\right)\left(\frac{r}{R}\right)^{\alpha }\right]\ .
\eea

For $\sigma =0$ the leading terms of $F_5$ and $F_6$, given above, become zero.
If $a(\l)$ is constant for all $\l$, these terms actually vanish identically. If
$a(\l)$ only asymptotes to a constant value, the leading behavior of $F_5$ and
$F_6$ is determined by the next-to-leading term in the asymptotics of $a(\l)$.
These will contribute in some particular cases, as we discuss below.

\begin{itemize}

\item $\sigma > 0$ and $\rho<4/3$. The tachyon EoM \eqref{Teomg} is dominated by
the terms $\propto T$, $T'T^2$, which have the coefficients $F_2$ and $F_5$,
respectively. Solving the tachyon from these terms gives the asymptotics
\be \label{constas}
 T(r) \sim T_0 - \frac{4 R\, e^{2A_c+\rho \l_c}}{3\alpha\sigma T_0}
\int_r^\infty d\hat r \exp\left[\left(\frac{3}{2}\rho-2\right)\left(\frac{\hat
r}{R}\right)^{\alpha }\right] \ .
\ee
Substituting this back to the full equation of motion, we see that with the
above constraints for $\sigma$ and $\rho$ it is indeed a solution. Further, the
factor $\exp\left(-a(\l)T^2\right)$ vanishes double-exponentially, which
confirms the decoupling of the tachyon from the other fields. In addition to the
trivial reparametrization symmetry, the only free parameter of the asymptotic
solution is $T_0$, which suggest that the singularity is of the ``good'' kind.
However, tachyon solutions that are regular in the IR have bulk flavor anomalies
which differ from those of QCD \cite{ckp,ikp}. Therefore, we discard this
option.

\item $\sigma > 0$ and $\rho=4/3$. The same terms continue to dominate, but the
asymptotic changes. We find instead
\be \label{Tsqrtas}
 T(r) \sim 2\sqrt{\frac{2 R}{3\alpha \sigma \f_0}}
e^{A_c+\frac{2}{3}\l_c}\sqrt{r-r_1}
\ee
where $r_1$ is a free parameter. One can again check that this is indeed a
solution, and that the tachyon decouples from $A$ and $\l$. The terms
proportional to $T'^2 T$ and $T'^3T^2$ are suppressed only by $r^{-\a}$, but
taking them into account results in a trivial factor multiplying the equation of
motion, so that the solution in Eq.~\eqref{Tsqrtas} is unchanged for any value
of $\a$. The asymptotics has only one free parameter, and the tachyon diverges
as $r \to \infty$, so this solution is acceptable.

\item $\sigma \le 0$ and $\rho<4/3-\sigma$. We find two different cases. The
asymptotics is qualitatively similar to \eqref{constas}, but arises in a
slightly different way.  The leading terms are those proportional to $T'$ and
$T$, corresponding to coefficients $F_1$ and $F_2$, respectively, as well as the
double derivative term $T''$. This term being leading, one might expect that the
asymptotics contains two free parameters, and is thus not of the good kind. This
is indeed the case for $\tau > 2+\sigma$. For $\tau\le 2+\sigma$ only one
parameter family of the asymptotic solutions is consistent with the other terms
being subleading. However, since the tachyon becomes constant in the IR, we have
the aforementioned problem with flavor anomalies, and hence we shall discard
this solution in any case.

\item $\sigma = 0$, $\rho=4/3$, and $\tau<10/3$. The leading terms are typically
those proportional to $T'^3$ and $T'^2T$. However, if $a(\l)$ is not constant,
there is an extra constraint from the next-to-leading term in the expansion of
$a(\l)$. If, for example, $a(\l)=a_0 +a_1/\l +\cdots$, we need to require
$\a>1$. Assuming that all constraints are met, we find the exponential behavior
\be
 T(r) \sim T_0 e^{C r}
\ee
where $T_0$ is a free parameter and
\be
  C=\frac{4 e^{A_c+ 4\l_c/3} R a_0}{ (10-3 \tau) \a \f_0 } \ .
\ee
This special case is the asymptotics that was discussed in the probe limit in
\cite{ihqcd}, and is acceptable. By inserting the expressions for $A_c$ and
$\l_c$ from Eqs.~\eqref{IRresA}, \eqref{IRresl}, and \eqref{Rrel}, the
coefficient simplifies to
\be
 C = \frac{2^{2+P} 3^{2-P}  a_0}{ (10-3 \tau )(1-P) \f_0 V_0 R} \ .
\ee

\item $\sigma = 0$, $\rho=4/3$, and $\tau>10/3$. The leading terms are
proportional to $T'$ and $T$, which results in the tachyon vanishing
asymptotically. In this case the factor $\exp\left(-a(\l)T^2\right)$ goes to one
rather zero, which suggest that the correct physical picture, as discussed in
the main text, cannot be achieved, even though the tachyon apparently decouples
in the IR.

\item $\sigma < 0$ and $\rho=4/3-\sigma$. The leading terms are again
proportional to $T'$ and $T$. For $\tau<10/3-\sigma/2$, the solution is
exponentially increasing,
\be
 T(r) \sim e^{C r}
\ee
with
\be
  C=\frac{4 e^{A_c+ 4\l_c/3} R a_0}{ (10-3 \tau-3\delta) \a \f_0 } \ .
\ee
The factor $\exp\left(-a(\l)T^2\right)$ vanishes in the IR limit if $\a<1$,
which is the expected behavior, so the solution is acceptable. For $\a>1$ or
$\tau>10/3-\sigma/2$ the factor goes to one instead, and it seems that the
correct physical picture cannot be obtained. The borderline case $\a=1$ has
either behavior depending on the values of other parameters.

\end{itemize}

In all the remaining cases, in particular for large $\rho$, the asymptotic
solution of Eq.~\eqref{Teomg} oscillates with increasing frequency
as $r \to \infty$. Therefore, the tachyon is apparently not decoupled from the
other fields, and the singularity discussed in the subsection does not exist.

In summary, we found good solutions only if $\rho=4/3$ and $\sigma \ge 0$, or
$\rho=4/3-\sigma$ and $\sigma < 0$. The latter case is included for $0<\a<1$,
and the former case we found some extra constraints at the endpoint $\sigma=0$,
which are detailed above. In the numerics we shall fix $P=1/2$ so that 
$\a = 2$, and use potentials with  $\rho=4/3$ and $\sigma \ge 0$, see
Appendix~\ref{AppPotentials}.

\subsection{Singularity at the IR fixed point with $T\equiv 0$} \label{TzeroIR}

As discussed in the main text, some of the solutions with identically vanishing
tachyon are expected to correspond to field theories where the chiral symmetry
is conserved. Potentials for Yang-Mills theory were discussed extensively in
\cite{ihqcd,gkmn}. Therefore, we will only consider the case of potentials
$V_{\rm eff}(\l)=V_g(\l)-xV_f(\l,0)$ which have a (single) maximum at some
$\l=\l_*$, interpreted as an infrared fixed point.

In the absence of the tachyon the space of solutions is one-dimensional. We
identify two distinct one-parameter families of solutions: one where the
solution bounces back to smaller couplings as the $\beta$-function $d\l/dA$ goes
to zero (a special case of the bounce-back solutions discussed above in
Sec.~\ref{bounceb}) and another where the $\beta$-function asymptotes as $d\l/dA
\sim -3 \l$ (a special case of the solutions discussed in Sec.~\ref{Xeq1as}).
These families are separated by a single solution where the $\beta$-function
terminates at zero at the maximum of the potential $\l=\l_*$.

The limiting solution has a singularity at $r=\infty$ where $A$ diverges and
$\l$ approaches $\l_*$ from below. We  expand the potential around $\l=\l_*$ as
\be
 V_{\rm eff}(\l) = V_0 +V_2 (\l-\l_*)^2 + \cdots
\ee
where $V_2$ is negative. The equations of motion are solved by
\bea \label{lexpT0}
 \l&=&\l_* -  \left(\frac{r-r_0}{R}\right)^{-\delta} + \cdots \\ \label{AexpT0}
 A&=& -\log(r-r_0) + A_0 +A_1 \left(\frac{r-r_0}{R}\right)^{-2\delta} + \cdots
\eea
where $A_0$ and $\delta$ are related to the IR AdS radius and the derivative of
the $\beta$-function, respectively, by
\bea
 \ell_\mathrm{IR}^2 &=& e^{2 A_0} = \frac{12}{V_0} \\
 \label{deltadef}
 \lim_{r\to\infty}\frac{1}{\l-\l_*}\frac{\l'(r)}{A'(r)} &=& \delta =
\sqrt{4-\frac{9 V_2 \l_*^2}{V_0}}-2 \ ,
\eea
and we also find
\be
 A_1 = -\frac{2 \delta}{9 (2 \delta -1) \l_*^2} \ .
\ee

\subsubsection{Generalization to $T \ll 1$} \label{AppMissedFP}

A generalization of the $T\equiv 0 $ case, which is of high interest to us, is
where the tachyon mass and the condensate are very small in the UV, such that
the tachyon remains small ($|T|\ll 1$) even as $\l$ approaches the fixed point
($\l_*-\l \ll 1$). However, for any nonzero tachyon profile, the tachyon will
eventually become $\mathcal{O}(1)$ at some high $r=r_\mathrm{IR}$, and drive the
flow away from the fixed point. Setting $r_0=0$, the above
formulas~\eqref{lexpT0} and~\eqref{AexpT0} then hold in the limit $R \ll r \ll
r_\mathrm{IR}$ (with the understanding that depending on the value of $\delta$,
the next-to-leading terms may be affected by the tachyon solution). This
approximation is useful in the quasiconformal, or ``walking'' region of
backgrounds for $x$ below, but close to the edge of the conformal window at
$x_c$. Notice that $R$ is the scale where the coupling starts to deviate
significantly from its fixed point value $\l_*$, and may be therefore identified
as the UV scale of the theory. Therefore we shall denote $R=r_\mathrm{UV}$ in
this case.

The tachyon profile can also be derived in this region. Inserting the solutions
of Eqs.~\eqref{lexpT0} and~\eqref{AexpT0} into the tachyon EoM~\eqref{e3}, and
taking $|T| \ll 1$, we get
\be
 T''(r) + \frac{3}{r} T'(r)  + \frac{2 a(\l_*)  \ell_\mathrm{IR}^2}{\f(\l_*)
r^2} T(r) =  0 \ .
\ee
There are two kinds of solutions depending on the value of the tachyon mass
term. As in Sec.~\ref{SecIRFPdim} we can define the dimension $\Delta$
\be
 \Delta(4-\Delta) = \frac{2 a(\l_*)  \ell_\mathrm{IR}^2}{\f(\l_*)} = \frac{24
a(\l_*) }{\f(\l_*) V_\mathrm{eff}(\l_*)}\ .
\ee
If the combination on the right hand side is less than 4, which corresponds to
$x>x_c$ we find two real roots $\Delta_\pm$. In this case the tachyon solution
is
\be \label{tachyonmonot}
 T(r) \sim T_0 \left(\frac{r}{r_\mathrm{IR}}\right)^{\Delta_-}
\ee
for $r_\mathrm{IR} \ll r \ll r_\mathrm{IR}$, where $\Delta_-$ is the smaller
root (unless we tuned the boundary conditions such that the quark mass is very
small, in which case the solution with $\Delta_- \to \Delta_+$ needs to be
included).

For  $x<x_c$ we have two complex roots $\Delta_\pm= 2 \pm i k$ instead, and the
tachyon behaves as
\be \label{tachyonosc}
 T(r) \sim T_0 \left(\frac{r}{r_\mathrm{IR}}\right)^{2} \sin \left(k\log
\frac{r}{r_\mathrm{IR}} + \phi\right) \ .
\ee
This oscillating solution is the root of the rich structure of backgrounds found
for $x<x_c$ in Sec.~\ref{SecBG}, which are also discussed below in
Appendix~\ref{AppBG}.

\section{Structure of the background as a function of $x$ and $T_0$}
\label{AppBG}

In this Appendix we will explain in detail how the phase structure of the
background solutions seen in Fig.~\ref{figuvbeh}, and in particular the region
with nearly conformal behavior, arises.

The structure in the plots is linked to the transition of the system from the UV
region, where the tachyon is small and the background is characterized  by the
potential $V_{\rm eff}(\l)=V_g(\l)-x V_{f0}(\l)$, to the IR region, where the
tachyon is large, and the background is characterized  by $V_g(\l)$.  First,
recall that $V_{\rm eff}(\l)$ has a maximum at some $\l=\l_*$ which depends on
$x$. For $x \to 0$ we find from Eqs.~\eqref{Vgnumdef} and~\eqref{V0numdef} that
$\l_* \to \infty$, whereas for $x \to 11/2$ we obtain $\l_* \to 0$. This maximum
suggests a presence of an IR fixed point of the $\beta$-function for the
coupling $\l$.

However, for a nontrivial tachyon profile the fixed point is not reached. When
approaching $\l=\l_*$ from the UV, the solution is driven away from the fixed
point as soon as the tachyon becomes large, $T \sim \mathcal{O}(1)$, and the
system enters the region where all EoMs are nontrivially coupled. The point
where this happens, is controlled by the normalization of the tachyon in the IR,
i.e., the value of $T_0$ (Assuming potentials of scenario I). The larger $T_0$,
the smaller is the value of $\l$ where the tachyon decouples.

 The solid blue curve in Fig.~\ref{figuvbeh} is the critical value of $T_0$
where the tachyon decouples at $\l=\l_*$. Actually, precisely at this curve the
UV asymptotics is that of Sec.~\ref{AppUVFP}, and the solution ends at the fixed
point. If $T_0$ is smaller than the critical value, the tachyon
will become small $\ll 1$ during the flow  toward the UV when we still have
$\l>\l_*$, so that the beta function corresponding to the effective potential
$V_\mathrm{eff}$ is positive. As the tachyon decouples, the beta function flow
approaches that defined by $V_\mathrm{eff}$.  Therefore, to the the region with
``standard'' UV behavior is not reached at all, but the solution bounces back at
a finite value of the coupling (see also Fig.~\ref{figxscan} below) where the
beta function (evaluated along the RG flow) crosses zero. The strong dependence
of the blue curve on $x$ is explained by the dependence of $\l_*$ on $x$: for
example as $x \to 0$, the fixed point moves to large values of $\l_*$, and the
critical tachyon IR value $T_0$ required for avoiding the bounce back approaches
zero.

The argument above is not rigorous, as it involves the location of the
decoupling of the tachyon which is not defined precisely. However, one should
notice that as the blue curve is approached from above, the system is on the
verge of reaching an IR fixed point so that the coupling freezes, i.e., it
evolves very slowly for a large range of $r$. Meanwhile, the tachyon grows
relatively fast with $r$. Therefore, the value of $\l$ where the tachyon
decoupling takes place becomes more and more precisely defined as the blue curve
is approached from above. This can also be seen in the numerical examples below
and in Section~\ref{SecBG}.

To understand how the red dashed curve arises, we need to study the tachyon
solutions in the UV region. Below the red curve the tachyon solution develops a
zero, as is required by the negative value of the quark mass. This zero actually
appears in the region, where (the absolute value of) the tachyon is still small.
If we continue on the plot toward lower values of $T_0$, the quark mass becomes
again zero, and then positive extremely close to the solid blue curve. This
happens as the tachyon develops a second zero in the UV. We can continue
further, and find solutions with an arbitrary number $n$ of zeroes (which are
very hard to construct numerically). See Fig.~\ref{mT0} (left) in the main text
for the qualitative behavior of the quark mass as the blue curve is approached.

 The oscillating behavior of the tachyon in the UV is linked to the violation of
the BF bound: when  $\Delta_\mathrm{IR}(4-\Delta_\mathrm{IR})$ is smaller than
4, the solutions for $\Delta_\mathrm{IR}$ are complex, which results in the
oscillations of the tachyon solution. We plotted the squared mass of the tachyon
at the IR fixed point in Fig.~\ref{figTmass}, where the solid thick blue curve
corresponds to the present choice of parameters.

As we approach the solid blue curve from within the contoured region in
Fig.~\ref{figuvbeh}, the background can be approximated  near the fixed point as
discussed above in Sec.~\ref{AppMissedFP}.
 By the definition of Eq.~\eqref{dimcond}, the BF bound is violated at the fixed
point for $x<x_c$. Therefore, as we approach the blue curve from above in this
region and the system is about to develop a fixed point, the tachyon necessarily
oscillates as soon as values of $\l$ close enough to $\l_*$ are reached. In this
case, the tachyon is well approximated by the solution of
Eq.~\eqref{tachyonosc},
\be \label{tachyonosc2}
 T(r) \sim T_0 \left(\frac{r}{r_\mathrm{IR}}\right)^{2} \sin \left(k\log
\frac{r}{r_\mathrm{IR}} + \phi\right) \ ,
\ee
where $k$ is fixed by the potentials,
\be
 k  = \sqrt{\frac{24 a(\l_*) }{\f(\l_*) V_\mathrm{eff}(\l_*)}-4} \ ,
\ee
but $\phi$ and $T_0$ are free parameters, which will be fixed by the boundary
conditions.
 For $x>x_c$, the BF bound is not violated at the fixed point, and therefore no
oscillations are expected, and the tachyon dependence is as in
Eq.~\eqref{tachyonmonot}. The (uppermost) curve of zero quark mass (the red
dashed one in the plots) essentially limits the region of oscillating tachyon:
for example on the left hand plot above this curve quark mass is positive (no
tachyon zeroes) and below it the mass is negative (one tachyon zero), see the
mass dependence in Fig.~\ref{mT0}. Notice that the red curve must therefore join
the blue curve at $x=x_c$.

As we approach the fixed point keeping $x$, the range $r_\mathrm{UV} \ll r \ll
r_\mathrm{IR}$ where Eq.~\eqref{tachyonosc2} holds grows without
limit.\footnote{We cannot prove analytically that the scaling region is
accessible, because this would require a description of the solution in the IR
region where tachyon is not decoupled. However, we may solve the EoMs starting
from the UV with arbitrary small quark masses and vevs, which are guaranteed to
enter the oscillating region, and have arbitrary many tachyon zeroes. The tricky
issue is, if we can tune the solutions such that it ends in the good IR
singularity after the tachyon finally grows large and the oscillations end in
the IR. Due to the oscillating nature of the solution, it is plausible that the
good IR singularity is found after any number of oscillations $n$, if it is
found, e.g., for small $n$. In the end this question is settled by the numerics,
which supports our expectations, and solutions with regular IR behavior are
indeed found for any $n$.} Here $r_\mathrm{UV}$ is the scale where the coupling
starts to deviate significantly from its fixed point value $\l_*$ as we follow
the flow toward the UV. According to Eq.~\eqref{tachyonosc2}, the number of
encountered tachyon zeroes is
\be
 \pi n \sim k\log \frac{r_\mathrm{IR}}{r_\mathrm{UV}} \ .
\ee

We may take one step further, and find the scaling of the quark mass and the
chiral condensate as $r_\mathrm{IR}/r_\mathrm{UV} \to \infty$ and  $n \to
\infty$. This can be done by matching the ``intermediate'' tachyon
solution~\eqref{tachyonosc2} with the UV (and IR) solutions (see
Sec.~\ref{secBKT} and Appendix~\ref{AppBKT} where we do the matching procedure
more carefully for $x \to x_c$). The tachyon is supposed to become large at $r
\sim r_\mathrm{IR}$, so $T_0 \sim 1$. On the other hand at $r\sim r_\mathrm{UV}$
we enter the standard UV region, where roughly $T \sim m r + \sigma r^3$.
Matching in the UV gives the typical sizes for $m$ and $\sigma$ for large $n$:
\be
  m r_\mathrm{UV} \sim \sigma  r_\mathrm{UV}^3 \sim
\left(\frac{r_\mathrm{UV}}{r_\mathrm{IR}}\right)^{2} \sim \exp\left(-\frac{ 2\pi
n}{k}\right) \ .
\ee
In particular, the maximal masses for which solutions with $n$ tachyon nodes
exist, or in other words the sizes of the bumps in Fig.~\ref{mT0} (left),
numbered from right to left, must obey this scaling law. As the mass scale
vanishes exponentially for $n \to \infty$, for any fixed $m\ne 0$ there are only
finite number of backgrounds as $n$ is limited from above, whereas for $m=0$ we
find an infinite tower of solutions.

At this point is good to remind that the solution with no tachyon nodes ($n=0$)
always has smaller the free energy than the solutions with $n>0$ (see
Sec.~\ref{SecFE}). Also, the $n=0$ solution is not found in the scaling region
where the system is close to having a fixed point in general. For $m=0$ this
solution (red dashed curve of Fig.~\ref{figuvbeh} enters the scaling region
(which is close to the solid blue curve in the same figure) only in the limit $x
\to x_c$, which will be discussed in detail in Sec.~\ref{secBKT}.

 We conclude with one more observation on the location of the curve of the
vanishing quark mass. The above discussion was relying on the dimension
$\Delta_\mathrm{IR}$ at the fixed point, and therefore
we could argue how the tachyon behaves in the limit where the background comes
arbitrarily close to the fixed point, i.e., as we approach the blue solid curve
from above in Fig.~\ref{figuvbeh}.
However, we can also present rough qualitative arguments on the behavior of the
curve further away from the solid blue curve. Because the tachyon is in any case
small in the interesting region, we can read directly from the linearized
tachyon EoM whether it oscillates or not. The EoM reads
\be \label{linTeom}
T''(r) +\left[3 A' + \l'\frac{d}{d\l} \log(\f(\l) V_{f0}(\l))\right]T'(r)  +
\frac{2 a(\l) e^{2A}}{\f(\l)} T(r) \simeq  0 \ .
\ee

Recall that the tachyon is decoupled in this region and evolves independently of
the other fields. Assuming decoupling the coefficients are essentially
independent of $T_0$ and can be solved directly from the potential $V_{\rm
eff}(\l) = V_g(\l)-xV_{f0}(\l)$.
$A$ is quite well approximated by $A \simeq -\log r$,\footnote{Explicit
expressions can be derived in some approximation schemes, e.g., as series in the
limit of small $\beta$-function.} the second term in the square brackets of the
coefficient of $T'$ is small, and therefore the essential term is, as in the
case of the fixed point, the ratio $a(\l)/\f(\l)$, which increases with $\l$. As
$\l$ grows, at some critical $\l_c$ the ratio becomes large enough, and the
tachyon starts to oscillate. In terms of the gamma function, this means that
$\gamma/T$ reaches the value of approximately $-2$. The oscillations do not take
place if tachyon grows large already for $\l<\l_c$ so that nonlinear terms
contribute in Eq.~\ref{linTeom}. Therefore, for the limiting solutions, the
tachyon becomes $\mathcal{O} (1)$ roughly at $\l=\l_c$. As the start of the
oscillations means that the quark mass goes to zero, this  mechanism also fixes
the location of the curve with zero quark mass (red dashed curve in
Fig.~\ref{figuvbeh}).

In summary, the solid blue curve of Fig.~\ref{figuvbeh} is stabilized by the
tachyon growing large at $\l=\l_*$, whereas the dashed red curve is stabilized
by the tachyon growing large at $\l=\l_c$. The two curves meet when $\l_c=\l_*$,
which gives an alternative way to formulate the definition of $x_c$.

It is interesting to compare the picture in our model to that arising in the
Dyson-Schwinger approach in the rainbow approximation (see, e.g.,
\cite{Appelquist:1988yc}). Also in this framework it is useful to define two
values of the coupling, corresponding to $\l_*$ and $\l_c$ above. The
definitions are similar as here: $\l_*$ is the zero of the $\beta$-function, and
$\l_c$ is the value of $\l$ where the anomalous dimension of the chiral
condensate reaches unity (so that $\Delta$ of Sec.~\ref{SecIRFPdim} equals two).

Indeed the latter definition corresponds to saturating the BF bound in the
present approach, which in the vicinity of the IR fixed point means the start of
the tachyon oscillations, matching with  our definition of $\l_c$. Similarly as
in our model, $\l_c=\l_*$ at the edge of the conformal window. A similar
description of the conformal transition was also found in the holographic model
of \cite{kutasov}.

\begin{figure}
\centering
\includegraphics[width=0.29\textwidth]{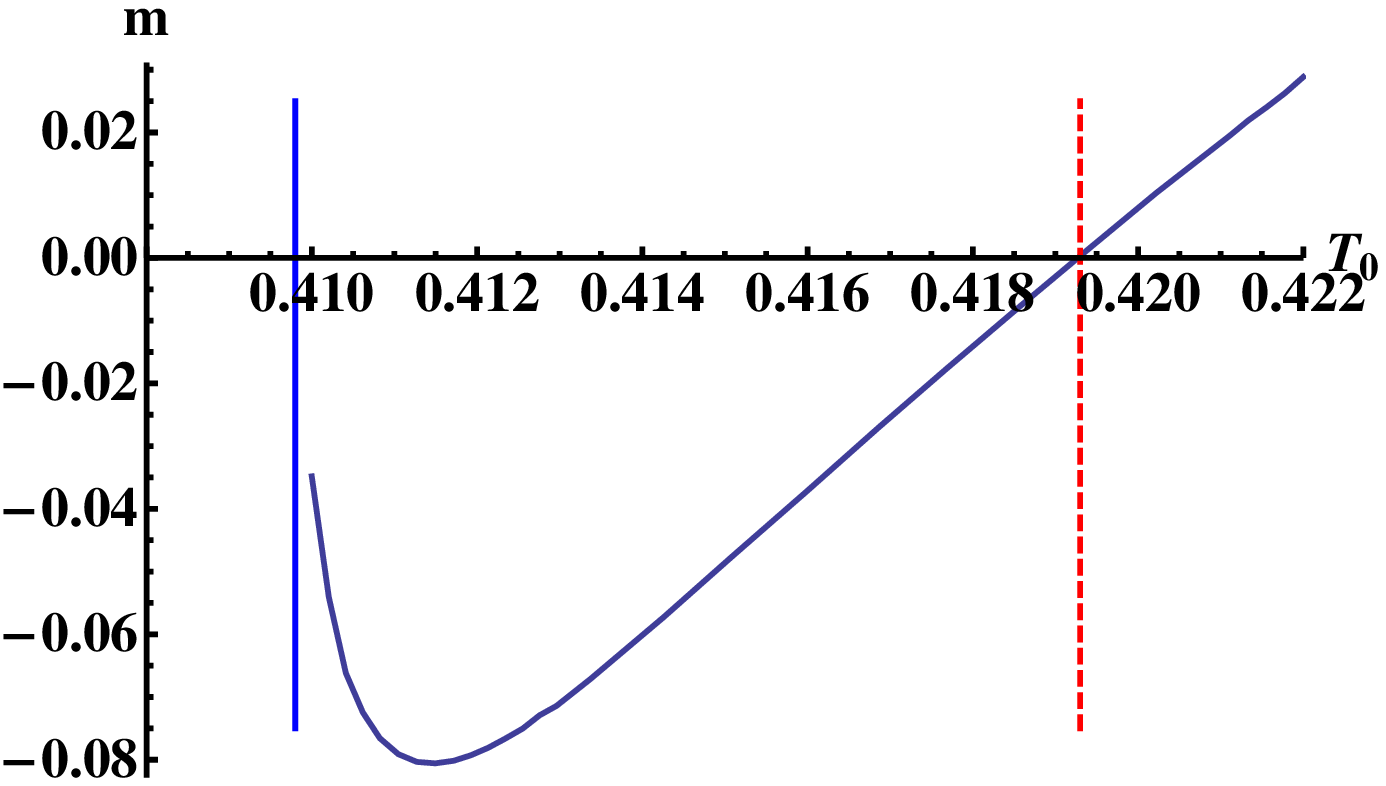}%
\hspace{3mm}
\includegraphics[width=0.31\textwidth]{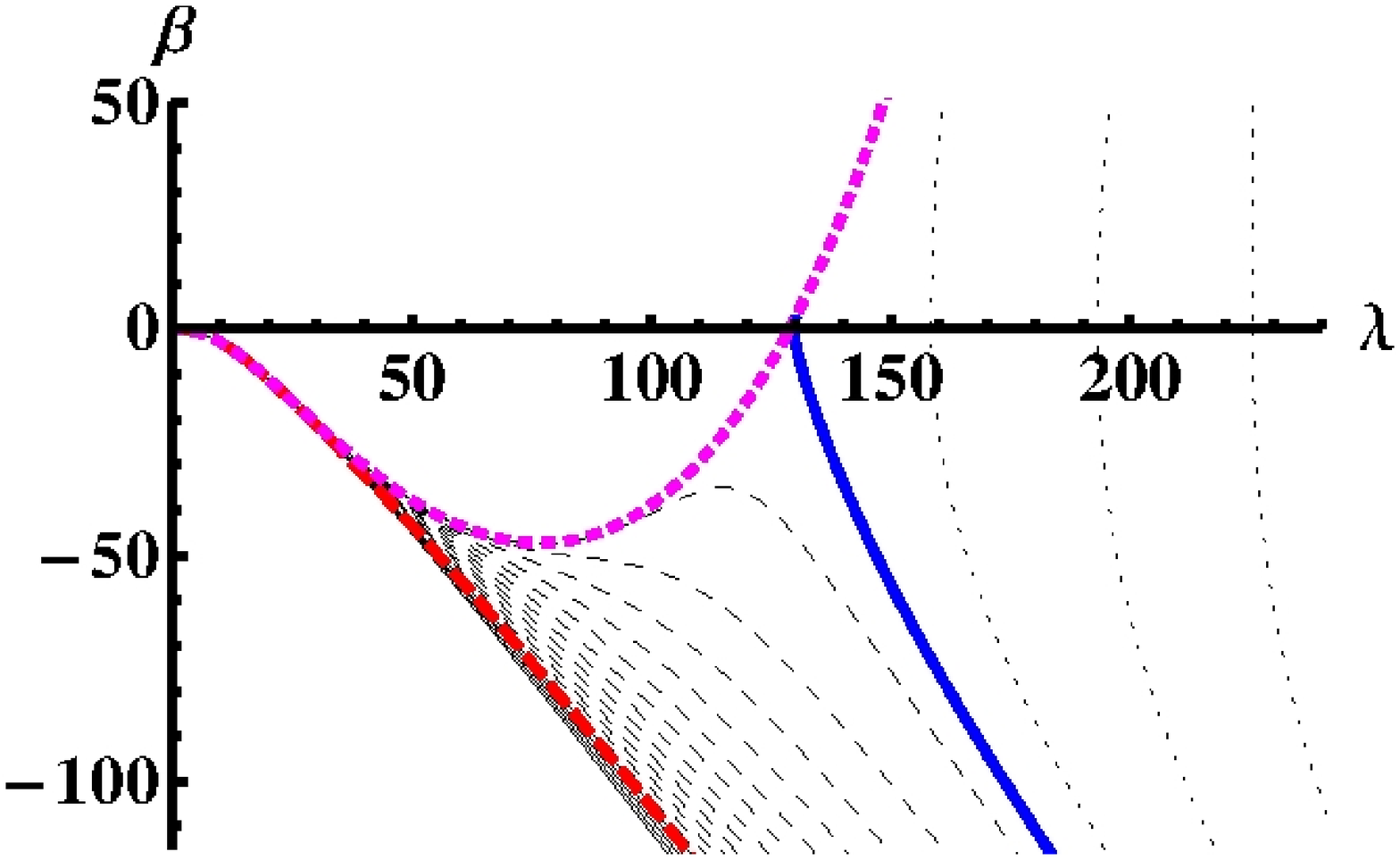}%
\hspace{3mm}
\includegraphics[width=0.31\textwidth]{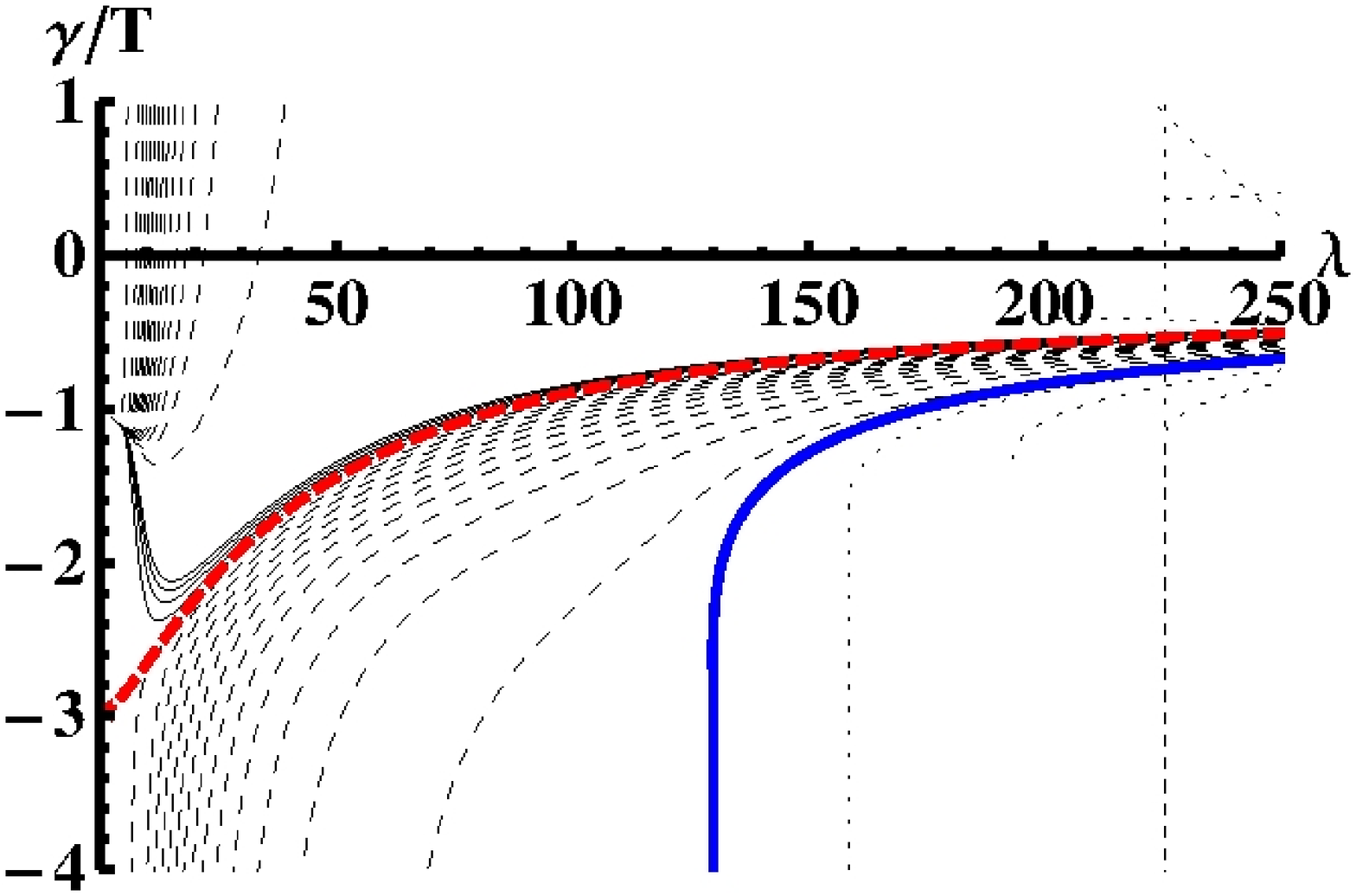}

\vspace{4mm}

\includegraphics[width=0.29\textwidth]{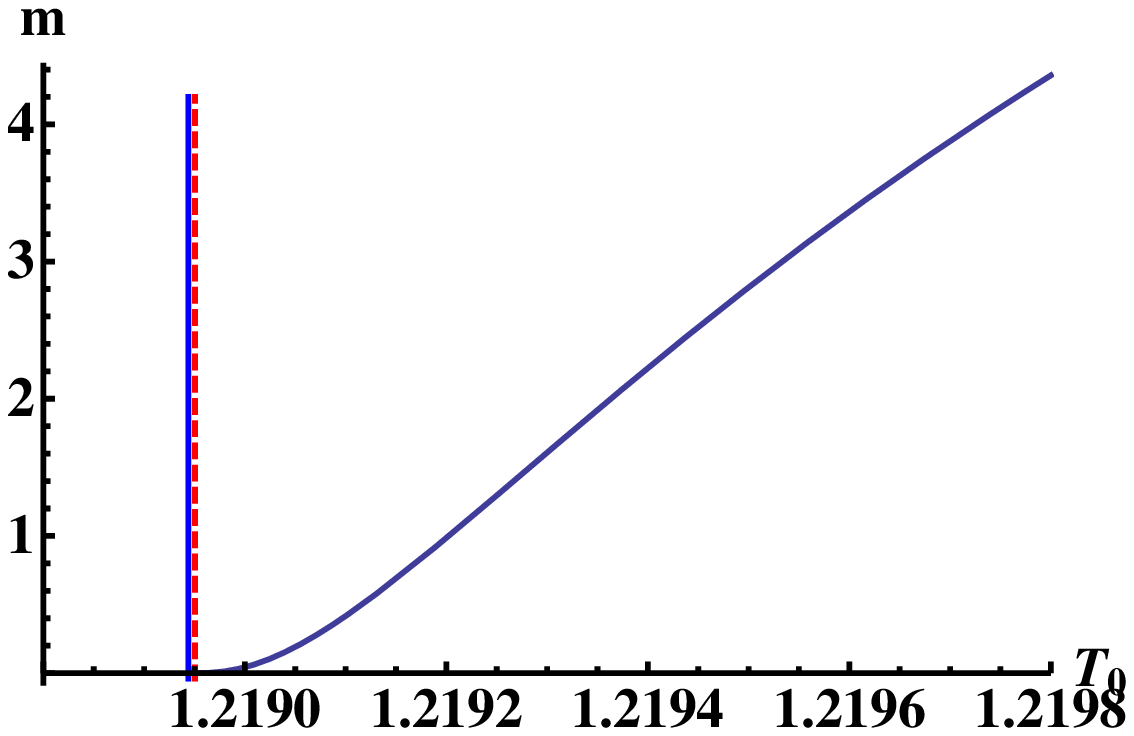}%
\hspace{3mm}
\includegraphics[width=0.31\textwidth]{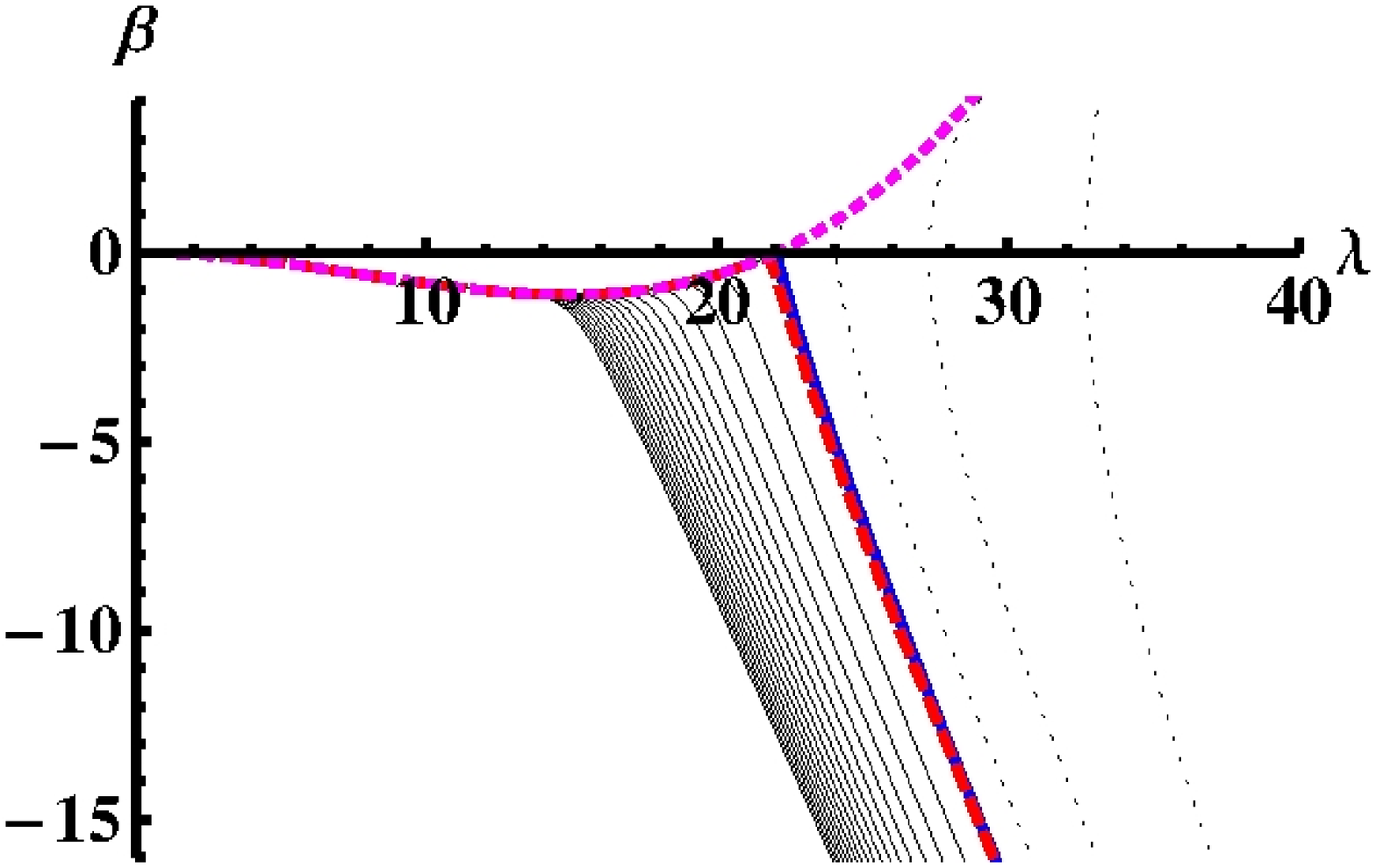}%
\hspace{3mm}
\includegraphics[width=0.31\textwidth]{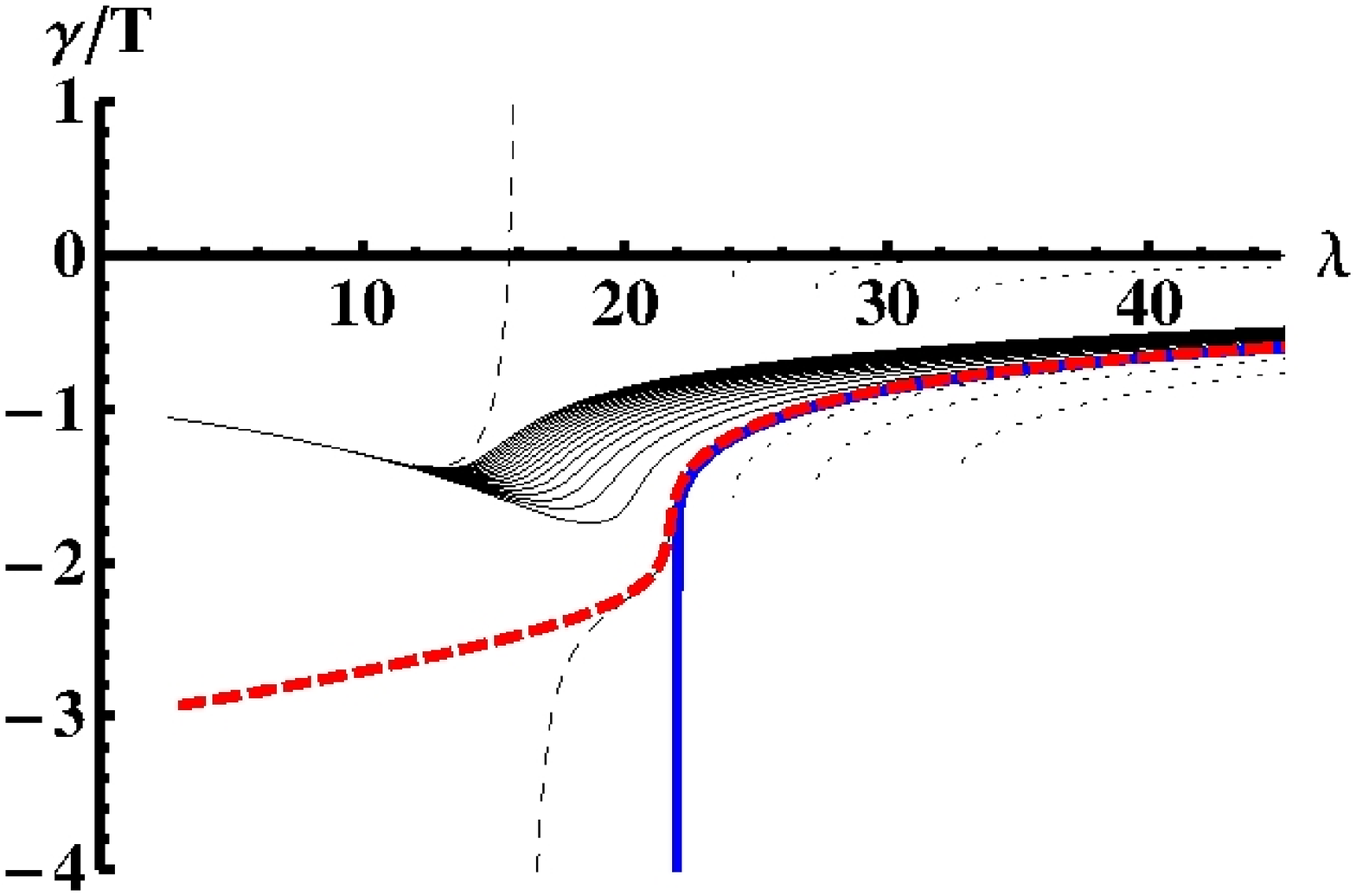}

\vspace{4mm}

\includegraphics[width=0.29\textwidth]{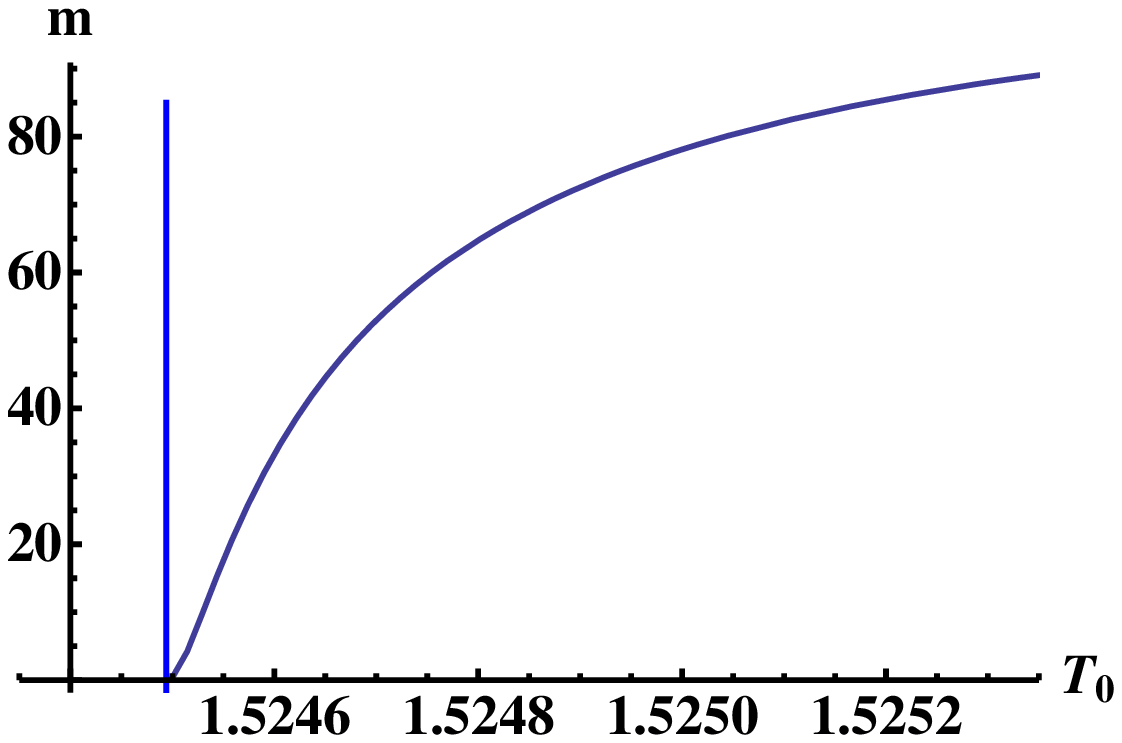}%
\hspace{3mm}
\includegraphics[width=0.31\textwidth]{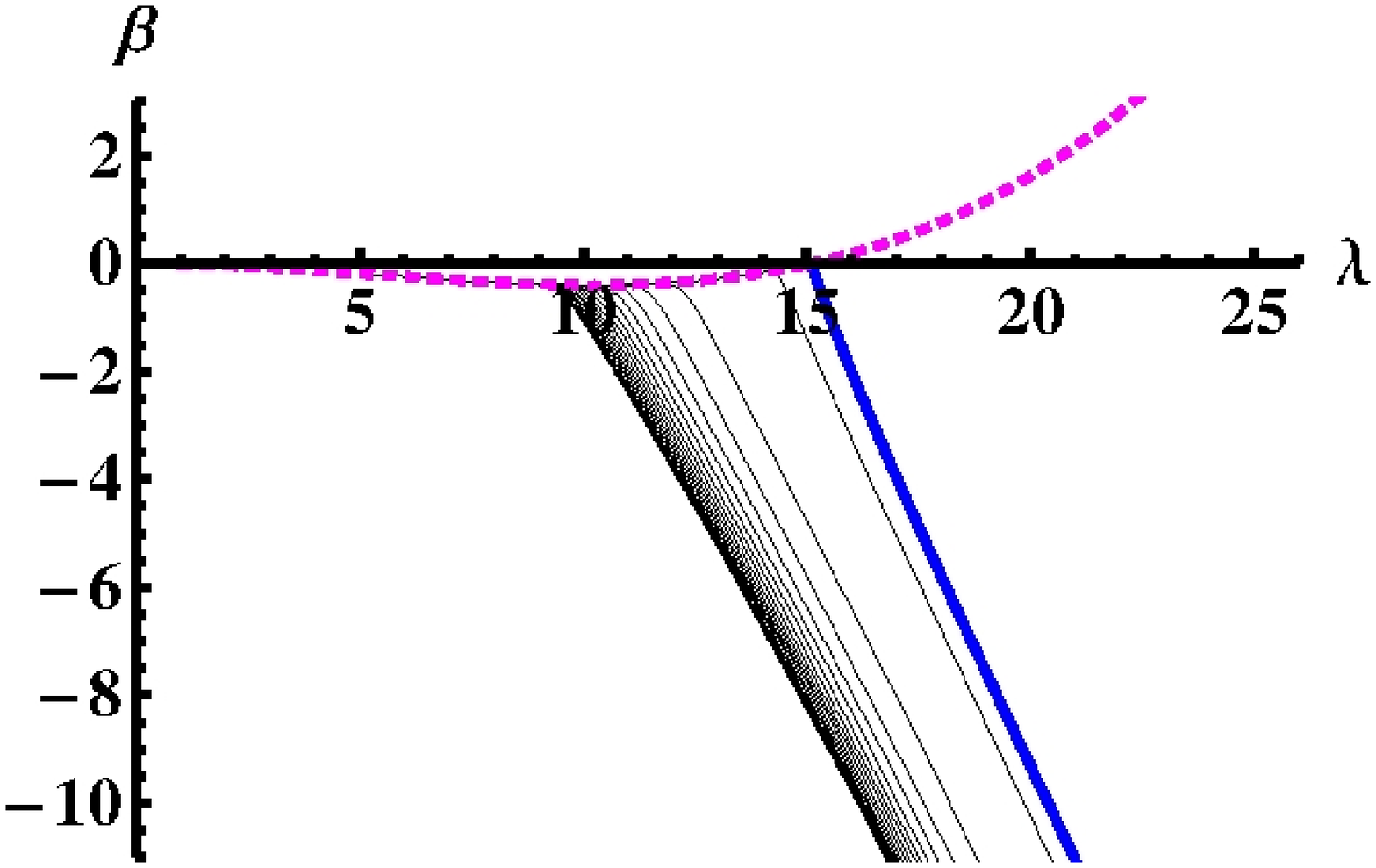}%
\hspace{3mm}
\includegraphics[width=0.31\textwidth]{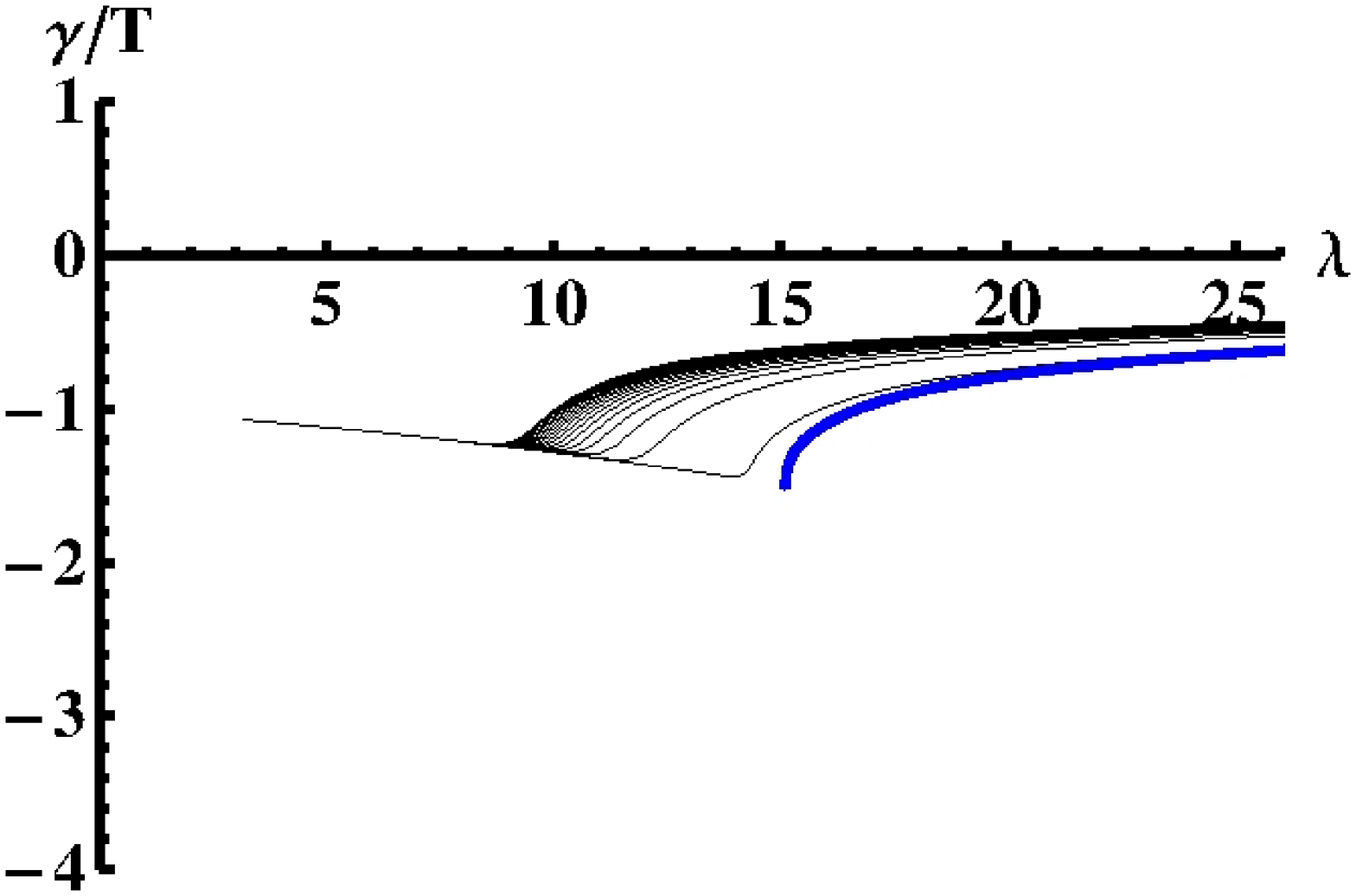}

\caption{The dependence of the background on $T_0$ near the red and blue lines
of Fig.~\protect\ref{figuvbeh}. Top row: $x=2$. Middle row: $x=3.9$. Bottom row:
$x=4.25$. Left column: the quark mass as a function of $T_0$. The vertical blue
solid and dashed red lines mark the solutions terminating at an UV fixed point,
and having $m=0$, respectively, as in Fig.~\protect\ref{figuvbeh}. Middle column:
$\beta$-functions $d\l/dA$ as $T_0$ is varied over the range of the mass plots
(left) with constant steps, shown as thin black curves. Dotted, dashed and solid
curves are $\beta$-functions for backgrounds with a bounce back towards the IR,
$m<0$, and $m>0$, respectively. The limiting cases between these behaviors are
the given by the thick blue solid curves, which terminate at an UV fixed point,
and the thick red dashed ones, which have $m=0$. The magenta dotted curve is the
$\beta$-function when the tachyon is completely decoupled, solved from
Eq.~\protect\eqref{Xde}. Right column: gamma functions  $1/T\times dT/dA$ as $T_0$ is
varied. The lines are marked as for the $\beta$-functions.}
\label{figxscan}
\end{figure}

\subsection{Numerical analysis}

We illustrate the  analysis above by studying the background numerically near
the red dashed and blue solid curves of Fig.~\ref{figuvbeh}.

We choose as reference values $x = 2$, $3.9$ and $4.25$, which have
qualitatively different behavior. For zero mass,  they correspond to field
theories with running, walking and IR conformal behavior of the coupling
constant. Fig.~\ref{figxscan} shows, for the above values of $x$, the quarks
mass in IR units (left column), the variation of the $\beta$-function (middle
column), and the variation of the gamma function (right column) as we scan over
a range of $T_0$ which includes the blue and red dashed curves of
Fig.~\ref{figuvbeh}.

For $x=2$ (top row) the quark mass (left column) is negative for a wide range of
$T_0$ between the zero mass solution (dashed red vertical line) and the critical
value of $T_0$ where the UV behavior of the solution changes such that the mass
is no longer defined (solid blue vertical line). The mass curve oscillates as we
approach the vertical blue line of critical $T_0$ as depicted in Fig.~\ref{mT0},
but the oscillations lie too close to line to be resolved at the used scale of
$T_0$. As $x$ is increased to $3.9$ (middle row), the zero mass solution is
driven very close to the critical $T_0$. For $x=4.25$ (bottom row), the
solutions with negative quark mass, and in particular the one with zero mass,
have disappeared.

In the the middle column we plot as the thin black curves the $\beta$-function
as $T_0$ is varied over the range of the left hand plots with a constant step
size. The blue thick curve is the limiting $\beta$-function that always
terminates at an UV fixed point and corresponds to the blue curve of
Fig.~\ref{figuvbeh}.
For $x=2$ all $\beta$-functions with positive quark mass (black thin solid
curves) are running, including the zero mass solution (red dashed thick curve),
whereas all walking solutions have negative quark mass (thin dashed
curves)\footnote{As mentioned above, there are also positive quark mass
solutions which basically overlap with the thick blue curve. Our resolution is
not enough to resolve these.}. The running curves are the lowest ones and all of
them basically overlap. The dotted thin lines are the $\beta$-functions with
bounce-back behavior in the UV, which occur for values of $T_0$ that are smaller
than the critical one. Increasing $x$ to $3.9$ (middle row), the zero mass
solution moves into the walking region, and approaches the critical one marked
with thick blue curve. As $x$ is further increased (bottom row), the zero mass
solution disappears by joining the blue curve.

Similar, but slightly more complicated behavior is seen for the gamma functions
(right column). The dashing and coloring have the same meaning as for the
$\beta$-functions. Notice that $\gamma/T$ asymptotes to $-3$ as $\l \to 0$ for
the zero mass solution (red thick dashed curve) whereas for the solutions with a
finite mass it asymptotes to $-1$, as expected. The solutions with negative
quark mass (thin dashed curves) have a zero of the tachyon, which shows up as a
pole in $\gamma/T$. The main change in the plots as $x$ increases from $2$ (top)
to $4.25$ (bottom) is the movement of the solution with the UV fixed point (blue
thick curve) towards smaller $\l$, as all solutions it ``crosses'' change
drastically. Otherwise the gamma functions (i.e., those solutions left of the
blue curve) are roughly independent of $x$.

Finally, we comment on the $m \to 0$ limit and discuss also
Fig.~\ref{figscalesep} in this same limit. For $x<x_c$ the background converges
smoothly to the $m=0$ one with chiral symmetry breaking in this limit, as also
seen from the $\beta$- and $\gamma$-functions on Fig.~\ref{figxscan} (top and
middle rows). Taking $m \to 0$ for $x<x_c$ in Fig.~\ref{figscalesep} defines the
enveloping curve of the fixed mass curves, which diverges for $x=x_c$. (We shall
discuss the $m=0$ case in more detail in Sec.~\ref{secBKT} below.) For $x>x_c$,
the scale ratio $\Lambda_\mathrm{UV}/\Lambda_\mathrm{IR}$ diverges for $m \to 0$
(see Fig.~\ref{figscalesep}), and we can consider two different limits. If we
keep $\Lambda_\mathrm{UV}$ fixed as $m \to 0$, the IR scale is driven to
infinity and the background converges pointwise to the one with identically
vanishing tachyon and an IR fixed point, discussed in Sec.~\ref{SecBGT0}. If we
instead keep the IR scale fixed, the background converges towards the one having
an UV fixed point of Sec.~\ref{AppUVFP} (see the $\beta$- and $\gamma$-functions
shown with the thick blue line on the bottom row of Fig.~\ref{figxscan}). Notice
that the backgrounds with vanishing tachyon and $x<x_c$ are not connected to the
backgrounds having a finite quark mass by any limiting procedure, which is in
line with our expectation that they are unphysical. Indeed we shall show in
Section~\ref{SecFE} that these solutions have larger free energy than the ones
with nontrivial tachyon and chiral symmetry breaking.

\section{Extracting UV coefficients from numerical solutions} \label{AppExtr}

\subsection{Extracting free energy differences} \label{AppExtrDE}

As pointed out in Section~\ref{SecBG}, we need to check numerically which one of
the two solutions with vanishing quark mass, the one with chiral symmetry
breaking or without, minimizes the free energy.
In order to do this we need to extract $\hat A$ (for fixed $\Lambda$) defined in
Eq.~\eqref{Alexps2} for both solutions and then calculate the energy difference
through Eq.~\eqref{DEres}.

The corrections involving $\hat A$ are highly suppressed $\sim r^4$ in the
region which is under perturbative control ($\log(r\Lambda)$ is small).
Extracting $\hat A$ directly from the numerical solutions of $A$ and $\l$ is
practically impossible, since it is difficult to require the two solutions to
have the same $\Lambda$ to a high enough precision. Therefore we study
variations in $X$, which is invariant in scalings of $r$. In principle we could
match the numerically extracted variation of $X$ to the correction term in
Eq.~(\ref{Xexpres}). This is doable (except for values of $x$ very close to the
critical one $x_c \simeq 3.9959$), but a large uncertainty in the value of $\hat
A$ still remains. Therefore we proceed as follows. We substitute $X_0+X_1$ in
the equation (\ref{Xde}), where $X_1$ is treated as a small perturbation. From
the linearized equation, we can solve $X_1$ exactly:
\be \label{X1exact}
 X_1(\l) = X_C \exp\left[\int _1^{\l }\frac{16 V(\hat \l) X_0(\hat \l)^3/\hat
\l+3 V'(\hat \l)+3 X_0(\hat \l)^2 V'(\hat \l)}{6 V(\hat \l) X_0(\hat \l)^2}
d\hat \l \right]
\ee
where $X_C$ is a constant. So, if $X_0$ is known, this equation gives $X_1$ in
Eq.~(\ref{Xexpdef}) to all orders in $\log (r \Lambda)$. Since the UV expansion
of $X_0$ is also known, we can use that to expand $X_1$ and to calculate the
relation between the constants $\hat A$ and $X_C$ by using Eq.~\eqref{X1UVexp}.
A straightforward calculation gives
\bea \label{hatACXrel}
 \hat A &=& \frac{3 V_1 \Lambda^4 X_C}{20}\exp\left[ - \frac{\log(9 V_1/8) (23
V_1^2 - 64 V_2)}{9 V_1^2}\right] \\
&&\times  \exp\Bigg\{ -\int _0^{1}\bigg[\frac{16 V(\hat \l) X_0(\hat \l)^3/\hat
\l+3 V'(\hat \l)+3 X_0(\hat \l)^2 V'(\hat \l)}{6 V(\hat \l) X_0(\hat \l)^2}
\nn\\\nn
&& -\left(\frac{32}{9 V_1 \hat \l^2}+\frac{14}{9 \hat \l}-\frac{64V_2}{9 V_1^2
\hat \l}\right)\bigg] d\hat \l  +\frac{32}{9 V_1} \Bigg\} \ .
\eea

To obtain the free energy difference between two solutions, we calculate
numerically the variation of $X$ and match with Eq.~(\ref{X1exact}), where we
use either of the two solutions to evaluate $X_0$ and the integral numerically.
Since $X_1$ given by Eq.~(\ref{X1exact}) is a good approximation already at
small $r$ ($\log(r \Lambda)$ does not need to be small) we can  obtain the
difference $\Delta X_C$ between the solutions to a good precision. The value is
then used to calculate numerically $\Delta \hat A$ and $\Delta {\cal E}$ through
equations~(\ref{DEres}) and~(\ref{hatACXrel}).

\subsection{Extracting the chiral condensate at $m=0$} \label{AppExtrsigma}

Let us then discuss how the value of the chiral condensate can be extracted from
a numerical solution to the differential equations.
Recall that our method for constructing the backgrounds requires tuning the
normalization of the tachyon in the IR ($T_0$) such that the quark mass
vanishes. This procedure is however limited by the numerical precision of the
solution, and therefore exactly zero quark mass cannot be obtained. As it turns
out, this makes direct extraction of the condensate value difficult, since the
approximately linear term $\sim m r$ of the tachyon dominates over the cubic
solution $\sim\sigma r^3$ in the deep UV. This is particularly problematic as
$x$ approaches $x_c$, since the ratio of the UV and IR energy scales grows,
pushing the asymptotic UV region to smaller $r$ while the condensate value
$\propto \sigma$ decreases.

Therefore, we employ a subtraction procedure. In the UV, the tachyon solution
(with the mass $m_1$ as small as can be achieved) can be written as
\be
 T_1(r) = m_1 T_m(r) +\sigma_1 T_\sigma(r) \ .
\ee
The UV expansions of the two (approximately) linearly independent solutions $T_m
\simeq \ell r(-\log( \Lambda r))^C$ and $T_\sigma(r) \simeq \ell r^3(-\log(
\Lambda r))^{-C}$ are given in Appendix~\ref{TUV}. We will remove the remaining
mass term by subtracting another solution
\be
 T_2(r) = m_2 T_m(r)+ \sigma_2 T_\sigma(r)
\ee
which is calculated at the same value of $x$ but for different $T_0$ such that
$m_2$ is about the same order as $m_1$. Since the mass terms dominate at very
small $r$, it is easy to extract $m_2/m_1$ to a high precision, and
construct\footnote{There is a subtlety here: the UV scales $\Lambda$ which
appear in Eq.~\eqref{TUVres} will be different for the two solutions. This needs
to be fixed, e.g., by scaling one of the solutions such that the corresponding
solutions for $\l$ match as $\l \to 0$.}
\be
 T_\delta(r) =  T_1(r)-\frac{m_1}{m_2}T_2(r) =\left( \sigma_1 - \frac{m_1
\sigma_2}{m_2}\right) T_\sigma(r) \ .
\ee
If $\sigma$ is analytic at $m=0$,
\be
 \sigma_i = \sigma^{(0)} + \sigma^{(1)} m_i +  \sigma^{(2)} m_i^2 + \cdots \ ,
\ee
we obtain
\be
T_\delta(r) =\left[ \sigma^{(0)} \left(1 - \frac{m_1}{m_2}\right) +
\sigma^{(2)} m_1(m_1-m_2)+\cdots  \right] T_\sigma(r) \ .
\ee
If the masses $m_i$ are sufficiently small,  the quadratic correction
$\propto\sigma^{(2)}$ as well as higher order terms can be neglected, and the
difference $T_\delta$ is proportional to $\sigma^{(0)}$ to a good precision. The
reliability of the subtraction procedure can be tested by varying, say, $m_2$
and checking that this does not affect the result.

Now $\sigma^{(0)}$ can extracted from $T_\delta$ in a straightforward manner.
However the result still has quite large error bars in particular for $x$ close
to $x_c$, because of numerical uncertainty in the deep UV. We do an additional
trick to remove this problem. The extracted  $T_\delta$ extends close enough to
the UV singularity to ensure that the tachyon is decoupled from $A$ and $\l$. On
the other hand, we can obtain a decoupled tachyon solution with zero quark mass
by solving first the background functions $A$ and $\l$ with $T \equiv 0$,
inserting these in the tachyon EoM, and solving that by shooting from the UV. We
can basically extend this solution as close to the UV singularity as we want.
Matching the two solutions in the region where both are reliable, we can extend
$T_\delta$ easily up to $\sim r^{-100}$ in the UV.

Finally, to extract $\sigma^{(0)}$  we consider the expansion of the tachyon at
small $\l$:
\bea
 \log T_\delta(\l)-\log \left(1 - \frac{m_1}{m_2}\right) -\log\ell  \simeq \log
\sigma^{(0)} + \log T_\sigma(\l) -\log\ell  &&\\\nn
= \log \frac{\sigma^{(0)}}{\Lambda^3} - \frac{8}{3 V_1 \l} +
\left(\frac{4}{3}\frac{h_1}{V_1} + \frac{39 V_1^2-64 V_2}{12 V_1^2}\right)
\log\frac{9 V_1\l}{8} + \cdots &&\ ,
\eea
which is obtained by using Eqs.~\eqref{TUVres} and \eqref{UVexpsapp}. The
coefficients $V_i$ and $h_1$ were defined in Eqs.~\eqref{Vhexps}, and
$\Lambda=\Lambda_\mathrm{UV}$ is the scale of the UV expansions. We match this
expansion with the extended  $T_\delta$, and extrapolate to $\l = 0$ to obtain
$\log(\sigma^{(0)}/\Lambda^3)$.

\section{Details on BKT scaling} \label{AppBKT}

In this Appendix we discuss some technical details on the BKT scaling.
First, there is a subtlety with matching the approximation~\eqref{Tirfpfin} with
the tachyon solution in the IR (for which we were unable to write analytic
approximations). Since the approximation contains a periodic function, the
boundary conditions at $r \simeq r_\mathrm{IR}$ will not fix the tachyon
uniquely: one can let the tachyon oscillate first, and do the matching only
after the oscillations have created $n$ zeroes, where $n=0,1,2,\ldots$.
Therefore, in Eq.~\eqref{sinargmatch} we should actually have (including the
solutions where the tachyon changes sign)
\be \label{logreq}
 \sqrt{\kappa (\l_*-\l_c)} \log\frac{r_\mathrm{IR}}{\hat r} \sim \pi n + {\cal
O}(1) \ .
\ee
An analogous result was found in \cite{kutasov} in a case that was analytically
more tractable. As in their case, we expect that the solution with no nodes
($n=0$) is the one that minimizes the free energy, as we already verified
numerically in Section~\ref{SecFE}.

One can actually do the matching procedure even more precisely, and in
particular derive results for $K$ and $\hat K$.
We need to study the approximation \eqref{Tirfpfin} more closely: actually it
can be matched with the other solutions only in the vicinity of the zeroes of
the sine function. At $r=r_\mathrm{IR}$ we have
\bea
 T(r_\mathrm{IR}) &\simeq& C_\mathrm{fp}
\left(\frac{r_\mathrm{IR}}{r_\mathrm{UV}}\right)^2  \sin\left(\sqrt{\kappa
(\l_*-\l_c)} \log\frac{r_\mathrm{IR}}{\hat r} + \hat  \phi\right) \\
 T'(r_\mathrm{IR}) &\simeq&  C_\mathrm{fp} \frac{r_\mathrm{IR}}{r_\mathrm{UV}^2}
\Big[2 \sin\left(\sqrt{\kappa (\l_*-\l_c)} \log\frac{r_\mathrm{IR}}{\hat r} +
\hat  \phi\right) \nn\\\nn
&&+\sqrt{\kappa (\l_*-\l_c)} \cos\left(\sqrt{\kappa (\l_*-\l_c)}
\log\frac{r_\mathrm{IR}}{\hat r} + \hat  \phi\right) \Big]
\eea
It appears that the cosine factor in the derivative can be neglected due to the
small factor $\sqrt{\kappa (\l_*-\l_c)}$ multiplying it. However, in this case
the ratio $T(r_\mathrm{IR})/T'(r_\mathrm{IR}) = 1/2r_\mathrm{IR}$ is fixed so we
cannot match both the function and its derivative with the (nontrivial) IR part
of the tachyon solution. Therefore, we indeed need to be close to a zero of the
sine,
\be
 \sin\left(\sqrt{\kappa (\l_*-\l_c)} \log\frac{r_\mathrm{IR}}{\hat r} + \hat
\phi\right) = \mathcal{O}(\sqrt{\l_*-\l_c}) \ ,
\ee
which brings in the required dependence of the solution on both $C_\mathrm{fp}$
and $\phi$.

Matching the solution \eqref{Tirfpfin} towards the UV is only possible close to
its node, as well\footnote{Proving this turns out to be more tricky than in the
IR, since the solutions \eqref{Tintermapp} and \eqref{Tirfpfin}  have the same
power behavior, and therefore they apparently join smoothly for large values of
the sine function. However, this is a fake effect due to the roughness of the
approximation \eqref{Tintermapp}: we have seen that in the UV the power law of
the tachyon actually changes from $2$ to $3$ within a range of $\log r$ which is
small with respect to $1/\sqrt{\l_*-\l_c}$. Therefore the same argument as in
the IR applies, and the sine function needs to be
$\mathcal{O}(\sqrt{\l_*-\l_c})$ in order the matching to work.}. This gives
\be
 \hat \phi = \mathcal{O}(\sqrt{\l_*-\l_c}) \ .
\ee
Consequently, the argument of the sine in the approximation~\eqref{Tirfpfin}
must change by a positive multiple of $\pi$ as we move from the UV to the IR, so
that Eq.~\eqref{logreq} is actually written as
\be \label{logreq2}
 \sqrt{\kappa (\l_*-\l_c)} \log\frac{r_\mathrm{IR}}{\hat r} \simeq \pi (n+1) \ .
\ee
where $n$ is the number of the zeroes of the tachyon solution\footnote{Our
arguments do not exclude zeroes of the tachyon solution near the endpoints of
the validity of the approximation~\eqref{Tirfpfin}, i.e., near $r=\hat r$ and
$r=r_\mathrm{IR}$. However it is reasonable to expect that  the approximation
joins the UV and IR solutions in a smooth manner, so that no extra nodes appear,
which can also be verified numerically.}.
Notice that Eqs.~\eqref{CfpUV} and~\eqref{Cfpfix} also receive extra factors
from the refined scaling argument. We have now
\be
  C_\mathrm{fp} \sim \sigma r_\mathrm{UV}^3  \frac{1}{\sqrt{\l_*-\l_c}} \sim
\left(\frac{r_\mathrm{UV}}{r_\mathrm{IR}}\right)^2 \frac{1}{\sqrt{\l_*-\l_c}}
\ee
instead of Eq.~\eqref{Cfpfix}. The additional square root factors cancel in the
result for $\sigma$, and they could in any case be neglected as a subleading
correction to the exponential scaling.

In order to write down the results for $K$ and $\hat K$ we recall that the
definitions of $\l_c(x)$ and $x_c$ read
\be
 G(\l_c(x),x) = 4 \ ; \qquad G(\l_*(x_c),x_c)= 4 \ ,
\ee
respectively, where $\l_*(x)$ was defined by $V_g'(\l_*(x))-x
V_{f0}'(\l_*(x))=0$. Near the critical point we have the two expansions
\be
 G(\l_*,x) \simeq 4 + \frac{\p }{\p \l}G(\l_c,x)(\l_*-\l_c) \simeq 4 +
\frac{d}{dx}G(\l_*(x),x)\big|_{x=x_c}(x-x_c)
\ee
where $\kappa = \frac{\p }{\p \l}G(\l_c,x)$.
By using Eq.~\eqref{logreq2} we finally see that
\be
 \frac{r_\mathrm{IR}}{r_\mathrm{UV}} \sim
\exp\left[\frac{K(n+1)}{\sqrt{\l_*-\l_c}}\right] \sim\exp\left[\frac{\hat
K(n+1)}{\sqrt{x_c-x}}\right]
\ee
where
\bea \label{Kres}
 K = \frac{\pi}{\sqrt{\kappa}} = \frac{\pi}{\sqrt{ \frac{\p }{\p \l}G(\l_c,x)}}
\ ; \qquad \hat K = \frac{\pi}{\sqrt{- \frac{d }{d x}G(\l_*(x),x)\big|_{x=x_c}}}
\ ,
\eea
and setting $n=0$ gives the result for the solution which has no tachyon zeroes
and lowest free energy.

Finally, let us notice the free energy scaling result for the solutions having
several tachyon zeroes. Following the arguments of Sec.~\eqref{SecFEBKT}, we see
that the energy difference between the solution with vanishing quark mass and
the solution with $n$ tachyon zeroes scales as
\be
 \frac{\Delta \mathcal{E}}{M^3N^2 V_4} \sim
r_\mathrm{UV}^{-4}\exp\left[-\frac{4K(n+1)}{\sqrt{\l_*-\l_c}}\right] \sim
r_\mathrm{UV}^{-4}\exp\left[-\frac{4 \hat K(n+1)}{\sqrt{x_c-x}}\right] \ .
\ee
Since we have numerically checked that $\Delta \mathcal{E}$ is positive, this
result verifies that the $n=0$ solution has the lowest free energy for $x \to
x_c$.

\end{document}